\documentclass[preprint,superscriptaddress,amsmath,amssymb,prd,aps,showpacs,floatfix,nofootinbib]{revtex4-1}
\usepackage{graphicx}
\usepackage{dcolumn}
\usepackage{bm}
\usepackage{mathrsfs}
\usepackage{hyperref}
\usepackage{amsmath}
\usepackage{amssymb}
\usepackage{float}
\usepackage{dcolumn}
\usepackage{multirow}

\begin{document}


\title{Masses of doubly heavy baryons in the Bethe-Salpeter equation approach}
\author{Qi-Xin Yu}
\email[E-mail: ]{yuqx@mail.bnu.edu.cn}
\affiliation{College of Nuclear Science and Technology,
Beijing Normal University, Beijing 100875, China}

\author{Xin-Heng Guo}
\email[Corresponding author, e-mail: ]{xhguo@bnu.edu.cn}
\affiliation{College of Nuclear Science and Technology,
Beijing Normal University, Beijing 100875, China}


\begin{abstract}
A doubly heavy baryon can be regarded as composed of a heavy diquark and a light quark. In this picture, we study the masses of the doubly heavy diquarkes in the Bethe-Salpeter (BS) formalism first, which are then used as one of the inputs in studying the masses of the doubly heavy baryons in the quark-diquark model. We establish the BS equations for both the heavy diquarks and the heavy baryons with and without taking the heavy quark limit, respectively. These equations are solved numerically with the kernel containing the scalar confinement and one-gluon-exchange terms. The mass of the doubly charmed baryon $\Xi_{cc}^{(\ast)}$ is obtained in both approaches, $3.60\sim3.65\,\rm GeV$ ($\Xi_{cc}^{(\ast)}$) under the heavy quark limit, $3.53\sim3.56\,\rm GeV$ for $\Xi_{cc}$ and $3.61\sim3.63\,\rm GeV$ for $\Xi_{cc}^\ast$ without taking the heavy quark limit. The masses of $\Xi_{bc}^\prime$, $\Xi_{bc}^{(\ast)}$, $\Xi_{bb}^{(\ast)}$, $\Omega_{cc}^{(\ast)}$, $\Omega_{bc}^\prime$, $\Omega_{bc}^{(\ast)}$ and $\Omega_{bb}^{(\ast)}$ are also predicted in the same way. We find that the corrections to the results in the heavy quark limit are about $-0.02\,\rm GeV\sim-0.11\,\rm GeV$ for the masses of the doubly heavy baryons.
\end{abstract}

\maketitle


\section{Introduction}
In the last few decades, heavy baryons which contain one heavy quark have been studied in many references, both theoretically and experimentally \cite{Ebert:1995fp,Mannel:1990vg,Isgur:1990pm,Roberts:2007ni,Albrecht:1992xa,Frabetti:1992bm,Guo:1996jj}. Whether the theoretical models for such baryons also work for baryons containing two or more heavy quarks needs to be investigated  \cite{Roberts:2007ni,Brambilla:2005yk,Ebert:2002ig,Kiselev:2001fw,Lewis:2001iz}. $\Xi^{++}_{cc}$ is the first doubly heavy baryon observed in experiments \cite{Aaij:2017ueg}. Although the SELEX Collaboration did report the observation of $\Xi^{+}_{cc}$ at a mass of $3519\pm2$ MeV years ago \cite{Mattson:2002vu,Ocherashvili:2004hi}, other collaborations like FOCUS \cite{Ratti:2003ez}, BaBar \cite{Aubert:2006qw} and Belle \cite{Chistov:2006zj}, on the other hand, failed to observe such a state with the properties SELEX observed. Recently, LHCb confirmed the existence of the doubly heavy baryon $\Xi^{++}_{cc}$, and the mass was measured as $3621.40\pm0.72(stat)\pm0.27(syst)\pm0.14(\Lambda^+_c)$ MeV \cite{Aaij:2017ueg}.

It is expected that more doubly heavy baryons will be observed in the further, hence theoretical studies of doubly heavy baryons are urgent. In this paper, we will study the masses of the doubly heavy baryons systematically. Since baryons consist of three quarks they are much more complicated than mesons. Because $bb$, $bc$ and $cc$\footnote{Another reason is that $cc$ have a relatively small radius, which makes it more appropriate to consider $cc$ as a diquark in the structure of doubly charmed baryon.} have good spin and isospin quantum numbers, we will regard them as diquarks in the baryon, where it can be treated as a single subatomic particle with which the third quark interacts via strong interaction. Using this quark-diquark model can greatly simplify the study of these baryons which are two-body bound states \cite{Lichtenberg:1982jp}.

The bound state equation we employ in our work is the Bethe-Salpeter (BS) equation, which has been extensively used to investigate the properties of heavy mesons and baryons in recent years \cite{Guo:1996jj,Guo:1998ef,Guo:2007qu,Weng:2010rb,Wang:2017rjs,Liu:2016wzh,Liu:2015qfa,Guo:2007mm,Guo:1998yj,Dai:1993np,Jain:1993qh,Dai:1993kt,Dai:1993kt}. The kernel for the BS equation is motivated by the potential model, which includes the scalar confinement and one-gluon-exchange terms in the form of the covariant instantaneous approximation \cite{Dai:1993kt,Dai:1993np,Guo:1996jj,Guo:2007mm,Guo:2007mm,Eichten:1978tg}, and this approximation is appropriate since the energy exchange between constituents inside the heavy baryons is expected to be small when we use the constituent quark masses in the BS equation. Thus we employ this approximation in this work for the purpose of simplifying our calculations. It should be noted that solving the BS equation in the Minkowski space has been done before by other researchers \cite{Carbonell:2010zw,dePaula:2016oct}, and we will also try to accomplish that by modifying our model in our future work. We will establish BS equations for doubly heavy diquarks first, then the BS equations for doubly heavy baryons. The masses of heavy diquarks will be solved out first and then used as an input to obtain the masses of doubly heavy baryons.

The masses of heavy baryons are studied with different approaches \cite{Sun:2016wzh,DeRujula:1975qlm,Roncaglia:1995az,Ebert:1996ec,Kiselev:2001fw,He:2004px,Roberts:2007ni,Weng:2010rb}, and most of them are done under the heavy quark limit, which works well for the doubly bottom diquark and doubly bottom baryons since $m_b$ is much larger than the QCD scale, $\Lambda_{QCD}$. However, one may expect lager $1/m_c$ corrections for heavy diquarks and baryons including the charm quark. Therefore, we will also establish the BS equations for doubly heavy diquarks and doubly heavy baryons without taking the heavy quark limit. We will compare the results in the two approaches.

This paper is organized as follows, In Sec.~\uppercase\expandafter{\romannumeral2}, the BS equations for the heavy diquarks will be presented under the heavy quark limit. In Sec.~\uppercase\expandafter{\romannumeral3}, we will establish the BS equations for the ground-state doubly heavy baryons under the heavy quark limit. Then, we will present our study for the heavy scalar and axial-vector diquarks without using the heavy quark limit, the formalism of which will be presented in Sec.~\uppercase\expandafter{\romannumeral4}, After that, the formalism for the doubly heavy baryons without taking the heavy quark limit will be presented in Sec.~\uppercase\expandafter{\romannumeral5}. Then in Sec.~\uppercase\expandafter{\romannumeral6}, we will show our results derived from the former sections. Finally, the summary will be given in Sec. \uppercase\expandafter{\romannumeral7}.
\section{Doubly heavy diquarks under the heavy quark limit}
\subsection{The BS equations for doubly heavy diquarks}
In the following, we will present the BS equations for the scalar diquark composed of $c$ and $b$ quarks, and axial-vector diquarks composed of $cc$, $bb$ and $bc$. First, the BS wave function for the scalar diquark composed of two fermions can be written as follows:
\begin{eqnarray}\label{sdi}
\chi(x_1,x_2)_{\alpha\beta}&=&\varepsilon^{ijk}\langle 0|T\psi_1(x_1)^{i}_{\alpha} \psi_2(x_2)^{j}_{\beta}|P_D,k\rangle\nonumber\\
                           &=&e^{-iP_DX}\int\frac{d^4p}{(2\pi)^4}e^{-ipx}\chi_{P_D}(p)_{\alpha\beta},
\end{eqnarray}
where $\psi_1(x_1)$ and $\psi_2(x_2)$ are the quark fields, $\alpha$ and $\beta$ are spinor indices, $i$, $j$, $k$ are color indices, $P_D$ is the momentum of the diquark, $X=\lambda_1x_1+\lambda_2x_2$ ($\lambda_1=\frac{m_1}{m_1+m_2}$, $\lambda_2=\frac{m_2}{m_1+m_2}$, $m_{1(2)}$ is the mass of the first (second) heavy quark) is the center-of-mass coordinate of the diquark, $x=x_1-x_2$ is the relative coordinate and $p$ represents the relative momentum between the two heavy quarks.

Furthermore, the momenta of two heavy quarks can be written with $P_D$ and $p$ as $p_1=\lambda_1P_D+p$, $p_2=\lambda_2P_D-p$, respectively. Define $p_l=v\cdot p$ and $p_t=p-p_lv$, where $v$ is the velocity of the diquark. The BS equation for the scalar diquark has the following matrix form \cite{Guo:2007qu,Weng:2010rb}:
\begin{equation}\label{bs}
  \widetilde{\chi}^{T}_{P_D}(p)=S(p_2)\int\frac{d^4q}{(2\pi)^4}[-\gamma^\mu \widetilde{\chi}^{T}_{P_D}(q)\gamma_\mu K^{(1g)}(p_t-q_t)+ \widetilde{\chi}^{T}_{P_D}(q)K^{(cf)}(p_t-q_t)]S(-p_1),
\end{equation}
where $S(p_2)$ and $S(p_1)$ are the propagators for two heavy quarks, and $\widetilde{\chi}_{P_D}(p)=C\chi_{P_D}(p)$ is the conjugate form of $\chi_{P_D}(p)$, $K^{(1g)}$ and $K^{(cf)}$ denote the one-gluon-exchange and confinement terms, respectively, which have the following forms in the diquark picture:
\begin{equation}\label{k1}
  \widetilde K^{(1g)}(p_t-q_t)=-\frac{i8\pi}{3}\frac{\alpha_s}{(p_t-q_t)^2+\mu^2},
\end{equation}
\begin{equation}\label{k2}
  \widetilde K^{(cf)}(p_t-q_t)=\frac{i4\pi\kappa^\prime}{[(p_t-q_t)^2+\mu^2]^2}-(2\pi)^3\delta^3(p_t-q_t)\int\frac{d^3k}{(2\pi)^3}\frac{i4\pi\kappa^\prime}{[k^2+\mu^2]^2},
\end{equation}
where $\alpha_s$ and $\kappa^\prime$ are coupling parameters in the meson case, and $\mu$ is introduced to avoid the infrared divergence in later numerical calculations, the limit $\mu\rightarrow0$ will be taken in the end of our calculations. As for the forms of propagators, they can be constructed as follows in the leading order of $1/m_Q$ $(Q=b,c)$ expansion:
\begin{equation}\label{pro1}
  S(p_1)=i\frac{m_1(1-v\!\!\!/)}{2\omega_1(\lambda_1M_D+p_l-\omega_1-i\epsilon)},
\end{equation}
\begin{equation}\label{pro2}
  S(p_2)=i\frac{m_2(1+v\!\!\!/)}{2\omega_2(\lambda_2M_D-p_l-\omega_2+i\epsilon)},
\end{equation}
where $M_D$ is the mass of the scalar diquark and $\omega_{1(2)}=\sqrt{m^2_{1(2)}-p^2_t}$. It can be seen from Eqs.~\eqref{bs} and \eqref{pro2} that $\chi_{P_D}^T(p)$ satisfies the condition
\begin{equation}\label{heavycondition}
  v\!\!\!/\chi_{P_D}^T(p)=\chi_{P_D}^T(p).
\end{equation}
Then, after considering the relation in Eq.~\eqref{heavycondition} and other restrictions from Lorentz convariance and parity transformation on the form of $\widetilde\chi^T_{P_D}$, we obtain the following form for the BS wave function of the scalar diquark:
\begin{equation}\label{swave}
  \widetilde{\chi}^{T}_{P_D}(p) = (1+v\!\!\!/)\gamma^5g_1,
\end{equation}
where $g_1$ is a scalar function of $p_t^2$ and $p_l$. Defining $\widetilde V^{(1g)}=-i\widetilde K^{(1g)}$, $\widetilde V^{(cf)}=-i\widetilde K^{(cf)}$, $\widetilde g_{1}=\int\frac{dp_l}{2\pi}g_{1}$, and substituting Eqs.~\eqref{pro1}-\eqref{swave} into Eq.~\eqref{bs}, we obtain the following equation for the BS wave function of the scalar diquark under the heavy quark limit:
\begin{equation}\label{g1}
  \widetilde g_{1}(p_t)=-\frac{m_1m_2}{\omega_1\omega_2(M_D-\omega_1-\omega_2)}\int\frac{d^3q_t}{(2\pi)^3}[\widetilde V^{(1g)}(p_t-q_t)+\widetilde V^{(cf)}(p_t-q_t)]\widetilde g_{1}(q_t).
\end{equation}
On the other hand, for the form of the BS wave function of the axial-vector diquark, we have
\begin{equation}\label{vwave}
  \widetilde{\chi}^{(r)T}_{P_D}(p) = (1+v\!\!\!/){\xi^{(r)}}\!\!\!\!\!\!\!\!/\,\,\,\,\,g_2,
\end{equation}
where $\xi^{(r)}_\mu$ is the polarization vector of the axial-vector diquark, $g_2$ is a scalar function of $p_t^2$ and $p_l$. Following the same technique in the scalar diquark case, we have the following equation for the BS wave function of the axial-vector diquark under the heavy quark limit:
\begin{equation}\label{g2}
  \widetilde g_2(p_t)=-\frac{m_1m_2}{\omega_1\omega_2(M_D-\omega_1-\omega_2)}\int\frac{d^3q_t}{(2\pi)^3}[\widetilde V^{(1g)}(p_t-q_t)+\widetilde V^{(cf)}(p_t-q_t)]\widetilde g_2(q_t),
\end{equation}
which shows that the BS wave function for the axial-vector diquark satisfies the same equation as that for the scalar diquark when taking the heavy quark limit.
\section{Doubly heavy baryons under the heavy quark limit}
In this section, we will construct the formalism for the doubly heavy baryons under the heavy quark limit.
\subsection{Baryons composed of a heavy scalar diquark and a light quark}
The BS wave functions of baryons composed of a heavy scalar diquark and a light quark, like $\Omega_{bc}$ and $\Xi_{bc}$, can be defined as
\begin{eqnarray}\label{scalarbaryon}
  \chi(x_1,x_2,P)&=&\langle 0|T\psi_l(x_1)\phi_D(x_2)|P,k\rangle \nonumber\\
                 &=&e^{-iPX}\int\frac{d^4k}{(2\pi)^4}e^{-ikx}\chi_{P}(k),
\end{eqnarray}
where $\psi_l(x_1)$ and $\phi_D(x_2)$ are the field operators of the light quark and the scalar diquark, respectively, $X=\eta_1x_1+\eta_2x_2$ ($\eta_1=\frac{m_l}{m_D+m_l},\eta_2=\frac{m_D}{m_D+m_l}$, $m_l$ and $m_D$ are the masses of the light quark and the heavy diquark, respectively.) is the center-of-mass coordinate of the baryon and $x$ is the relative coordinate, $P$ is the total momentum of the baryon, while $k$ is the relative momentum between the heavy scalar diquark and the light quark. Then, $k_1=\eta_1P+k$, $k_2=\eta_2P-k$ are the momenta of the light quark and the heavy scalar diquark, respectively. For the BS equation of the doubly heavy baryon, we have
\begin{equation}\label{bsequation1}
  \chi_P(k)=S_l(k_1)\int\frac{d^4q}{(2\pi)^4}G(P,k,q)\chi_P(q)S_D(k_2),
\end{equation}
where $G(P,k,q)$ is the interaction kernel, $S_l$ and $S_D$ are the propagators for the light quark and scalar heavy diquark,  respectively, which can be expressed as the following forms under the heavy quark limit:
\begin{equation}\label{lightquark}
  S_l(k_1)=\frac{i}{2\omega_l}\left[\frac{v\!\!\!/\omega_l+k_t\!\!\!\!/+m_l}{\eta_1M+k_l-\omega_l+i\epsilon}+\frac{v\!\!\!/\omega_l-k_t\!\!\!\!/-m_l}{\eta_1M+k_l+\omega_l-i\epsilon}\right],
\end{equation}
\begin{equation}\label{scalardiquark}
  S_D(k_2)=\frac{i}{2\omega_D(\eta_2M-k_l-\omega_D+i\epsilon)},
\end{equation}
where we define the variables $k_l=v\cdot k$, $k_t=k-k_lv$, $\omega_l=\sqrt{m_l^2-k_t^2}$, $\omega_D=\sqrt{m_D^2-k_t^2}$, and $M$ is the mass of the doubly heavy baryon. For the kernel $G(P,k,q)$, we will take the form that has been used in Ref.~\cite{Weng:2010rb}:
\begin{equation}\label{kernel1}
  -iG(P,k,q)=I\otimes IV_1(k,q)+\gamma_\mu\otimes\Gamma^\mu V_2(k,q),
\end{equation}
where the vertex $\Gamma^\mu$ has the form: $\Gamma^\mu=(k_2^\mu+k_2^{\prime\mu})F_S(Q^2)$ with $F_S(Q^2)=\frac{\alpha_{seff}Q_0^2}{Q^2+Q_0^2}$ describing the interaction between diquarks and the gluon ($Q_0$ is a parameter which freezes $F_S(Q^2)$ as $Q^2\to0$), and $V_1$ and $V_2$ are the scalar confinement and one-gluon-exchange terms that have the following forms in the covariant instantaneous approximation:
\begin{equation}\label{v1}
  \widetilde V_1(k_t-q_t)=\frac{8\pi\kappa}{[(k_t-q_t)^2+\mu^2]^2}-(2\pi)^3\delta^3(k_t-q_t)\int\frac{d^3r}{(2\pi)^3}\frac{8\pi\kappa}{(r^2+\mu^2)^2},
\end{equation}
\begin{equation}\label{v2}
  \widetilde V_2(k_t-q_t)=-\frac{16\pi}{3}\frac{\alpha^2_{seff}Q_0^2}{\left[(k_t-q_t)^2+\mu^2\right]\left[(k_t-q_t)^2+Q_0^2\right]},
\end{equation}
where $\alpha_{seff}$ and $\kappa$ are the coupling parameters in the baryon case, and $\mu$ is introduced to avoid infrared divergence in numerical calculations. It should be noted that the confinement parameter in the baryon case is different from that in the diquark or the meson cases, so we replace $\kappa^\prime$ with $\kappa$. Since $\kappa^\prime$ is around $0.2\,\rm GeV^2$ in the meson case \cite{Eichten:1978tg,Dai:1993np}, and considering $\Lambda_{QCD}$ is the only parameter related to confinement, we expect $\kappa$ around $0.04\,\rm GeV^3$ due to the relation $\kappa\sim\Lambda_{QCD}\kappa^\prime$ \cite{Guo:1996jj}.

One the other hand, after writing down all the possible Dirac structures and considering the Lorentz and P-parity transformations, we have the following form for $\chi_P(k)$:
\begin{equation}\label{generalform}
  \chi_P(k)=(h_1+k_t\!\!\!\!/\,h_2)u(v,s),
\end{equation}
where $u(v,s)$ is the Dirac spinor of the doubly heavy baryon with the velocity $v$ and the helicity $s$, which satisfies the relation $v\!\!\!/u(v,s)=u(v,s)$, and $h_{1(2)}$ is a scalar function with respect to $k_t^2$ and $k_l$. Substituting Eqs.~\eqref{lightquark}, \eqref{scalardiquark}, \eqref{kernel1} and \eqref{generalform} into Eq.~\eqref{bsequation1}, we obtain the coupled integrating equations as follows,
\begin{align}\label{h1}
  \widetilde h_1(k_t)=&\frac{1}{4\omega_l\omega_D(-M+\omega_l+\omega_D)}\int\frac{d^3q_t}{(2\pi)^3}\left\{\left[(m_l+\omega_l)(\widetilde V_1+2\omega_D\widetilde V_2)-(k_t^2+k_t\cdot q_t)\widetilde V_2\right]\widetilde h_1(q_t)\right.\nonumber\\
          &+\left[(\widetilde V_1-2\omega_D\widetilde V_2)k_t\cdot q_t-(m_l+\omega_l)(k_t\cdot q_t+q_t^2)\widetilde V_2-k_t\cdot q_t\widetilde V_2\right]\widetilde h_2(q_t)\left.\right\},
\end{align}
\begin{align}\label{h2}
  \widetilde h_2(k_t)=&\frac{1}{4\omega_l\omega_D(-M+\omega_l+\omega_D)}\int\frac{d^3q_t}{(2\pi)^3}\left\{\left[\widetilde V_1+2\omega_D\widetilde V_2+(\omega_l-m_l)\frac{k_t\cdot q_t+k_t^2}{k_t^2}\widetilde V_2\right]\widetilde h_1(q_t)\right.\nonumber\\
          &+\left[(m_l-\omega_l)(\widetilde V_1-2\omega_D\widetilde V_2)\frac{k_t\cdot q_t}{k_t^2}-\widetilde V_2q_t^2\right]\widetilde h_2(q_t)\left.\right\},
\end{align}
where we have defined $\widetilde h_{1(2)}=\int\frac{dq_l}{2\pi}h_{1(2)}$, and $\widetilde V_{1(2)}=V_{1(2)}|_{k_l=q_l}$. The following equations based on lorentz invariance have been used while obtaining the above two coupled integrals:
\begin{equation}\label{t1}
  \int\frac{d^4q}{(2\pi)^4}q_t^\mu f=f_1v^\mu+f_2k_t^\mu,
\end{equation}
and
\begin{equation}\label{t2}
  \int\frac{d^4q}{(2\pi)^4}q_t^\mu q_t^\nu f=a_1g^{\mu\nu}+a_2v^\mu v^\nu+a_3v^\mu k_t^\nu+a_4k_t^\mu v^\nu+a_5k_t^\mu k_t^\nu,
\end{equation}
where $f$ is a random function of $k^2$, $q^2$ and $k_t\cdot q_t$. By multiplying appropriate vectors or tensors on both sides of Eqs.~\eqref{t1} and \eqref{t2}, we can easily obtain that
\begin{eqnarray}\label{t3}
  f_2&=& \int\frac{d^4q}{(2\pi)^4}\frac{k_t\cdot q_t}{k_t^2}f, \nonumber \\
  a_1&=& \int\frac{d^4q}{(2\pi)^4}\frac{k_t^2q_t^2-(k_t\cdot q_t)^2}{2k_t^2}f, \nonumber \\
  a_2&=& \int\frac{d^4q}{(2\pi)^4}\frac{(k_t\cdot q_t)^2-k_t^2q_t^2}{2k_t^2}f, \nonumber \\
  a_5&=& \int\frac{d^4q}{(2\pi)^4}\frac{3(k_t\cdot q_t)^2-k_t^2q_t^2}{2k^4_t}f, \nonumber\\
  f_1&=& a_3=a_4=0.
\end{eqnarray}
\subsection{Baryons composed of a heavy axial-vector diquark and a light quark}
Now we consider the doubly heavy baryons composed of a heavy axial-vector diquark and a light quark. First, we give the definition of the BS wave function for such a baryon:
\begin{align}\label{avbaryon}
  \chi_P^\mu(y_1,y_2,P)&=\langle 0|\psi_l(y_1)A^\mu_D(y_2)|P\rangle \nonumber\\
  &=e^{-iPY}\int\frac{d^4q}{(2\pi)^4}\chi^\mu_P(q)e^{-iqy},
\end{align}
where $A^\mu_D$ denotes the axial-vector field and the BS equation for the baryon in momentum space has the following form:
\begin{equation}\label{bsforavbaryon}
  \chi^\mu_P(k)=S_l(k_1)\int\frac{d^4q}{(2\pi)^4}G_{\rho\nu}(P,k,q)\chi^\nu_P(q)S_D^{\mu\rho}(k_2),
\end{equation}
where $S_D^{\mu\rho}(k_2)$ is the propagator of the heavy axial-vector diquark, which can be written as the following form under the heavy quark limit ($m_D\to\infty$):
\begin{equation}\label{avpropagator}
  S^{\mu\rho}_D(k_2)=-\frac{i(g^{\mu\rho}-v^\mu v^\rho)}{2\omega_D(\eta_2M-k_l-\omega_D+i\epsilon)},
\end{equation}
where $\mathcal O(1/m_D)$ terms in the propagator are omitted due to the heavy quark limit. On the other hand, motivated by the potential model \cite{Eichten:1978tg}, the following form for the kernel $G_{\rho\nu}$ will be used \cite{Weng:2010rb}
\begin{equation}\label{avkernel}
  iG_{\rho\nu}(P,k,q)=g_{\rho\nu}I\otimes IV_1+\gamma^\mu\otimes\Gamma_{\mu\rho\nu}V_2,
\end{equation}
where $\Gamma_{\mu\rho\nu}=\left[g_{\rho\nu}(k_2+k_2^\prime)_\mu-(k_{2\nu}g_{\mu\rho}+k_{2\rho}^\prime g_{\mu\nu})\right]F_A(Q^2)$ is the coupling vertex of heavy diquarks and a gluon with $F_A(Q^2)$ being the form factor describing the effect of the diquark structure, we will take $F_A(Q^2)=F_S(Q^2)$ for simplicity \cite{Anselmino:1987vk,Kroll:1988cd}. Meanwhile, we have the following form for $\chi_P^\mu(k)$ under the heavy quark limit:
\begin{equation}\label{avbs}
  \chi^\mu_P(k)=(b_1+k\!\!\!/_tb_2)B^\mu(v),
\end{equation}
where $b_{1(2)}$ is a scalar function of $k_t^2$, $q_t^2$ and $k_t\cdot q_t$, $B^\mu(v)$ is the spinor of the doubly heavy baryon, which has the following explicit forms:
\begin{eqnarray}\label{b}
  B^\mu_1&=&\frac{1}{\sqrt 3}(\gamma_\mu+v_\mu)\gamma_5u(v),\nonumber\\
  B^\mu_2&=&u_\mu(v),
\end{eqnarray}
for spin-$\frac{1}{2}$ and spin-$\frac{3}{2}$ baryons, respectively, $u(v)$ denotes the Dirac spinor, and $u_\mu(v)$ is the Rarita-Schwinger spinor. Furthermore, it can be shown that $B^\mu_m(m=1,2)$ satisfies the conditions \cite{Guo:1998ef}:
\begin{equation}\label{condition}
  v\!\!\!/B^\mu_m(v)=B^\mu_m(v),\quad\quad v_\mu B^\mu_m(v)=0,\quad\quad\gamma_\mu B^\mu_2(v)=0.
\end{equation}

Substituting Eqs.~\eqref{lightquark}, \eqref{avpropagator}, \eqref{avkernel} and \eqref{avbs} into Eq.~\eqref{bsforavbaryon}, and integrating $q_l$ in the integral equations, the terms like $\int\frac{d^4q}{(2\pi)^4}q_t^\mu f$ and $\int\frac{d^4q}{(2\pi)^4}q_t^\mu q_t^\nu$ can be transformed to the desired forms with the help of Eqs.~\eqref{t1}-\eqref{t3}. After all of these steps and some calculations, we arrive at:
\begin{align}\label{b1}
  \widetilde b_1(k_t)=&\frac{1}{4\omega_l\omega_D(-M+\omega_l+\omega_D)}\int\frac{d^3q_t}{(2\pi)^3}\bigg\{\left[(m_l+\omega_l)(\widetilde V_1+2\omega_D\widetilde V_2)-(k_t\cdot q_t)\widetilde V_2\right]\widetilde b_1(q_t)\nonumber\\
          &+\Big[(\widetilde V_1-2\omega_D\widetilde V_2)(k_t\cdot q_t)-(m_l+\omega_l)(k_t\cdot q_t+q_t^2)\widetilde V_2\nonumber\\
          &+(m_l-\omega_l)(k_t\cdot q_t)\widetilde V_2+3(m_l+\omega_l)\frac{k_t^2q_t^2-(k_t\cdot q_t)^2}{2k_t^2}\widetilde V_2\Big]\widetilde b_2(q_t)\bigg\},
\end{align}
\begin{align}\label{b2}
  \widetilde b_2(k_t)=&\frac{1}{4\omega_l\omega_D(-M+\omega_l+\omega_D)}\int\frac{d^3q_t}{(2\pi)^3}\Bigg\{\Big[\widetilde V_1+2\omega_D\widetilde V_2+(\omega_l-m_l)\frac{2k_t^2+k_t\cdot q_t}{k_t^2}\widetilde V_2\Big]\widetilde b_1(q_t)\nonumber\\
          &+\Big[(m_l+\omega_l)\frac{k_t\cdot q_t}{k_t^2}\widetilde V_1-2\omega_D(m_l-\omega_l)\frac{k_t\cdot q_t}{k_t^2}\widetilde V_2-q_t^2\widetilde V_2-\frac{k_t^2q_t^2-(k_t\cdot q_t)^2}{2k_t^2}\widetilde V_2\Big]\widetilde b_2(q_t)\Bigg\},
\end{align}
where $\widetilde b_{1(2)}(k_t)$ is defined as $\widetilde b_{1(2)}(k_t)=\int\frac{d^4q}{(2\pi)^4}b_{1(2)}(k)$.
\section{Doubly heavy diquarks without taking the heavy quark limit}
\subsection{The BS equation for the doubly scalar diqaurk}
In this section, we will present the BS equation for a scalar diquark without taking the heavy quark limit $m_Q\to\infty$. To obtain the form of the BS wave function for the scalar diquark with the masses of heavy quarks being finite, we first write down the general decomposition of the following Dirac fields \cite{Guo:2007qu}:
\begin{align}\label{dirac}
  \bar \psi^{ci}_\alpha\psi^j_\beta =&\frac{1}{4}\Big[I(\bar \psi^{ci}\psi^j)+\gamma^\mu(\bar \psi^{ci}\gamma_\mu\psi^j)+\frac{1}{2}\sigma^{\mu\nu}(\bar \psi^{ci}\sigma_{\mu\nu}\psi^j)\big. \nonumber\\
   &+ \gamma^5(\bar \psi^{ci}\gamma_5\psi^j)-\gamma^5\gamma^\mu(\bar \psi^{ci}\gamma_5\gamma_\mu\psi^j)\big.\Big]_{\beta\alpha}.
\end{align}
Then from the definition of the BS wave function of the scalar diquark, and imposing the P-parity transformation on the BS wave function corresponding to each term on the right-hand-side of Eq.~\eqref{dirac}, we have
\begin{equation}\label{scalarexpansion}
  \widetilde\chi_{P_D}(p)_{\alpha_1\alpha_2}=\left[\gamma_5d_1+\gamma_5\gamma_\mu(P^\mu_Dd_2+p_t^\mu d_3)/M_D+(-i)\gamma_5\sigma_{\mu\nu}P^\mu_Dp^\nu_td_4/M_D^2\right]_{\alpha_2\alpha_1},
\end{equation}
where $d_i$ $(i=1,2,3,4)$ are Lorentz scalar functions of $p_t^2$ and $p_l$.

Now, we will check if these four functions are independent of each other. To achieve this goal, we introduce the projecting operators \cite{Chang:2003ua}:
\begin{eqnarray}
  \Lambda_1^\pm &=& \frac{v\!\!\!/\omega_1\pm (p\!\!\!/_t+m_1)}{2\omega_1}, \nonumber\\
  \Lambda_2^\pm &=& \frac{v\!\!\!/\omega_2\pm (-p\!\!\!/_t+m_2)}{2\omega_2},
\end{eqnarray}
It can be shown that these projecting operators satisfy the conditions: $\Lambda_i^\pm v\!\!\!/\Lambda_i^\pm=\Lambda_i^\pm$, $\Lambda_i^\pm v\!\!\!/\Lambda_i^\mp=0$, $i=1,2$. With these operators, we can express the full propagators of quarks as follows:
\begin{equation}\label{p1}
  S_F(p_1)=i\left[\frac{\Lambda_1^+}{\lambda_1M_D+p_l-\omega_1+i\epsilon}+\frac{\Lambda_1^-}{\lambda_1M_D+p_l+\omega_1-i\epsilon}\right],
\end{equation}
\begin{equation}\label{p2}
  S_F(p_2)=i\left[\frac{\Lambda_2^+}{\lambda_2M_D-p_l-\omega_2+i\epsilon}+\frac{\Lambda_2^-}{\lambda_2M_D-p_l+\omega_2-i\epsilon}\right].
\end{equation}
Besides, we have the following equations for the projecting operators:
\begin{equation}\label{c1}
\Lambda_1^\pm v\!\!\!/S(p_1)=\left\{ \begin{aligned}
    \frac{i\Lambda_1^+}{\lambda_1M_D+p_l-\omega_1+i\epsilon}, \\
    \frac{i\Lambda_1^-}{\lambda_1M_D+p_l+\omega_1-i\epsilon},
\end{aligned} \right.
\end{equation}
and
\begin{equation}\label{c2}
\Lambda_2^\pm v\!\!\!/S(p_2)=\left\{ \begin{aligned}
    \frac{i\Lambda_2^+}{\lambda_2M_D-p_l-\omega_2+i\epsilon}, \\
    \frac{i\Lambda_2^-}{\lambda_2M_D-p_l+\omega_2-i\epsilon}.
\end{aligned} \right.
\end{equation}
Now, by multiplying a certain combination of \eqref{c1} and \eqref{c2} on both sides of Eq.~\eqref{bs}, one obtains the following constrains:
\begin{equation}\label{cons1}
  \Lambda_2^-v\!\!\!/\widetilde\chi_{P_D}^T(p)\mathcal C(\Lambda_1^+v\!\!\!/)^T\mathcal C^{-1}=0,
\end{equation}
\begin{equation}\label{cons2}
  \Lambda_2^+v\!\!\!/\widetilde\chi_{P_D}^T(p)\mathcal C(\Lambda_1^-v\!\!\!/)^T\mathcal C^{-1}=0,
\end{equation}
which may help reduce the number of independent scalar functions in Eq.~\eqref{scalarexpansion}. We find $d_3=0$, $d_4=-\frac{M_D}{m}d_2$ when $m_1=m_2=m=m$ ($\omega_1=\omega_2=\omega$).

However, when the masses of the two quarks in the diquark are different, we find no relation among these four scalar wave functions. Substituting Eqs.~\eqref{scalarexpansion}, \eqref{p1}, \eqref{p2}, $K^{(cf)}=iV^{(cf)}$ and $K^{(1g)}=iV^{(1g)}$ into Eq.~\eqref{bs}, and completing the integration $\int\frac{dq_l}{2\pi}$ on both sides of Eq.~\eqref{bs}, we obtain four coupled integral equations:
\begin{align}\label{d1}
  \widetilde d_1(p_t) =& -\frac{1}{2\omega_1\omega_2[(\omega_1+\omega_2)^2-M_D^2]}\int\frac{d^3q_t}{(2\pi)^3}\Big\{(\omega_1+\omega_2)\nonumber\\
   &\times(-\omega_1\omega_2+p_t^2-m_1m_2)(\widetilde V^{(cf)}+4\widetilde V^{(1g)})\widetilde d_1(q_t)\nonumber\\
   &+M_D(\omega_2m_1+\omega_1m_2)(\widetilde V^{(cf)}-2\widetilde V^{(1g)})\widetilde d_2(q_t)\nonumber\\
   &+\frac{1}{M_D}(\omega_1+\omega_2)(m_2-m_1)(\widetilde V^{(cf)}-2\widetilde V^{(1g)})(p_t\cdot q_t)\widetilde d_3(q_t)\nonumber\\
   &-(\omega_1+\omega_2)(p_t\cdot q_t)\widetilde V^{(cf)}\widetilde d_4(q_t)\Big\},
\end{align}
\begin{align}\label{d2}
  \widetilde d_2(p_t) =& -\frac{1}{2\omega_1\omega_2[(\omega_1+\omega_2)^2-M_D^2]}\int\frac{d^3q_t}{(2\pi)^3}\Big\{M_D(\omega_2m_1+\omega_1m_2)\nonumber\\
   &\times(\widetilde V^{(cf)}+4\widetilde V^{(1g)})\widetilde d_1(q_t)\nonumber\\
   &+(\omega_1+\omega_2)(-\omega_1\omega_2-p_t^2-m_1m_2)(\widetilde V^{(cf)}-2\widetilde V^{(1g)})\widetilde d_2(q_t)\nonumber\\
   &+(\omega_1-\omega_2)(\widetilde V^{(cf)}-2\widetilde V^{(1g)})(p_t\cdot q_t)\widetilde d_3(q_t)\nonumber\\
   &+\frac{1}{M_D}(\omega_1+\omega_2)(m_1+m_2)(p_t\cdot q_t)\widetilde V^{(cf)}\widetilde d_4(q_t)\Big\},
\end{align}
\begin{align}\label{d3}
  \widetilde d_3(p_t) =& -\frac{1}{2\omega_1\omega_2[(\omega_1+\omega_2)^2-M_D^2]}\int\frac{d^3q_t}{(2\pi)^3}\Big\{M_D(\omega_1+\omega_2) \nonumber\\
   &\times(m_2-m_1)(\widetilde V^{(cf)}+4\widetilde V^{(1g)})\widetilde d_1(q_t)\nonumber\\
   &+M_D^2(\omega_1-\omega_2)(\widetilde V^{(cf)}-2\widetilde V^{(1g)})\widetilde d_2(q_t)\nonumber\\
   &+\frac{(\omega_1+\omega_2)(\omega_1\omega_2+p_t^2-m_1m_2)}{p_t^2}(\widetilde V^{(cf)}-2\widetilde V^{(1g)})(p_t\cdot q_t)\widetilde d_3(q_t)\nonumber\\
   &-\frac{M_D(\omega_1m_2-\omega_2m_1)}{p_t^2}(p_t\cdot q_t)\widetilde V^{(cf)}\widetilde d_4(q_t)\Big\},
\end{align}
\begin{align}\label{d4}
  \widetilde d_4(p_t) =& -\frac{1}{2\omega_1\omega_2[(\omega_1+\omega_2)^2-M_D^2]}\int\frac{d^3q_t}{(2\pi)^3}\Big\{-M_D^2(\omega_1+\omega_2)\nonumber\\
   &\times(\widetilde V^{(cf)}+4\widetilde V^{(1g)})\widetilde d_1(q_t)\nonumber\\
   &+M_D(\omega_1+\omega_2)(m_1+m_2)(\widetilde V^{(cf)}-2\widetilde V^{(1g)})\widetilde d_2(q_t)\nonumber\\
   &+\frac{M_D(\omega_2m_1-\omega_1m_2)}{p_t^2}(\widetilde V^{(cf)}-2\widetilde V^{(1g)})(p_t\cdot q_t)\widetilde d_3(q_t)\nonumber\\
   &+\frac{(\omega_1+\omega_2)(\omega_1\omega_2-p_t^2-m_1m_2)}{p_t^2}(p_t\cdot q_t)\widetilde V^{(cf)}\widetilde d_4(q_t)\Big\},
\end{align}
where we have defined $\widetilde d_i(p_t)=\int\frac{d^4q}{(2\pi)^4}d_i(p)$ $(i=1,2,3,4)$. It can be shown that Eqs.~\eqref{d1} and \eqref{d2} are consistent with the equations in Ref.~\cite{Guo:2007qu} when the masses of the two quarks are equal.

\subsection{The BS equation for doubly axial-vector diquarks}
In this subsection, we will present the formalism for the axial-vector diquark. The general form for the BS equation of the axial-vector diquark is similar to that of the scalar diquark:
\begin{equation}\label{bsforadi}
 \widetilde{\chi}^{(r)T}_{P_D}(p)=S_F(p_2)\int\frac{d^4q}{(2\pi)^4}[-\gamma^\mu \widetilde{\chi}^{(r)T}_{P_D}(q)\gamma_\mu K^{(1g)}(p_t-q_t)+ \widetilde{\chi}^{(r)T}_{P_D}(q)K^{(cf)}(p_t-q_t)]S_F(-p_1),
\end{equation}
where $r$ stands for the index of the polarization vector of the axial-vector diquark. The BS equation for the axial-vector diquark in momentum space has the form:
\begin{align}\label{avgeneralform}
  \widetilde\chi_{P_D}^{(r)T}(p)&=\epsilon^{ijk}\langle0|T\psi_\alpha^{(ci)}(p_2)\psi_\beta^{(j)}(p_1)|1^+(P_D)\rangle\nonumber\\
  &=\Gamma^\rho\varepsilon_\rho^{(r)},
\end{align}
where $\Gamma^\rho$ indicates the Dirac structure of $\widetilde \chi_{P_D}^{(r)T}$, and $\varepsilon_\rho^{(r)}$ is the $r-th$ polarization state of the axial-vector diquark, which satisfies the relation $\varepsilon_\rho^{(r)}P^\rho_D=0$. In general, Dirac field operators for the axial-vector diquark can be decomposed into several terms like the scalar diquark case, and the BS wave function corresponding to each term in Eq.~\eqref{dirac} can be further expanded as a linear combination of $p_t$, $P_D$ and $g^{\mu\rho}$. After writing down all the terms contained in $\Gamma^\rho$, we use the constrains $\varepsilon_\rho^{(r)}P_D^\rho=0$, P-parity and the Lorentz transformations to simplify the form of $\chi_{P_D}^{(r)T}$. Finally, we have the full expansion of $\chi_{P_D}^{(r)T}$:
\begin{align}\label{avdi}
   \chi_{P_D}^{(r)T}(p)=&\bigg[\frac{1}{M_D}p_t^\rho H_1+\gamma_\mu\Big(\frac{1}{M_D^2}p_t^\rho P_D^\mu H_2+\frac{1}{M_D^2}p_t^\rho p_t^\mu H_3+g^{\mu\rho}H_7\Big)+\frac{1}{M_D^2}\gamma_5\gamma^\mu\varepsilon^{\mu\rho\alpha\beta}p_{t\alpha}P_{D\beta}H_5  \nonumber\\
    &-i\sigma_{\mu\nu}\Big(\frac{1}{M_D^3}p_t^\rho p_t^\mu P_D^\nu H_4+\frac{1}{M_D}g^{\rho\mu}P_D^\nu H_6+\frac{1}{M_D}g^{\rho\mu}p_t^\nu H_8\Big)\bigg]\varepsilon_\rho^{(r)},
\end{align}
where $H_i$ $(i=1,\cdots,8)$ are scalar functions. When the masses of the two heavy quarks are equal, we follow the same routine used in the scalar diquark case to reduce the number of the scalar functions. Consequently, we have the constrains for these scalar functions $H_i$ $(i=1,\cdots,8)$:
\begin{eqnarray}\label{constrain2}
   H_2 &=&  H_8=0, \nonumber\\
   H_1 &=& -\frac{p_t^2 H_3+M_D^2 H_7}{m M_D}, \nonumber\\
   H_6 &=& \frac{m}{M_D} H_5,
\end{eqnarray}
where the masses of the two quarks in the diqaurk are equal $(m_1=m_2=m)$. Therefore, in this case, in the eight scalar functions there are only four independent wave functions $H_i$ $(i=3,4,5,7)$.

With the help of the constrains in Eq.~\eqref{constrain2}, we are able to obtain four integral equations for scalar functions in the case of the axial-vector diquark:
\begin{align}\label{H3}
  \widetilde H_3(p_t) =&-\frac{1}{2\omega p_t^4(m_D^2-4\omega^2)}\int\frac{d^3q_t}{(2\pi)^3}\Big\{mM_D\big[p_t^2q_t^2-3(p_t\cdot q_t)^2\big]\widetilde V^{(cf)}\widetilde H_4(q_t)\nonumber \\
   &+4M_D^2p_t^2\big[(p_t^2+(p_t\cdot q_t))\widetilde V^{(cf)}+2(p_t^2-2(p_t\cdot q_t))\widetilde V^{(1g)}\big]\widetilde H_7(q_t)\nonumber\\
   &+2M_D^2p_t^2(p_t\cdot q_t)(\widetilde V^{(cf)}-2\widetilde V^{(1g)})\widetilde H_5(q_t)\nonumber\\
   &+\Big[\big(-2\omega^2p^2_tq_t^2+4p_t^2q_t^2(p_t\cdot q_t)+2(2m^2+\omega^2)(p_t\cdot q_t)^2\big)\widetilde V^{(cf)}\nonumber\\
   &+4\big(-\omega^2p_t^2q_t^2-4p_t^2q_t^2(p_t\cdot q_t)+(2m+\omega^2)(p_t\cdot q_t)^2\big)\widetilde V^{(1g)}\Big]\widetilde H_3(q_t)\Big\},
\end{align}
\begin{align}\label{H4}
  \widetilde H_4(p_t) =&\frac{1}{2m\omega p_t^4(m_D^2-4\omega^2)}\int\frac{d^3q_t}{(2\pi)^3}\Big\{2m\big[m^2p_t^2q_t^2-(m^2+2\omega^2)(p_t\cdot q_t)^2\big]\widetilde V^{(cf)}\widetilde H_4(q_t)\nonumber \\
   &+4m^2M_Dp_t^2\big[(p_t^2+p_t\cdot q_t)\widetilde V^{(cf)}-2(p_t\cdot q_t)\widetilde V^{(1g)}\big]\widetilde H_5(q_t)\nonumber\\
   &+2m^3p_t^2(p_t\cdot q_t)(\widetilde V^{(cf)}-4\widetilde V^{(1g)})\widetilde H_7(q_t)\nonumber\\
   &+M_D\Big[\big(-2m^2p^2_tq_t^2+2p_t^2q_t^2(p_t\cdot q_t)+3m^2(p_t\cdot q_t)^2\big)\widetilde V^{(cf)}\nonumber\\
   &+2\big(-m^2p_t^2q_t^2-4p_t^2q_t^2(p_t\cdot q_t)+3m^2(p_t\cdot q_t)^2\big)\widetilde V^{(1g)}\Big]\widetilde H_3(q_t)\Big\},
\end{align}
\begin{align}\label{H5}
  \widetilde H_5(p_t) =&\frac{1}{2M_D\omega p_t^2(m_D^2-4\omega^2)}\int\frac{d^3q_t}{(2\pi)^3}\Big\{2m\big[-p_t^2q_t^2+(p_t\cdot q_t)^2\big]\widetilde V^{(cf)}\widetilde H_4(q_t)\nonumber \\
   &+2M_D^3p_t^2\big[\widetilde V^{(cf)}+2\widetilde V^{(1g)}\big]\widetilde H_7(q_t)\nonumber\\
   &-4M_Dp_t^2\big[(m^2+p_t\cdot q_t)\widetilde V^{(cf)}-2(p_t\cdot q_t)\widetilde V^{(1g)}\big]\widetilde H_5(q_t)\nonumber\\
   &+M_D\big[(p_t^2q_t^2-(p_t\cdot q_t)^2)(\widetilde V^{(cf)}+2\widetilde V^{(1g)})\big]\widetilde H_3(q_t)\Big\},
\end{align}
\begin{align}\label{H7}
  \widetilde H_7(p_t) =&\frac{1}{2M_D^2\omega p_t^2(m_D^2-4\omega^2)}\int\frac{d^3q_t}{(2\pi)^3}\Big\{mM_D\big[p_t^2q_t^2-(p_t\cdot q_t)^2\big]\widetilde V^{(cf)}\widetilde H_4(q_t)\nonumber \\
   &-4M_D^2p_t^2\omega^2\big[\widetilde V^{(cf)}+2\widetilde V^{(1g)}\big]\widetilde H_7(q_t)\nonumber\\
   &+2M_D^2p_t^2\big[(m^2+p_t\cdot q_t)\widetilde V^{(cf)}-2(p_t\cdot q_t)\widetilde V^{(1g)}\big]\widetilde H_5(q_t)\nonumber\\
   &-2\omega^2\big[(p_t^2q_t^2-(p_t\cdot q_t)^2)(\widetilde V^{(cf)}+2\widetilde V^{(1g)})\big]\widetilde H_3(q_t)\Big\},
\end{align}
where $\widetilde H_i(p_t)=\int\frac{dq_l}{2\pi}H_i(p)$.

On the other hand, for the doubly heavy axial-vector diquark composed of two quarks with different masses, $e.g.$, $bc$ with $J^P=1^+$, we also have similar relations for the scalar functions in the BS wave function for the axial-vector diquark $bc$ without taking the heavy quark limit:
\begin{equation}\label{co1}
 \widetilde H_1=-\frac{(\omega_1+\omega_2)(p_t^2\widetilde H_3+M_D^2\widetilde H_7)}{M_D(m_1\omega_2+m_2\omega_1)},
\end{equation}
\begin{equation}\label{co2}
 \widetilde H_2=-\frac{2p_t^2(m_1-m_2)\left[m_1^2\widetilde H_4+m_1(m_2\widetilde H_4+M_D\widetilde H_5)+m_2M_D\widetilde H_5-\omega_1(\omega_2+\omega_2)\widetilde H_4\right]}{M_D\left[4p_t^4-(5m_1^2+2m_2^2+4\omega_1\omega_2)+3m_1^2p_t^2\right]},
\end{equation}
\begin{equation}\label{co3}
  \widetilde H_6=\frac{p_t^2\left[2(m_1^2-m_2^2)m_1M_D\widetilde H_5-2\omega_1(m_1+m_2)(\omega_1+\omega_2)M_D\widetilde H_5\right]}{M_D\left[4p_t^4-(5m_1^2+2m_2^2+4\omega_1\omega_2)+3m_1^2p_t^2\right]},
\end{equation}
\begin{equation}\label{co4}
  \widetilde H_8=\frac{(m_2^2-m_1^2)M_D\widetilde H_7}{(\omega_1+\omega_2)(m_1\omega_2+m_2\omega_1)}.
\end{equation}
In this case, we can follow the same routine used before, substituting Eqs.~\eqref{co1}-\eqref{co4} into Eq.~\eqref{avdi}, which is then substituted into the BS equation for the axial-vector diquark. We will not present the explicit forms of the integral equations for the scalar functions here because the calculations are long and demanding, we will present the numerical results for the axial-vector diquark $bc$ in Sec.~\uppercase\expandafter{\romannumeral6} instead.

It is noteworthy that there are four independent scalar functions in both scalar and axial-vector diquark cases, which is consistent with the physical picture since each of these two quarks has two degrees of freedom. In contrast, there is only one independent scalar function in the scalar and axial-vector diquark cases when the heavy quark limit is applied.
\section{Doubly heavy baryons without taking the heavy quark limit}
\subsection{Baryons composed of a doubly heavy scalar diquark and a light quark}
In this subsection, we continue to use the variables defined in Sec.~\uppercase\expandafter{\romannumeral3}. First, we can start from writing down the general form of the BS equation for baryons comprised of a heavy scalar diquark and a light quark, which has the same form as Eq.~\eqref{bsequation1}:
\begin{equation}\label{60}
  \chi_P(k)=S_l(k_1)\int\frac{d^4q}{(2\pi)^4}G(P,k,q)\chi_P(q)S_D^\prime(-k_2),
\end{equation}
except that for the propagator of the scalar diquark we will not take heavy quark limit. In this case, we have the following form for the propagator of the heavy scalar diqaurk:
\begin{equation}\label{61}
  S_D^\prime(-k_2)=\frac{i}{2\omega_D}\left[\frac{1}{\eta_2M-k_l-\omega_D+i\epsilon}-\frac{1}{\eta_2M-k_l+\omega_D-i\epsilon}\right].
\end{equation}
For the form of the BS wave function $\chi_P(k)$, after considering the constrain on the Dirac spinor of the baryon $u(v,s)$ $(v\!\!\!/u(v,s)=u(v,s))$ we write down all the possible combinations of Dirac structures and momentum $(P, k_t)$,
\begin{equation}\label{62}
  \chi_P(k)=(Y_1+Y_2\gamma_5+Y_3\gamma_5k\!\!\!/_t+Y_4k\!\!\!/_t+Y_5\sigma_{\mu\nu}\varepsilon^{\mu\nu\alpha\beta}k_{t\alpha}k_{t\beta})u(v,s),
\end{equation}
where $Y_i$ $(i=1\cdots5)$ are scalar wave functions of $k_t^2$, $k_l$ and $k_t\cdot q_t$. Furthermore, each term in the expansion of $\chi_P(k)$ transforms exactly in the way that $\chi_P(k)$ transforms under P-parity and Lorentz transformations, which can help us simplify the form of $\chi_P(k)$. Finally, we arrive at
\begin{equation}\label{63}
  \chi_P(k)=(A_1+A_2k\!\!\!/_t)u(v,s),
\end{equation}
following the same routine we have used in the last few sections. $A_1$ and $A_2$ are scalar functions of $k_t^2$, $k_l$ and $k_t\cdot q_t$. Define $\widetilde A_{1(2)}=\int\frac{dq_l}{2\pi}A_{1(2)}$. Then, after we substitute Eqs.~\eqref{lightquark}, \eqref{kernel1}, \eqref{61} and \eqref{63} into Eq.~\eqref{60}, we obtain two coupled equations for the scalar functions for the baryons composed of a doubly scalar diqaurk and a light quark as follows:
\begin{align}\label{A1}
  \widetilde A_1(k_t) =&\frac{1}{4\omega_D\omega_l[(\omega_D+\omega_l)^2-M^2]}\int\frac{d^3q_t}{(2\pi)^3}\Bigg\{\bigg[\Big(2M\omega_l+2m_l(\omega_D+\omega_l)\Big)\widetilde V_1\nonumber \\
   &+\Big(4Mm_l\omega_D+3M(k_t^2+k_t\cdot q_t)+(\omega_D+\omega_l)(4\omega_D\omega_l-5(k_t^2-k_t\cdot q_t))\Big)\widetilde V_2\bigg]\widetilde A_1(q_t)\nonumber\\
   &+\bigg[2(\omega_D+\omega_l)(k_t\cdot q_t)\widetilde V_1\nonumber\\
   &+2\Big((m_l(\omega_D+\omega_l)+\omega_lM)q_t^2-(k_t\cdot q_t)(M(2\omega_D+\omega_l)+m_l(\omega_D+\omega_l))\Big)\widetilde V_2\bigg]\widetilde A_2(q_t)\Bigg\},
\end{align}
\begin{align}\label{A2}
  \widetilde A_2(k_t) =&\frac{1}{4\omega_D\omega_l[(\omega_D+\omega_l)^2-M^2]}\int\frac{d^3q_t}{(2\pi)^3}\Bigg\{\bigg[2(\omega_D+\omega_l)\widetilde V_1\nonumber \\
   &+\Big(2M(2\omega_D+\omega_l)+2M\frac{k_t\cdot q_t}{k_t^2}\omega_l-2m_l\frac{k_t^2+k_t\cdot q_t}{k_t^2}(\omega_D+\omega_l)\Big)\widetilde V_2\bigg]\widetilde A_1(q_t)\nonumber\\
   &+\bigg[2\Big(M\omega_l\frac{k_t\cdot q_t}{k_t^2}-m_l(\omega_D+\omega_l)\frac{k_t\cdot q_t}{k_t^2}\Big)\widetilde V_1\nonumber\\
   &+2\Big(-M\omega_D\omega_l\frac{k_t\cdot q_t}{k_t^2}+m_l\omega_D(\omega_D+\omega_l)\frac{k_t\cdot q_t}{k_t^2}-(\omega_D+\omega_l)(k_t\cdot q_t+q_t^2)\Big)\widetilde V_2\bigg]\widetilde A_2(q_t)\Bigg\}.
\end{align}
It can be seen from these two equations that the corrections are generated from both the $\widetilde V_1$ and $\widetilde V_2$ terms compared with the situation under the heavy quark limit, which will be the reason why the masses calculated from Eqs.~\eqref{A1} and \eqref{A2} are different from those obtained through Eqs.~\eqref{h1} and \eqref{h2}. One can prove that Eqs.~\eqref{h1} and \eqref{h2} can be obtained by taking the heavy quark limit $(m_Q\to\infty)$.
\subsection{Baryons composed of a heavy axial-vector diquark and a light quark}\label{import}
The BS equation formalism for the baryons comprised of a heavy axial-vector diquark and a light quark without taking the heavy quark limit will be established in this subsection. In this situation, the degenerate states like $\{\Xi_{cc}, \Xi_{cc}^\ast\}$ in the heavy quark limit will spilt into two states with different masses: spin-$\frac{1}{2}$ and spin-$\frac{3}{2}$. We will construct different forms of BS wave functions for spin-$\frac{1}{2}$ and spin-$\frac{3}{2}$ states of these doubly heavy baryons, respectively. The most difficult thing in deriving the BS equation for this type of baryons, like $\Xi_{cc}^{++}$, is to find the correct form of the wave function for them without the help from the heavy quark symmetry. In the heavy quark limit, the internal dynamics is determined by the light degrees of freedom due to the $SU(2)_s\times SU(2)_f$ symmetry, then we can easily decompose the wave function into the following form \cite{Guo:1998ef}:
\begin{equation}\label{66}
  \chi_P^\mu(k)=\langle0|T\psi A_\mu|P\rangle=u(v)\xi_\nu\zeta^{\mu\nu}(v,p),
\end{equation}
where $\psi$ and $A_\mu$ represent the light quark and the axial-vector diquark fields, respectively, $u(v)$ denotes the Dirac spinor of the light quark, $\xi_\nu$ stands for the polarization vector of $A_\mu$, and $u(v)\xi_\nu$ can be decomposed into spin-$\frac{1}{2}$ and spin-$\frac{3}{2}$ baryons $B_{m\nu}$ $(m=1,2)$. The tensor $\zeta^{\mu\nu}$ includes all the dynamics and contains two independent scalar functions in Eq.~\eqref{avbs}. When the masses of the heavy quarks are finite the form of Eq.~\eqref{66} still holds due to the LSZ reduction formula \cite{Peskin:1995ev}. However, $\zeta^{\mu\nu}$, which includes internal dynamics, becomes more complicated in this scenario and will be different for spin-$\frac{1}{2}$ and spin-$\frac{3}{2}$ doubly heavy baryons. After taking into account all the possible Dirac structures and the combinations of Dirac gamma matrices and momenta $(k_t, P)$, and imposing the constrains from P-parity and Lorentz transformations we find that the BS wave function for all the spin-$\frac{1}{2}$ doubly heavy baryons, $\chi_P^\mu(k)_{1/2}$, can be written as the following form:
\begin{equation}\label{spin1}
  \chi_P^{\mu}(k)_{1/2}=\big(E_1v^\mu\gamma_5+E_2\gamma^\mu\gamma_5+E_3k_t^\mu\gamma_5+E_4v^\mu k\!\!\!/_t\gamma_5+E_5\gamma^\mu k\!\!\!/_t\gamma_5+E_6k_t^\mu k\!\!\!/_t\gamma_5\big)u,
\end{equation}
where the scalar functions $E_i$ $(i=1,\cdots,6)$ may have different dimensions. As before, we define
\begin{equation}\label{68}
  \widetilde E(k_t)=\int\frac{dk_l}{2\pi}E(k),\,\,\,\quad\quad i=1,\cdots,6.
\end{equation}

As for the propagator of the axial-vector diquark, we need its full form,
\begin{equation}\label{69}
  S_D^{\prime\mu\rho}(-k_2)=-i\frac{g^{\mu\rho}-\left[(\eta_2M-k_l)^2v^\mu v^\rho+(k_l-\eta_2M)(v^\mu k_t^\rho+k_t^\mu v^\rho)+k_t^\mu k_t^\rho\right]/m_D^2}{(\eta_2M-k_l+\omega_D-i\epsilon)(\eta_2M-k_l-\omega_D+i\epsilon)}.
\end{equation}
It can be seen that in comparison with the propagator under the heavy quark limit, Eq.~\eqref{avpropagator}, Eq.~\eqref{69} contains higher order corrections in $\mathcal O(1/m_D)$. On the other hand, Eq.~\eqref{p1} for the light quark propagator remains unchanged. The kernel applied in this case also includes the scalar confinement and one-gluon-exchange terms \cite{Guo:1996jj,Guo:1998ef,Dai:1993kt,Jin:1992mw},
\begin{equation}\label{70}
  iG^{\rho\mu}=g^{\rho\mu}I\otimes IV_1+\gamma_\mu\otimes\Gamma^{\mu\rho\nu}V_2,
\end{equation}
where $\Gamma^{\mu\rho\nu}$ is the vertex of a gluon with two axial-vector diquarks, which has the following form:
\begin{equation}\label{71}
\Gamma^{\mu\rho\nu}=\big[(k_2+k_2^\prime)^\mu g^{\rho\nu}-(k_2^\nu g^{\mu\rho}+k_2^{\prime\rho} g^{\mu\nu})\big]F_A(Q^2),
\end{equation}
where $g_s$ is the coupling constant for strong interaction in the vertex, and $F_A(Q^2)$ is the form factor describing the inner structure of the axial-vector diquark.

Using the same techniques as in the last few sections, we obtain coupled integral equations for $\widetilde E_i$ $(i=1,\cdots,6)$ for spin-$\frac{1}{2}$ doubly heavy baryons composed of a heavy axial-vector diquark and a light quark, which explicit forms are listed in Appendix A, where we have used Eqs.~\eqref{t1}, \eqref{t2} and \eqref{t3} in the derivation of those equations. On top of that, we have also used the following equations on the grounds of Lorentz invariance:
\begin{align}\label{t4}
   \int\frac{d^4q}{(2\pi)^4}q_t^\mu q_t^\nu q_t^\rho f(k^2,q^2,k\cdot q) &=x_1g^{\mu\nu}v^\rho+x_2g^{\mu\nu}k_t^\rho+x_3v^\mu v^\nu v^\rho+g_4v^\mu v^\nu k_t^\rho\nonumber \\
   &+x_5v^\mu k_t^\nu v^\rho+x_6v^\mu k_t^\nu k_t^\rho+x_7v^\nu k_t^\mu v^\rho+x_8v^\nu k_t^\mu k_t^\rho\nonumber\\
   &+x_9k_t^\mu k_t^\nu v^\rho+x_{10}k_t^\mu k_t^\nu k_t^\rho,
\end{align}
where $x_i$ $(i=1,\cdots,10)$ are scalar functions of $k_t^2$ and $k_l$, and it can be shown that
\begin{eqnarray}
  x_2 &=& \int\frac{d^4q}{(2\pi)^4}\frac{k_t^2q_t^2(k_t\cdot q_t)-(k_t\cdot q_t)^3}{3k_t^4}f(k^2,q^2,k\cdot q), \nonumber\\
  x_4 &=& \int\frac{d^4q}{(2\pi)^4}\frac{(k_t\cdot q_t)^3-k_t^2q_t^2(k_t\cdot q_t)}{3k_t^4}f(k^2,q^2,k\cdot q), \nonumber\\
  x_{10} &=& \int\frac{d^4q}{(2\pi)^4}\frac{(k_t\cdot q_t)^3}{k_t^6}f(k^2,q^2,k\cdot q), \nonumber\\
  x_1 &=& x_3=x_5=x_6=x_7=x_8=x_9=0.
\end{eqnarray}

In the same way, considering all the possible Dirac structures, the combinations of Dirac gamma matrices and momenta $(k_t, P)$, and the restrictions from P-parity and Lorentz transformations, we obtain the explicit form of BS wave function for spin-$\frac{3}{2}$ doubly heavy baryons, which has the following form:
\begin{align}\label{spin3}
  \chi_P^{\mu}(k)_{3/2}=&\big(C_1g^\mu_\nu+C_2k\!\!\!/_tg^\mu_\nu+C_3v^\mu k_{t\nu}+C_4k_t^\mu k_{t\nu}+C_5k\!\!\!/_tk_t^\mu k_{t\nu}\nonumber\\
  &+C_6k\!\!\!/_tv^\mu k_{t\nu}+C_7\gamma^\mu k_{t\nu}+C_8\gamma^\mu k\!\!\!/_tk_{t\nu}\big)u^\nu,
\end{align}
where $C_i$ ($i=1,\cdots,8$) are scalar functions of $k_t^2$, $q_l$ and $k_t\cdot q_t$, which correspond to eight different Dirac structures. Using the same techniques as before, we obtain the coupled integral equations for $\widetilde C_i$ $(\widetilde C_i=\int\frac{dk_l}{2\pi}C_i)$ $(i=1,\cdots,8)$ for spin-$\frac{3}{2}$ doubly heavy baryons, and the explicit expressions are given in Appendix.~\ref{appb}. One can see that Eqs.~\eqref{e1}-\eqref{e6} and Eqs.~\eqref{C11}-\eqref{C18} are much more complicated than Eqs.~\eqref{b1} and \eqref{b2}. However, most of the terms in Eqs.~\eqref{e1}-\eqref{e6} and Eqs.~\eqref{C11}-\eqref{C18} are higher order terms in $1/m_Q$ expansion. One can obtain Eqs.~\eqref{b1} and \eqref{b2} from Eqs.~\eqref{e1}-\eqref{e6} or Eqs.~\eqref{C11}-\eqref{C18} if taking the heavy quark limit.
\section{Numerical result}
In this section, we will solve all the coupled integral equations emerged in previous sections numerically. The integration region in each integral will be discretized into $n$ pieces, with $n$ being sufficiently large. In this way, the integral equation will be converted into an $n\times n$-matrix equation, and the scalar wave functions of each equation will now be regarded as an $n$-dimensional vector. It can be shown that the integral equations, for instance, Eqs.~\eqref{e1}-\eqref{e6}, can be illustrated as:
\begin{displaymath}
\left(\begin{array}{c}\widetilde E_1(k_t)\\ \widetilde E_2(k_t)\\ \widetilde E_3(k_t)\\ \widetilde E_4(k_t)\\ \widetilde E_5(k_t)\\ \widetilde E_6(k_t)\end{array}\right)=\left(
\begin{array}{cccccc}
e_{11}&e_{12}&e_{13}&e_{14}&e_{15}&e_{16}\\
e_{21}&e_{22}&e_{23}&e_{24}&e_{25}&e_{26}\\
e_{31}&e_{32}&e_{33}&e_{34}&e_{35}&e_{36}\\
e_{41}&e_{42}&e_{43}&e_{44}&e_{45}&e_{46}\\
e_{51}&e_{52}&e_{53}&e_{54}&e_{55}&e_{56}\\
e_{61}&e_{62}&e_{63}&e_{64}&e_{65}&e_{66}
\end{array}\right)\left(\begin{array}{c}\widetilde E_1(q_t)\\ \widetilde E_2(q_t)\\ \widetilde E_3(q_t)\\ \widetilde E_4(q_t)\\ \widetilde E_5(q_t)\\ \widetilde E_6(q_t)\end{array}\right)
\end{displaymath}
where $\widetilde E_i$ $(i=1,\cdots,6)$ is an $n$-dimensional vector, and $e_{ij}$ $(i,j=1,\cdots,6)$ is an $n\times n$ matrix function of $k_t$ and $q_t$. To make it more clear, we will take Eq.~\eqref{g1} as an example here. In Eq.~\eqref{g1}, the integral equation can be rewritten in the following way once we integrated the angular parts,
\begin{equation}\label{newone}
g^\prime_1(p_t)=\int d|q_t|M(|p_t|,|q_t|)g^\prime_1(|q_t|),
\end{equation}
where $|p_t|$ and $|q_t|$ are the absolute values for the relative momentum between quarks and diquarks in the initial and final states, respectively. Since we choose to integrate $|q_t|$ in Eq.~\eqref{newone} with the Gaussian quadrature rule, we need to convert the Gaussian integration nodes into physical values for $|q_t|$, which can be done using the following equation
\begin{eqnarray}
|q_t|=\epsilon+w\log\left[1+y\frac{1+t}{1-t}\right],
\end{eqnarray}
where $t$ is the ordinary Gaussian integration nodes ranging from $-1$ to $1$, $\epsilon$ is a parameter introduced to avoid divergence in numerical calculations, $w$ and $y$ are parameters used in controlling the slope of wave functions and finding the proper solutions for these functions. One can then obtain the numerical results of the $g^\prime_1(p_t)$ by requiring the eigenvalue of the eigenvalue equation to be $1$. Similar methods can be applied to evaluate other integral equations in our work.
\subsection{Doubly heavy diqaurks}
In this subsection, the numerical results based on the theoretical framework of doubly heavy diquarks will be presented. From the potential model and some former works on the BS analysis, $\kappa^\prime$ is usually taken to be around $0.20\,\rm GeV^2$ for the diquark system \cite{Guo:1996jj,Guo:2007qu,Weng:2010rb,Yu:2006ty,Dai:1993np}. For example, in Ref.~\cite{Weng:2010rb}, the authors used $\kappa^\prime=0.20\,\rm GeV^2$ to obtain the masses of heavy diquarks. In the present work, we will let $\kappa^\prime$ vary in a reasonable range, $0.18-0.20\,\rm GeV^2$, to see to what extent it will affect the masses of the doubly heavy diquarks $cc$, $bb$ and $cb$. In our work, we will employ the values of coupling parameters ($\alpha_s$ and $\kappa^\prime$) from Ref.~\cite{Jin:1992mw} and the heavy quark masses ($m_b=5.02$\,GeV, $m_c=1.58$\,GeV) from Ref.~\cite{Dai:1993np} for the calculation of the masses of the doubly heavy diquarks. These parameters, $e.g.$, $\kappa^\prime$ and $\alpha_s$ were obtained through the investigations of heavy mesons and heavy quarkonia and led to results consistent with the experimental data \cite{Jin:1992mw,Eichten:1979ms}.
\begin{table*}[h!]
\renewcommand\arraystretch{0.9}
\centering
\caption{\vadjust{\vspace{-5pt}}The masses of doubly heavy diquarks with and without taking the heavy quark limit. (all the masses are given in units $\rm GeV$)}\label{1}
\begin{tabular*}{\textwidth}{@{\extracolsep{\fill}}cccccccc}
\hline
 $\kappa^\prime(\rm GeV^2)$    &$\alpha_s$&                             &       &  $cc(1^+)$  &  $bc(0^+)$  &   $bc(1^+)$ &$bb(1^+)$     \\
\hline
\multirow{3}*{0.18}            &&                &$m_Q\to\infty$&3.40&6.79& 6.79&10.07  \\
                               &0.39&                  &$m_Q$ finite&3.37&6.70& 6.75&10.05  \\
                               &&                 &Corrections&-0.03&-0.09&-0.04&-0.02   \\
\hline
\hline
\multirow{3}*{0.20}            &&                 &$m_Q\to\infty$   &3.42&6.80& 6.80&10.08  \\
                               &0.40&                  &$m_Q$ finite&3.37&6.70&6.75&10.05  \\
                               &&                  &Corrections&-0.05 &-0.10 &-0.05 &-0.03   \\
\hline
\hline
\end{tabular*}
\end{table*}
\begin{figure}[h!]
  \centering
  \includegraphics[width=0.70\textwidth]{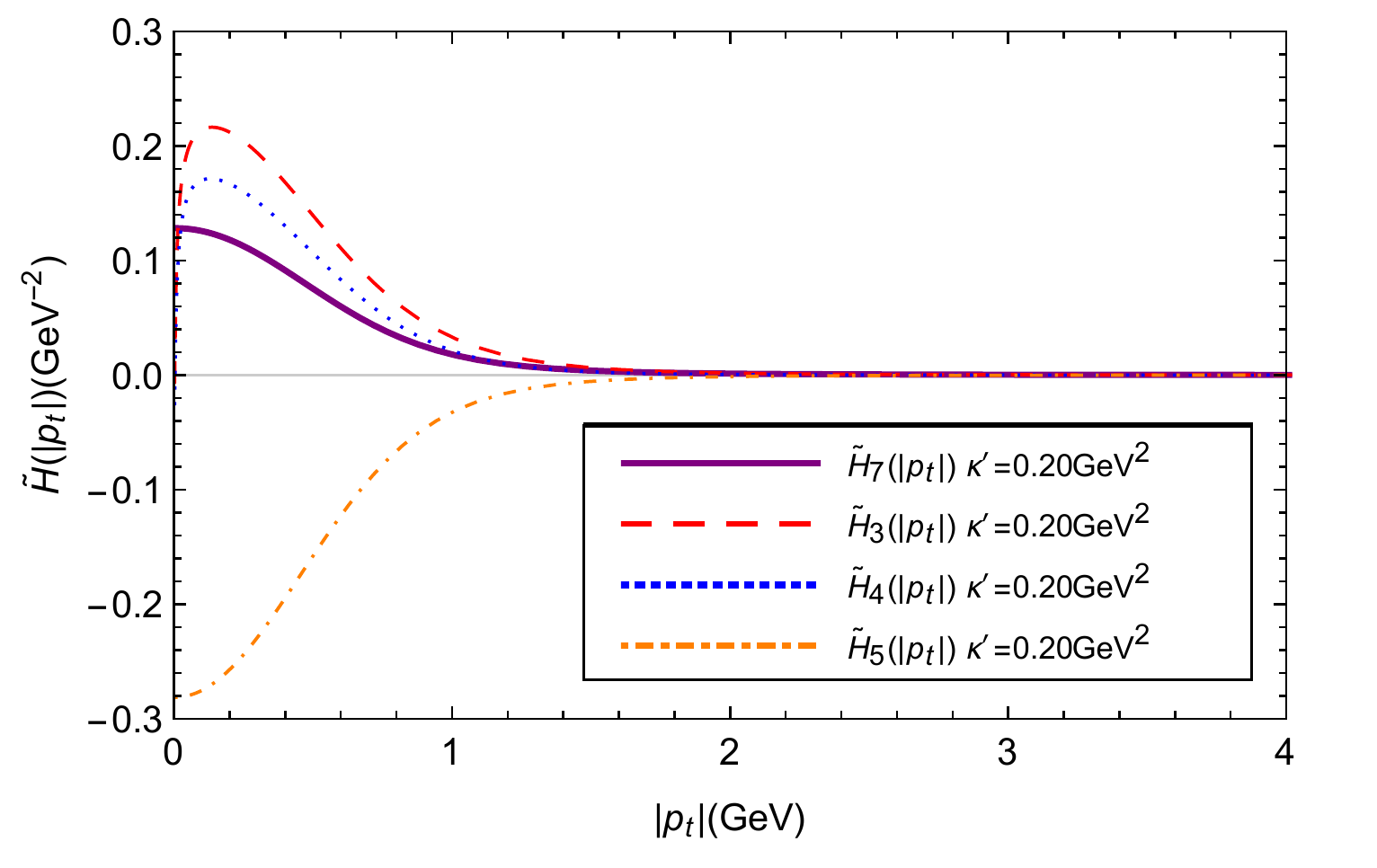}\\
  \caption{The BS scalar wave functions of the axial-vector diquark $cc$ without taking the heavy quark limit.}\label{fig1}
\end{figure}
\begin{figure}[h!]
 \centering
  \includegraphics[width=0.70\textwidth]{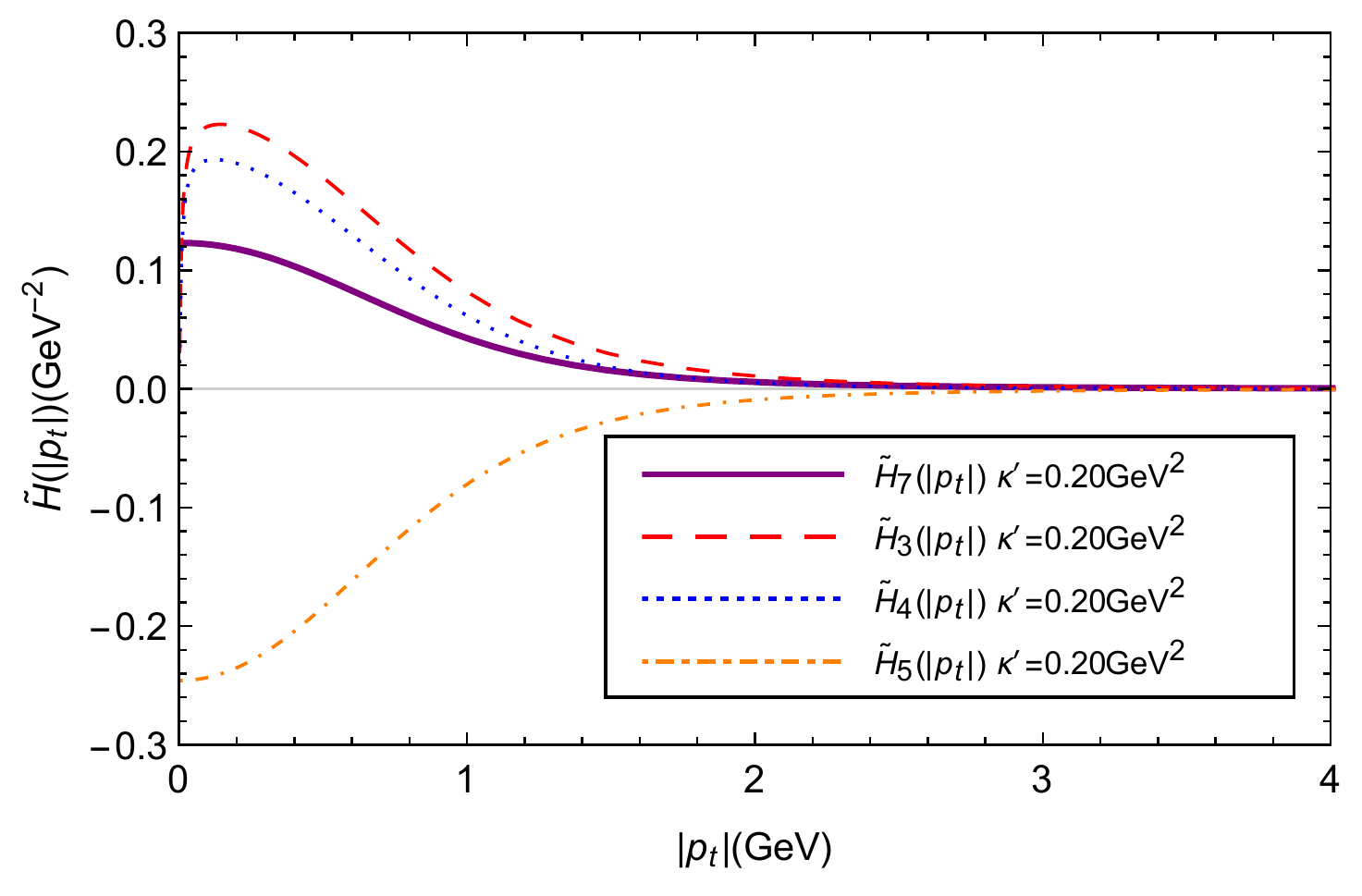}\\
  \caption{The BS scalar wave functions of the axial-vector diquark $bc$ without taking the heavy quark limit.}\label{fig2}
\end{figure}
\begin{figure}[h!]
 \centering
  \includegraphics[width=0.70\textwidth]{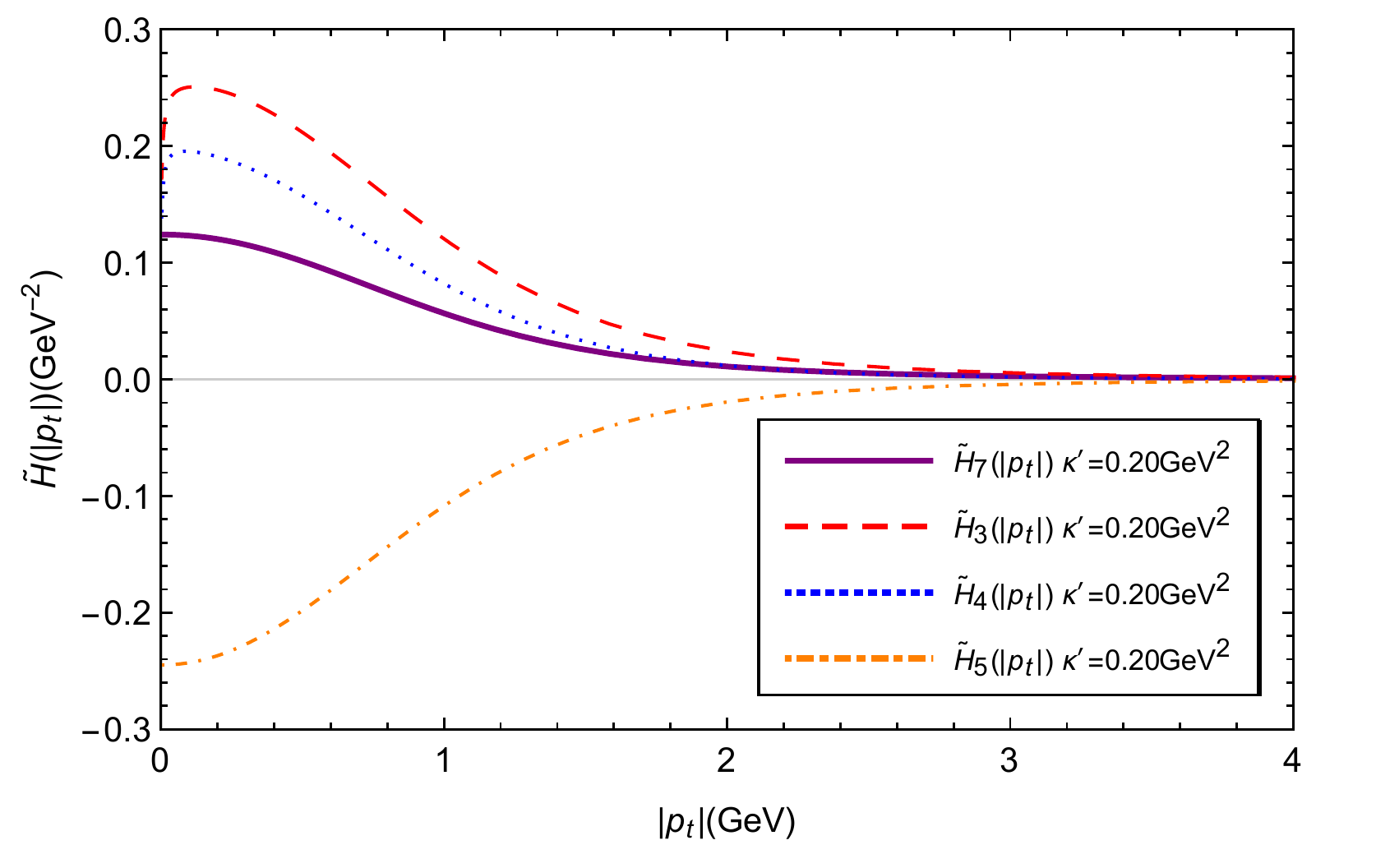}\\
  \caption{The BS scalar wave functions of the axial-vector diquark $bb$ without taking the heavy quark limit.}\label{fig3}
\end{figure}
\begin{figure}[h!]
 \centering
  \includegraphics[width=0.70\textwidth]{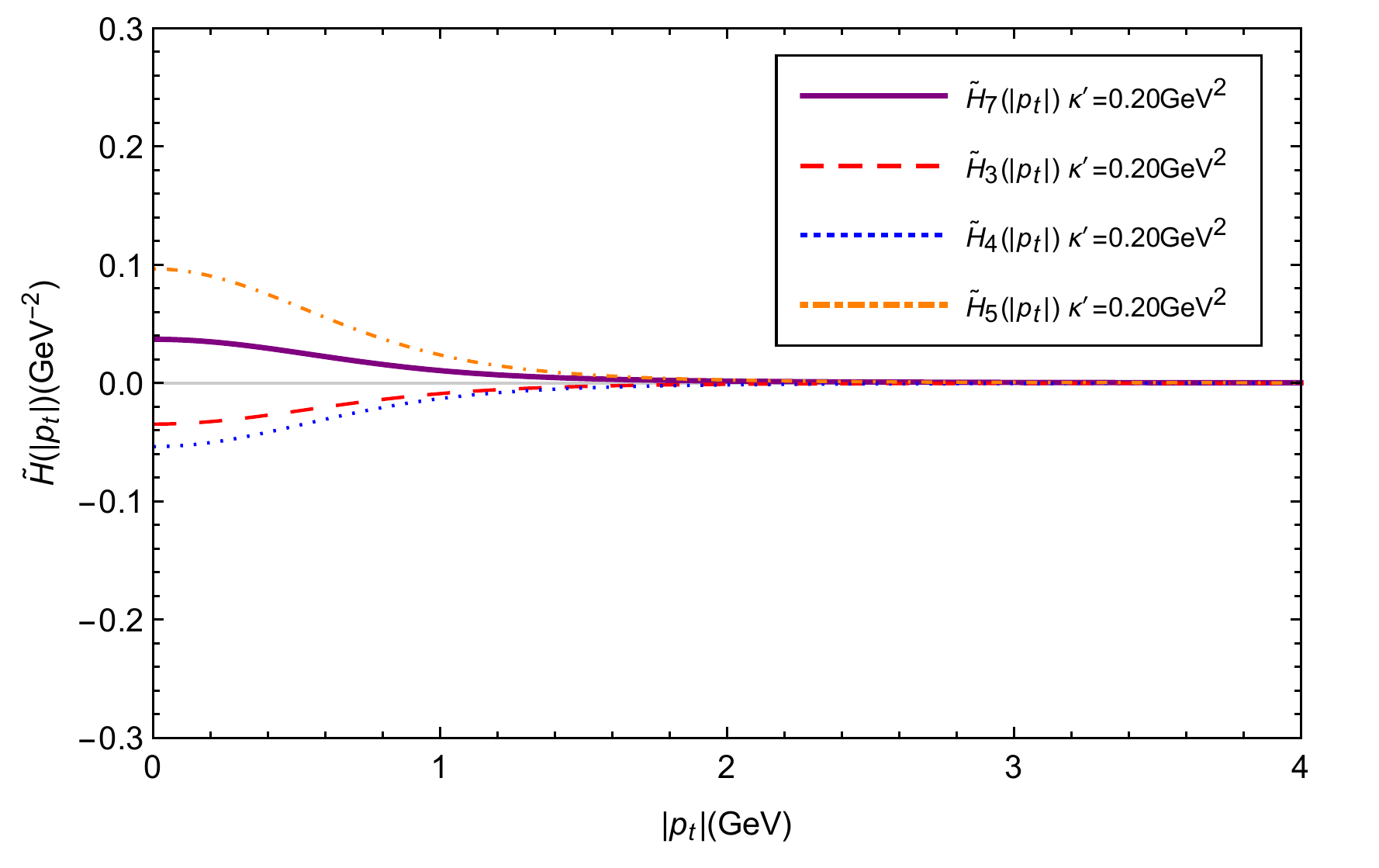}\\
  \caption{The BS scalar wave functions of the scalar diquark $bc$ without taking the heavy quark limit.}\label{fig4}
\end{figure}
\begin{figure}[h!]
 \centering
  \includegraphics[width=0.70\textwidth]{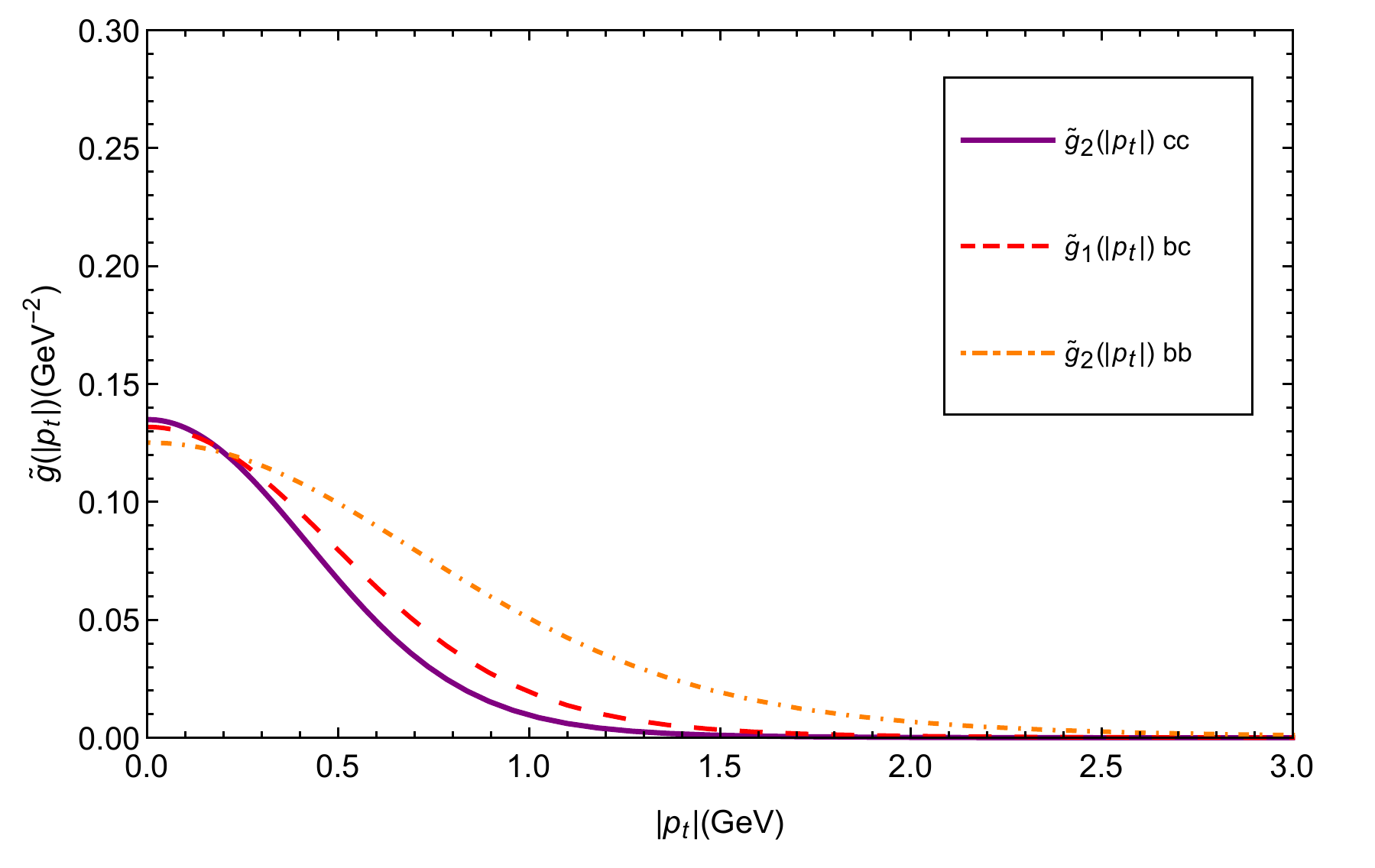}\\
  \caption{The BS scalar wave functions of the diquarks $cc$, $bc$ and $bb$ under the heavy quark limit with $\kappa^\prime=0.20\,\rm GeV^2$.}\label{fig5}
\end{figure}

In Table~\ref{1}, we present the results for the masses of diquarks under the heavy quark limit, and their counterparts with finite masses for heavy quarks. It can be seen from the table that the values under the heavy quark limit are generally bigger than their counterparts, especially in the scalar diqaurk case, where the correction is about $0.10\,\rm GeV$. The corrections to the masses of diquarks show a mild increase with the increase of $\kappa^\prime$. Furthermore, the data obtained in the heavy quark limit is less independent on the parameter $\kappa^\prime$. In contrast, the masses of diquarks calculated without taking the heavy quark limit remain almost unchanged with the variation of the parameter $\kappa^\prime$. On top of that, we find that the corrections to the masses of heavy diquarks are all negative, which suggests that the contribution from $1/m_Q$ and higher order terms cause a reduction in the masses of the diquarks.

In Figs.~\ref{fig1}, \ref{fig2}, \ref{fig3}, \ref{fig4} and \ref{fig5}, we plot all the BS scalar wave functions of the diquarks. For axial-vector diquarks $cc$, $bc$ and $bb$ in Figs.~\ref{fig1}, \ref{fig2} and \ref{fig3}, the shapes of the BS wave functions are quite similar, but the scalar wave functions of the diquarks decrease more rapidly as the diquark mass increases. This is because the radius of the diquark decreases with the increase of the diquark mass, leading to a larger average transverse momentum within the diquark \cite{Liu:2016wzh}. Meanwhile, we plot the scalar wave functions for the scalar diquark $bc$ in Fig.~\ref{fig4}, where the scalar functions decrease more rapidly compared with the axial-vector $bc$ plotted in Fig~\ref{fig2}. In Fig.~\ref{fig5}, we plot the BS scalar wave functions of $cc$, $bc$ and $bb$ diquarks under the heavy quark limit, where the shapes of the wave functions are steeper when the masses of the diquarks are smaller due to the same reason as that in the case of finite heavy quark masses \cite{Liu:2016wzh}. It should be noted that the axial-vector and scalar $bc$ diquarks have the same BS scalar functions as shown in Eqs.~\eqref{g1} and \eqref{g2}. Obviously, all the wave functions decrease to zero when $|p_t|$ is larger than about $3.0\,\rm GeV$, which is resulted from the confinement interaction.
\subsection{Doubly heavy baryons}
Now, we will use the obtained results for masses of the doubly heavy diquarks in last subsection to calculate the masses of the doubly heavy baryons in two scenarios: $m_Q\to\infty$ and $m_Q$ being finite. First, we will find the values of the coupling parameters $\kappa$ and $\alpha_{seff}$, and this is achieved by fitting the experimental data of baryons in these two scenarios (with and without taking the heavy quark limit), respectively. We take $m_s=0.458$\,\rm GeV, $m_{u,d}=0.350$\,\rm GeV in fitting the experimental data \cite{Dai:1993np}.

In the baryon case, the interaction kernel is different from that in the meson case. In general, the confinement term is still due to scalar interaction, thus the form of $\widetilde V_1$ is still the same as that in the meson case, only the confinement parameter $\kappa^\prime$ needs to be replaced with $\kappa$, which is the confinement parameter in the baryon case. We take $\kappa$ to be in the range $0.02\,\rm GeV^3\sim0.08\,\rm GeV^3$ due to the relation between $\kappa^\prime$ and $\kappa$ $(\kappa\simeq\mathcal O(\Lambda_{QCD})\kappa^\prime)$ \cite{Guo:1996jj}. Meanwhile, since the diquark is not a point-like particle, we introduce the form factor $F_{S/A}(Q^2)=\frac{\alpha_{seffQ_0^2}}{Q^2+Q_0^2}$ to describe the vertex between two diquarks and the gluon, where $Q_0^2$ is a parameter which freezes $F(Q^2)$ when $Q^2$ is small. By analyzing the electromagnetic form factors of the proton \cite{Lepage:1980fj,Liu:2016wzh}, $Q_0^2=3.2\,\rm GeV^2$ can result in a good agreement with the experimental data \cite{Anselmino:1987vk}.

There exists a constraint between the parameters $\kappa$ and $\alpha_{seff}$ when we solve the integral equations numerically. In this way, we can obtain the corresponding value of $\alpha_{seff}$ for each $\kappa$ in the range of $0.02\sim0.08\,\rm GeV^3$. By fitting the experimental data for the masses of $\Lambda_Q$, $\Xi_Q^\prime$, $\Sigma_Q^{(\ast)}$, $\Xi_Q^{(\ast)}$ and $\Omega_Q^{(\ast)}$ (baryons composed of a light diquark and a heavy quark) form the PDG \cite{Patrignani:2016xqp}, the values of $\alpha_{seff}$ without taking the heavy quark limit can be determined by Eqs.~\eqref{A1} and \eqref{A2}, where we need the masses of the light diquarks in fitting the experimental data of $\Lambda_Q$ or other baryons. We can obtain the masses of light axial-diquarks by solving Eqs.~\eqref{H3}-\eqref{H7}, and the results are $M_D=0.97\sim0.98\,\rm GeV$ for $uu$ and $dd$ diquarks, $M_D=1.19\sim1.21\,\rm GeV$ for $ss$ diquark, where the coupling parameters $\kappa^\prime$ is taken to be $0.18\sim0.20\,\rm GeV^2$. For the light scalar diquark $ud$, we also take $\kappa^\prime=0.18\sim0.20\,\rm GeV^2$ and we obtain $M_D=0.78\sim0.80\,\rm GeV$ for its mass. With these results for the diquark masses, we can obtain the values for $\alpha_{seff}$ by fitting the experimental data for the masses of $\Lambda_Q$, $\Xi_Q^\prime$, $\Sigma_Q^{(\ast)}$, $\Xi_Q^{(\ast)}$ and $\Omega_Q^{(\ast)}$ in the two different scenarios ($m_Q\to\infty$ and $m_Q$ is finite). In Table~\ref{2}, we give the values of $\alpha_{seff}$ without taking the heavy quark limit through solving Eqs.~\eqref{A1}-\eqref{A2}, Eqs.~\eqref{e1}-\eqref{e6} and Eqs.~\eqref{C11}-\eqref{C18} numerically, where we use the experimental data from PDG for the masses of the heavy baryons during the fitting to obtain the values of $\alpha_{seff}$, which are $2.29\,\rm GeV$ for $\Lambda_c$, $2.58\,\rm GeV$ for $\Xi_c^\prime$, $2.45\,\rm GeV$ for $\Sigma_c$, $2.52$ for $\Sigma_c^\ast$, $2.47\,\rm GeV$ for $\Xi_c$, $2.65\,\rm GeV$ for $\Xi_c^\ast$, $2.70\,\rm GeV$ for $\Omega_c$, $2.77\,\rm GeV$ for $\Omega_c^\ast$, $5.62\,\rm GeV$ for $\Lambda_b$, $5.94\,\rm GeV$ for $\Xi_b^\prime$, $5.81\,\rm GeV$ for $\Sigma_b$, $5.83$ for $\Sigma_b^\ast$, $5.79\,\rm GeV$ for $\Xi_b$, $5.95\,\rm GeV$ for $\Xi_b^\ast$, $6.05\,\rm GeV$ for $\Omega_b$. It should be noted that we include all these known heavy baryons containing $c$ and $b$ quarks to obtain $\alpha_{seff}$ without taking the heavy quark limit during the fitting, and we only take the ones containing $b$ quark when we fit for the values of $\alpha_{seff}$ in the heavy quark limit for the reason that $1/m_b$ corrections are very small.
\begin{table*}[h!]
\renewcommand\arraystretch{0.9}
\centering
\caption{\vadjust{\vspace{-5pt}}The values of $\alpha_{seff}$ in the baryons with $\kappa$ ranging from $0.02\,\rm GeV^3$ to $0.08\,\rm GeV^3$ when $m_Q$ is finite.}\label{2}
\begin{tabular*}{\textwidth}{@{\extracolsep{\fill}}cccccc}
\hline
 $\kappa(\,\rm GeV^3)$    &0.02&0.04&0.06&0.08 \\
\hline
 $\alpha_{seff}$    &$0.79\sim0.85$&$0.81\sim0.86$&$0.83\sim0.88$&$0.86\sim0.89$ \\
\hline
\hline
\end{tabular*}
\end{table*}

On the other hand, the values of $\alpha_{seff}$ under the heavy quark limit can be obtained by using the BS equations in Refs.~\cite{Guo:1996jj,Weng:2010rb} to fit the experimental data of $\Lambda_Q$, $\Xi_Q^\prime$, $\Sigma_Q^{(\ast)}$, $\Xi_Q^{(\ast)}$ and $\Omega_Q^{(\ast)}$ \cite{Patrignani:2016xqp}. In this way, we can obtain the proper $\alpha_{seff}$ values, and we assume these parameters obtained from baryons composed of a light diquark and a heavy quark can also be applied to doubly heavy baryons since the strong interaction is flavor independent \cite{Guo:1996jj,Guo:1998ef,Weng:2010rb}. Following the same techniques in obtaining the values in Table~\ref{2}, we have the values of $\alpha_{seff}$ for heavy baryons under the heavy quark limit in Table~\ref{3}.
\begin{table*}[h!]
\renewcommand\arraystretch{0.9}
\centering
\caption{\vadjust{\vspace{-5pt}}The values of $\alpha_{seff}$ in the baryons with $\kappa$ ranging from $0.02\,\rm GeV^3$ to $0.08\,\rm GeV^3$ under the heavy quark limit.}\label{3}
\begin{tabular*}{\textwidth}{@{\extracolsep{\fill}}cccccc}
\hline
 $\kappa(GeV^3)$    &0.02&0.04&0.06&0.08 \\
\hline
 $\alpha_{seff}$    &$0.77\sim0.81$&$0.79\sim0.83$&$0.81\sim0.85$&$0.80\sim0.86$ \\
\hline
\hline
\end{tabular*}
\end{table*}

With the numerical values for the parameter $Q_0^2$ in the form factor and the masses of diquarks $cc$, $bb$ and $bc$ obtained before, there is only one parameter, $\kappa$, left free in our model for doubly heavy baryons. By solving the integral equations for each doubly heavy baryon numerically, we obtain the results for the masses of doubly heavy baryons under the heavy quark limit and the results for spin-$1/2$ and spin-$3/2$ baryons respectively, without taking the heavy quark limit. The results are illustrated in Table~\ref{4} when $\kappa$ varies between $0.02$ and $0.08\,\rm GeV^3$.
\begin{table*}[h!]
\renewcommand\arraystretch{1.2}
\centering
\caption{\vadjust{\vspace{-5pt}}The masses of doubly heavy baryons with and without taking the heavy quark limit, and the comparisons with other approaches. All the masses are in units of GeV.}\label{4}
\begin{tabular*}{\textwidth}{@{\extracolsep{\fill}}ccccccccccc}
\hline
 & \multicolumn{2}{c}{$m_Q\to\infty$ (our results)} & \multicolumn{2}{c}{$m_Q$ finite (our results)} & \cite{Roberts:2007ni}&\cite{Weng:2010rb}&\cite{Roncaglia:1995az}&\cite{Ebert:1996ec}&\cite{He:2004px}&\cite{Weng:2018mmf}\\
\hline
$\kappa(\rm GeV^3)$&0.02&0.08&0.02&0.08& & & & & &\\
\hline
$\Xi_{cc}$&                3.63$\pm$0.02 &3.62$\pm$0.02    &3.55$\pm$0.01&3.54$\pm$0.01    &3.676   &3.540     &3.66  &3.66  &3.520  &3.633  \\
\hline
$\Xi_{cc}^\ast$&           3.63$\pm$0.02 &3.62$\pm$0.02    &3.62$\pm$0.01&3.62$\pm$0.01    &3.753   &          &3.74  &3.81  &3.630  &3.696  \\
\hline
$\Xi_{bc}$&                6.99$\pm$0.02 &6.98$\pm$0.03    &6.90$\pm$0.01&6.88$\pm$0.01    &7.011   &6.840     &6.99  &6.95  &6.838  &6.922\\
\hline
$\Xi_{bc}^\prime$&         7.01$\pm$0.02 &7.00$\pm$0.02    &6.98$\pm$0.01&6.97$\pm$0.02    &7.047   &          &7.04  &7.00  &7.028  &6.950  \\
\hline
$\Xi_{bc}^\ast$&           7.01$\pm$0.02 &7.00$\pm$0.02    &6.99$\pm$0.02&6.98$\pm$0.02    &7.074   &          &7.06  &7.02  &6.986  &6.973  \\
\hline
$\Xi_{bb}$&                10.31$\pm$0.01 &10.29$\pm$0.02  &10.25$\pm$0.01&10.25$\pm$0.01  &10.340  &10.090    &10.34 &10.23 &10.272 &10.169 \\
\hline
$\Xi_{bb}^\ast$&           10.31$\pm$0.01 &10.29$\pm$0.02  &10.27$\pm$0.01&10.26$\pm$0.01  &10.367  &          &10.37 &10.28 &10.337 &10.189\\
\hline
$\Omega_{cc}$&             3.73$\pm$0.02 &3.72$\pm$0.02    &3.64$\pm$0.01&3.63$\pm0.01$    &3.815   &3.635     &3.74  &3.76  &3.619  &3.732\\
\hline
$\Omega_{cc}^\ast$&        3.73$\pm$0.02 &3.72$\pm$0.02    &3.71$\pm$0.02&3.69$\pm$0.02    &3.876   &          &3.82  &3.89  &3.721  &3.802 \\
\hline
$\Omega_{bc}$&             7.09$\pm$0.01 &7.07$\pm$0.02    &6.97$\pm$0.01&6.96$\pm$0.01    &7.136   &6.945     &7.06  &7.05  &6.941  &7.011 \\
\hline
$\Omega_{bc}^\prime$&      7.10$\pm$0.01 &7.08$\pm$0.01    &7.05$\pm$0.02&7.05$\pm$0.02    &7.165   &          &7.09  &7.09  &7.116  &7.047 \\
\hline
$\Omega_{bc}^\ast$&        7.10$\pm$0.01 &7.08$\pm$0.01    &7.07$\pm$0.01&7.06$\pm$0.02    &7.187   &          &7.12  &7.12  &7.077  &7.066 \\
\hline
$\Omega_{bb}$&             10.37$\pm$0.01 &10.36$\pm$0.01  &10.34$\pm$0.01&10.33$\pm$0.01  &10.454  &10.185    &10.37 &10.32 &10.369 &10.259 \\
\hline
$\Omega_{bb}^\ast$&        10.37$\pm$0.01 &10.36$\pm$0.01  &10.35$\pm$0.01&10.35$\pm$0.01  &10.486  &          &10.40 &10.36 &10.429 &10.268  \\
\hline
\hline
\end{tabular*}
\end{table*}

In Table~\ref{4}, our results are different from the previous results in Ref.~\cite{Weng:2010rb} due to different form factors applied. Furthermore, the coupling parameters $\kappa$ and $\alpha_{seff}$ are obtained by fitting the known experimental data, and the masses of diquarks are also obtained within the BS formalism. The mass of $\Xi_{cc}$ is obtained as $3.60\sim3.65\,\rm GeV$ under the heavy quark limit, which is in a good agreement with the experimental data $3621.40\pm0.72\pm0.27\pm0.14\,\rm MeV$ \cite{Aaij:2017ueg}. Meanwhile, the mass for $\Xi_{cc}$ obtained without taking the heavy quark limit is $3.53\sim3.56\,\rm GeV$, which is about $0.10\,\rm GeV$ smaller than the result obtained under the heavy quark limit and also consistent with the experimental data. As for $\Xi_{cc}^\ast$, which is degenerate with $\Xi_{cc}$ under the heavy quark limit, we obtain $3.61\sim3.63\,\rm GeV$ for its mass without taking the heavy quark limit. Considering the big mass difference between the data obtained by SELEX ($3519\pm2\,\rm MeV$ for $\Xi_c^+$) and LHCb ($3621.40\pm0.72\pm0.27\pm0.14\,\rm MeV$ for $\Xi_{cc}^{++}$) (where the masses of those two states are supposed to be very close due to very small isospin split between $\Xi_c^+$ and $\Xi_{cc}^{++}$), and based on our theoretical results listed in Table~\ref{4}, we therefore suggest that those two states obtained in experiments may be spin-$1/2$ and spin-$3/2$ states of $\Xi_{cc}$, respectively. Of course, this will need further confirmation from the measurement of spin-parity of $\Xi_{cc}$.

In general, we can see from Table~\ref{4} that corrections to the results in the heavy quark limit gradually decrease as the masses of doubly heavy baryons increase. For example, the mass of $\Omega_{cc}$ is calculated to be about $3.73\pm0.02\,\rm GeV$ under the heavy quark limit and $3.64\pm0.01\,\rm GeV$ without taking the heavy quark limit, which gives the correction of about $0.10\,\rm GeV$. However, in the case of $\Omega_{bb}$ or $\Xi_{bb}$, the corrections are basically $0.03\sim 0.06\,\rm GeV$. It is consistent with the expectation of the heavy quark effective theory and shows the importance of corrections to those in the heavy quark limit when dealing with doubly heavy baryons containing $c$-quark. On the other hand, it can be seen from the results that in both scenarios the results have little dependency on the coupling parameter $\kappa$, and the results obtained under the heavy quark limit are generally larger than those in the case of $m_Q$ being finite, which is similar to the results we obtained for the doubly heavy diquarks in the previous subsection.

\begin{figure}[h!]
 \centering
  \includegraphics[width=0.65\textwidth]{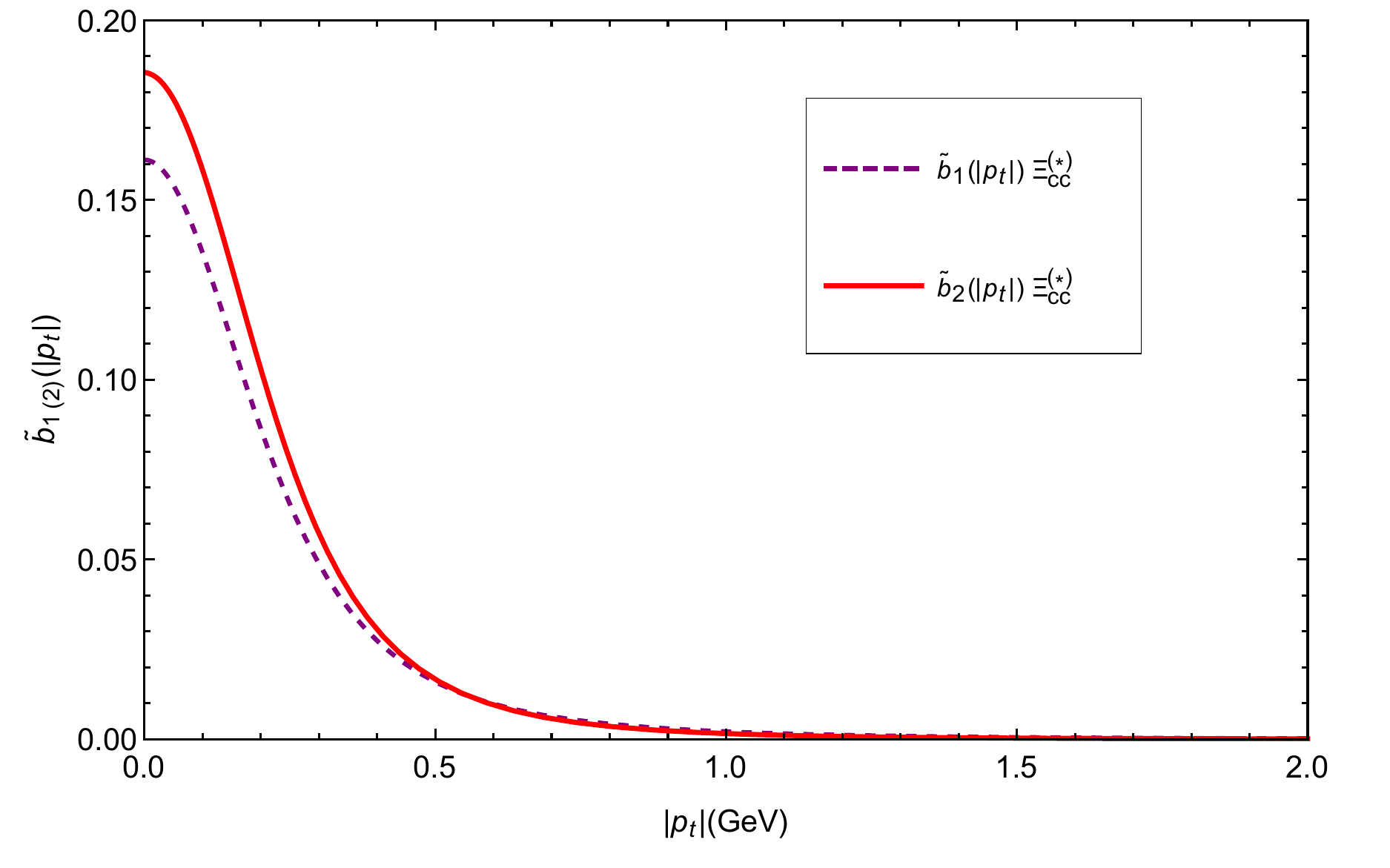}\\
  \caption{The BS wave functions of $\Xi_{cc}^{(\ast)}$ under the heavy quark limit with $\kappa=0.02\,\rm GeV^3$.}\label{fig6}
\end{figure}

\begin{figure}[h!]
 \centering
  \includegraphics[width=0.65\textwidth]{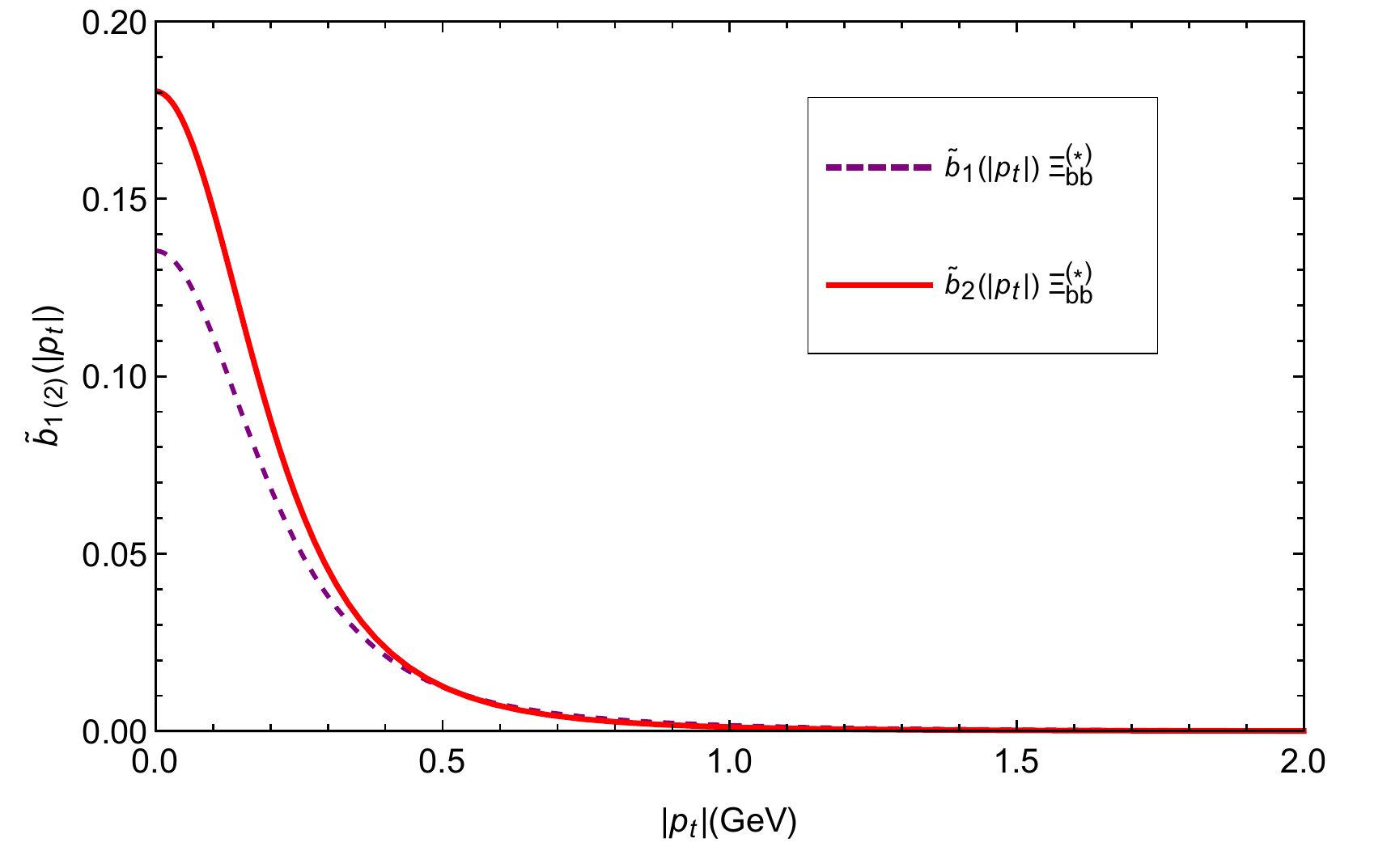}\\
  \caption{The BS wave functions of $\Xi_{bb}^{(\ast)}$ under the heavy quark limit with $\kappa=0.02\,\rm GeV^3$.}\label{fig7}
\end{figure}

\begin{figure}[h!]
 \centering
  \includegraphics[width=0.65\textwidth]{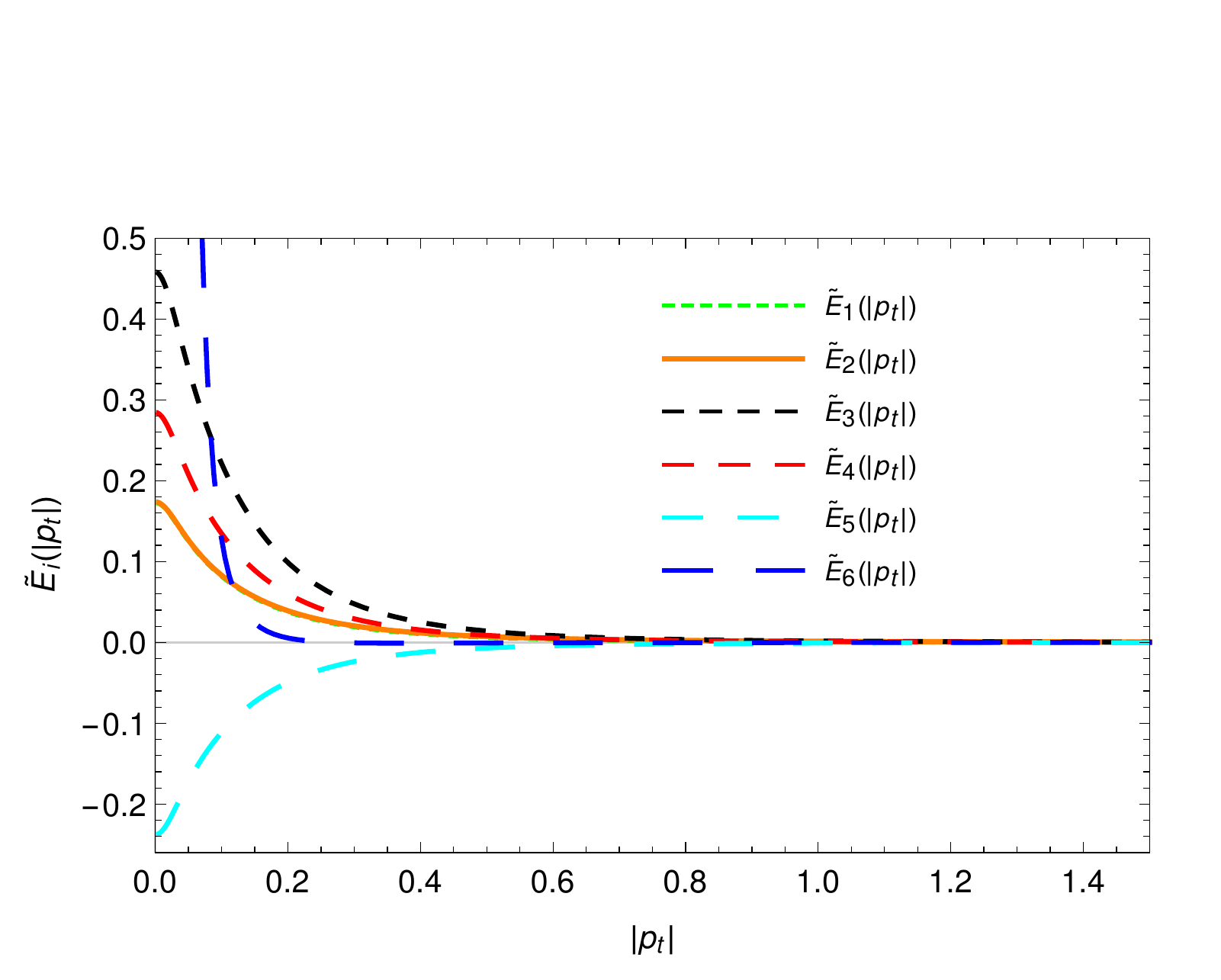}\\
  \caption{The BS wave functions of $\Xi_{cc}$ (spin-$1/2$) without taking the heavy quark limit when $\kappa=0.02\,\rm GeV^3$ ($\widetilde E_6(|p_t|)$ is 6305 when $|p_t|$ is 0).}\label{fig8}
\end{figure}

\begin{figure}[h!]
 \centering
  \includegraphics[width=0.65\textwidth]{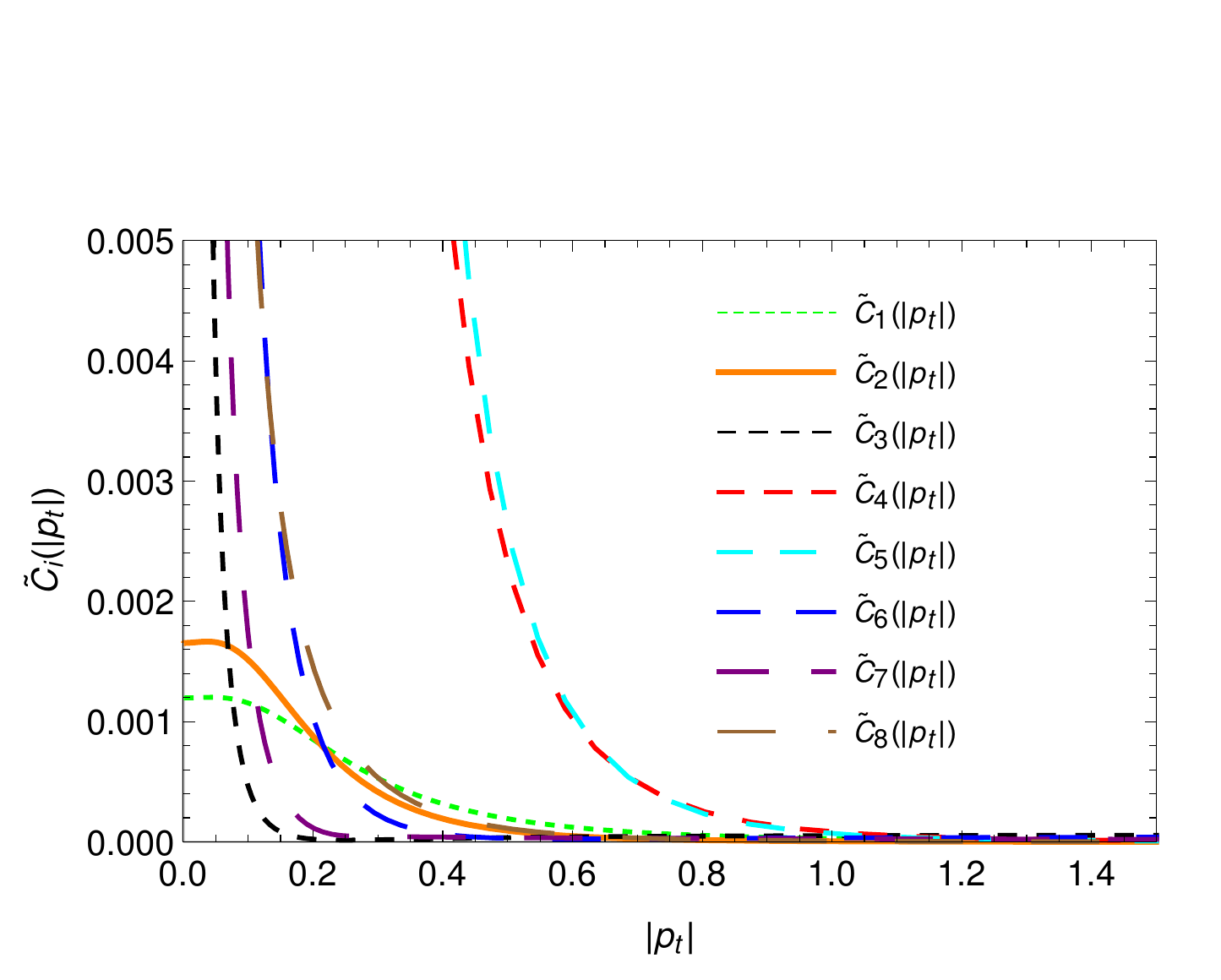}\\
  \caption{The BS wave functions of $\Xi_{cc}^\ast$ (spin-$3/2$) without taking the heavy quark limit when $\kappa=0.02\,\rm GeV^3$ ($\widetilde C_3(|p_t|)$, $\widetilde C_4(|p_t|)$, $\widetilde C_5(|p_t|)$, $\widetilde C_6(|p_t|)$, $\widetilde C_7(|p_t|)$ and $\widetilde C_8(|p_t|)$ are 14, 3203, 4703, 102, 31 and 70 respectively when $|p_t|$ is 0).}\label{fig9}
\end{figure}

\begin{figure}[h!]
 \centering
  \includegraphics[width=0.65\textwidth]{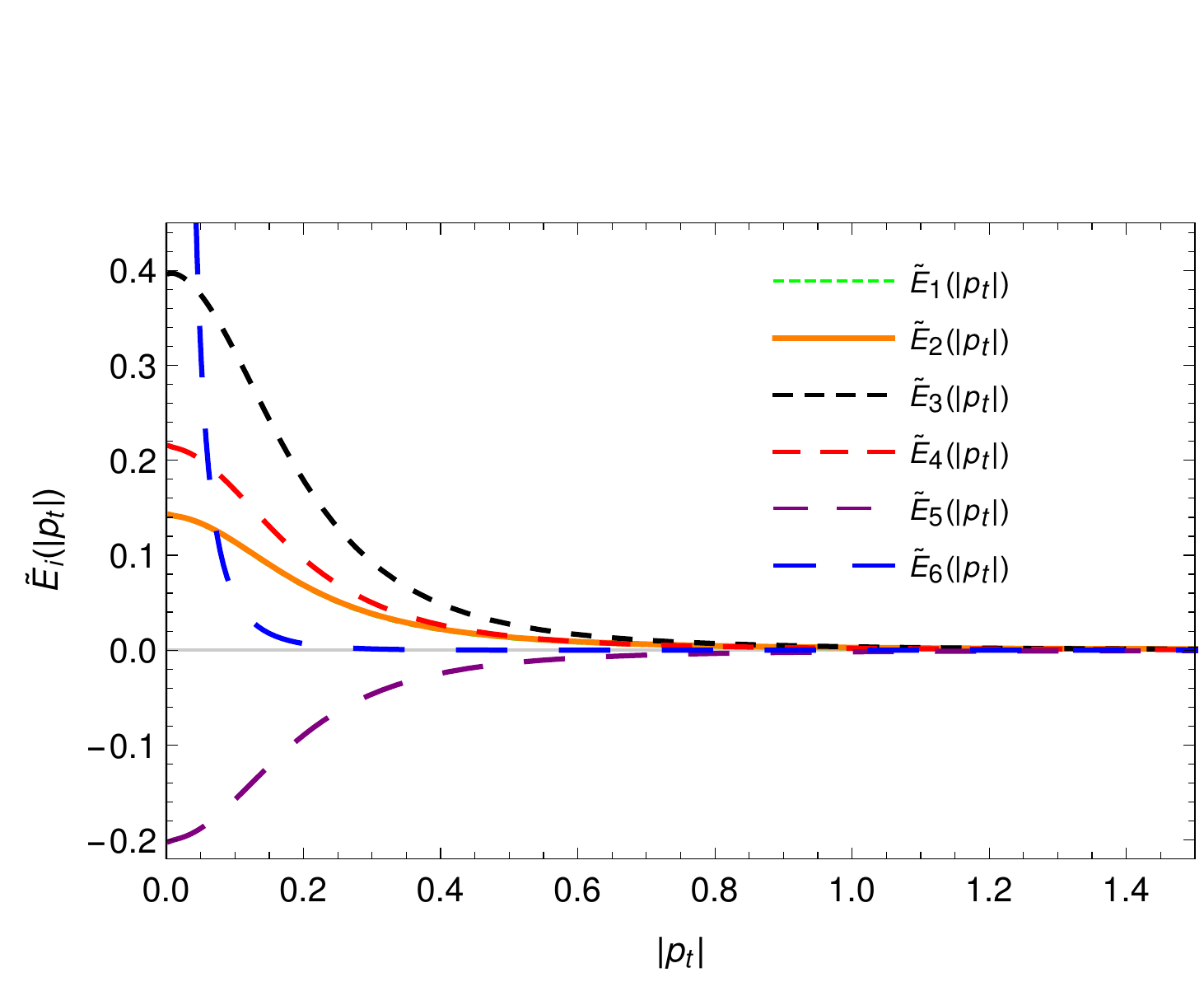}\\
  \caption{The BS wave functions of $\Xi_{bb}$ (spin-$1/2$) without taking the heavy quark limit when $\kappa=0.02\,\rm GeV^3$ ($\widetilde E_6(|p_t|)$ is 5774 when $|p_t|$ is 0).}\label{fig10}
\end{figure}

\begin{figure}[h!]
 \centering
  \includegraphics[width=0.65\textwidth]{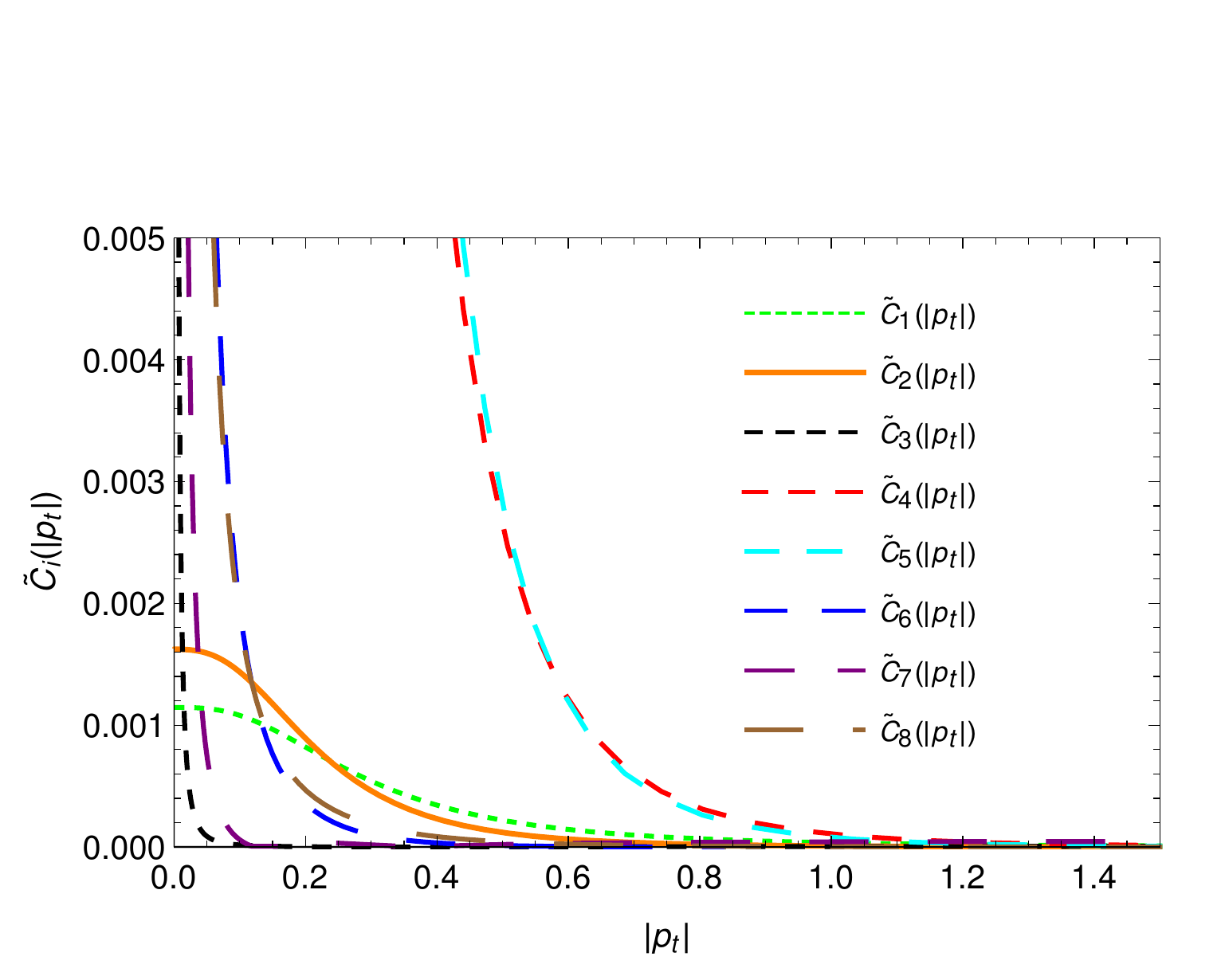}\\
  \caption{The BS wave functions of $\Xi_{bb}^\ast$ (spin-$3/2$) without taking the heavy quark limit when $\kappa=0.02\,\rm GeV^3$ ($\widetilde C_3(|p_t|)$, $\widetilde C_4(|p_t|)$, $\widetilde C_5(|p_t|)$, $\widetilde C_6(|p_t|)$, $\widetilde C_7(|p_t|)$ and $\widetilde C_8(|p_t|)$ are 0.3, 3274, 4687, 21, 2 and 17 respectively when $|p_t|$ is 0).}\label{fig11}
\end{figure}

In Figs.~\ref{fig6}-\ref{fig10}, we plot some of the BS wave functions of doubly heavy baryons as examples. In Figs.~\ref{fig6} and \ref{fig7}, we plot the scalar wave functions of $\Xi_{cc}^{(\ast)}$ and $\Xi_{bb}^{(\ast)}$ under the heavy quark limit with $\kappa$ fixed at $0.02\,\rm GeV^3$. It can be seen form these figures that wave functions of $\Xi_{cc}^{(\ast)}$ decrease more rapidly than those of $\Xi_{bb}^{\ast}$, which is quite similar to the situation for heavy diquarks. Also, it can be seen form the figures that when the heavy quarks have finite masses (Figs.~\ref{fig8} and \ref{fig9}) the curves of $\Xi_{cc}$ are clearly steeper than those of $\Xi_{bb}$ due to the mass difference between $b$ and $c$ quarks. In Figs.~\ref{fig8} and \ref{fig9}, we can see some of the curves, for example, $\widetilde E_6(|p_t|)$ (about $10^5$ times bigger than $\widetilde E_3(|p_t|)$ at zero relative momentum) of $\Xi_{cc}$ is significantly bigger in the vicinity of zero relative momentum and decreases much more rapidly than the rest of the scalar functions, and six scalar functions ($\widetilde C_3$-$\widetilde C_8$) in Fig.~\ref{fig9} are extremely large when $|p_t|$ approaches zero. The exact values of these scalar functions are given below the figures. In Figs.~\ref{fig10} and \ref{fig11}, we also plot the scalar functions for $\Xi_{bb}$ and $\Xi_{bb}^\ast$ without taking the heavy quark limit, which are similar to but broader than those in Figs.~\ref{fig8} and \ref{fig9} due to smaller radii of $\Xi_{bb}$ and $\Xi_{bb}^\ast$.

In our calculations where the heavy quark has finite mass, $\Xi_{cc}^\ast$ lies about $70\sim80\,\rm MeV$ above $\Xi_{cc}$, which is very close to the splitting between $\Sigma_c$ and $\Sigma_c^\ast$ ($64.44\,\rm MeV$ \cite{Patrignani:2016xqp}). This is consistent with the Gell-Mann$-$Okubo (GMO) mass relation \cite{Johnson:1976is,Weng:2018mmf}
\begin{equation}\label{GMO1}
M_{\Xi_{cc}^\ast}-M_{\Xi_{cc}}=M_{\Sigma_c^\ast}-M_{\Sigma_c}.
\end{equation}
The similar relation is the following with $u$, $d$ quarks being replaced by $s$ quark:
\begin{equation}\label{GMO2}
M_{\Omega_{cc}^\ast}-M_{\Omega_{cc}}=M_{\Omega_c^\ast}-M_{\Omega_c}.
\end{equation}
From Table~\ref{4}, we have the splitting around $60\sim70 \,\rm MeV$ on the left-hand side, which is also relatively close to the one between $\Omega_c^\ast$ and $\Omega_c$ ($70.7\,\rm MeV$). From the splitting information extracted here, we are able to analyze some decay possibilities. For example, since the splitting between two doubly charmed baryons $\Xi_{cc}^\ast$ and $\Xi_{cc}$ (and $\Omega_{cc}^\ast$ and $\Omega_{cc}$) is not big enough for the $\pi$ meson emission of $\Xi_{cc}^\ast$ $(\Omega_{cc}^\ast)$, $\Xi_{cc}^\ast$ $(\Omega_{cc}^\ast)$ can only transfer to $\Xi_{cc}$ $(\Omega_{cc})$ via $\gamma$ emission.

Similarly, we have the following relation for doubly bottomed baryons:
\begin{equation}\label{GMO3}
M_{\Xi_{bb}^\ast}-M_{\Xi_{bb}}= M_{\Sigma_b^\ast}-M_{\Sigma_b}.
\end{equation}
The splitting on the right-hand side of Eq.~\eqref{GMO3} is $20.2\,\rm MeV$ \cite{Patrignani:2016xqp}, in comparison to our result, about $10\sim20\,\rm MeV$, on the left-hand side, which is very close to the one on the right. Moreover, it is noted that the splitting between doubly bottomed baryons is much smaller than that between doubly charmed baryons, which is due to the fact that $b$ quark is much heavier than $c$ quark, and the hyperfine splitting is reciprocal to the mass of the heavy quark.
\section{Summary and discussion}
It is expected that more doubly heavy baryons will be observed in the future experiments, so theoretical studies for doubly heavy baryons are urgent at present. In this paper, we studied the masses of doubly heavy baryons in the BS formalism. We derived the BS equations for different diquark and baryon systems in the covariant instantaneous approximation, with and without taking the heavy quark limit, respectively. Then we discretized the integral equations and solved the eigenvalue equations numerically with the kernel containing the color confinement and one-gluon-exchange terms. We first calculated the masses of the doubly heavy diquarks $cc$, $bc$ and $bb$, and gave the corrections to the masses in the heavy quark limit. After that, by fitting the experimental data for some known heavy baryons, we obtained the proper values for the parameter $\alpha_{seff}$ in the two scenarios (with and out taking the heavy quark limit) corresponding to $\kappa$ ranging from $0.02\,\rm GeV^3$ to $0.08\,\rm GeV^3$, where the masses of the light diquarks ($uu$, $ud$, $ss$) were calculated from BS equations for them and were taken as the inputs during the fitting.

We then used the masses of the doubly heavy diquarks and the coupling parameters as the inputs to calculate the masses of doubly heavy baryons. We obtained the masses of $\Xi_{cc}^{(\ast)}$, which are $3.60\sim3.65\,\rm GeV$ in the heavy quark limit, $3.53\sim3.56\,\rm GeV$ for $\Xi_{cc}$ and $3.61\sim3.63\,\rm GeV$ for $\Xi_{cc}^\ast$ when $c$ quark has finite mass. These results are consistent with the data from SELEX ($3519\pm2\,\rm MeV$ for $\Xi_{cc}^+$) and LHCb ($3621.40\pm0.72\pm0.27\pm0.14\,\rm MeV$ for $\Xi_{cc}^{++}$) \cite{Aaij:2017ueg,Mattson:2002vu,Ocherashvili:2004hi}. Furthermore, with the increase in the masses of the doubly heavy baryons, we found the corrections to the results in the heavy quark limit become smaller. Besides, we found the results obtained with $m_Q$ being finite depend hardly on parameters and less sensitive to the variation of $\alpha_{seff}$ in our model for the masses of both diquarks and baryons, and because of this, our predictions for these doubly heavy baryons could help locate more doubly heavy baryons in the experiments with more accuracy. On top of that, in the situation where the heavy quarks have finite masses, we obtained the hyperfine splittings between spin-$1/2$ and spin-$3/2$ doubly heavy baryons, which satisfy the GMO mass relations quite well.
\section{ACKNOWLEDGEMENT}
One of us (Qi-Xin Yu) would like to thank Dr.~M.-H.~Weng and Dr.~L.-L.~Liu for helpful discussions, he is also indebted to Gustavo Hazel for the help with programming. This work was supported by the National Natural Science Foundation of China (Projects No. 11575023, No. 11775024 and No. 11805153).
\begin{appendix}\label{appendix}
\section{The explicit expressions for coupled integral equations for $\widetilde E_i$ $(i=1,\cdots,6)$ for spin-$\frac{1}{2}$ doubly heavy baryons}\label{appa}
In order to show the order of the corrections obviously and compare different BS equations while taking and not taking the heavy quark limit, we present explicit formulas here for the BS equations in Subsection.~\ref{import}. It should be noted that the following lengthy equations can be significantly reduced down to just a few terms if the heavy quark limit is taken, same for
the equations in Appendix.~\ref{appb}\footnote{The integrals equations in both Appendix.~\ref{appa} and \ref{appb} had been checked thoroughly, and programs can be provided upon requested.}
\begin{align*}
\widetilde E_1(k_t)=&\frac{-1}{4\omega_l\omega_D(M^2-(\omega_l+\omega_D)^2)m_D^2}\int\frac{d^3q_t}{(2\pi)^3}\bigg\{\Big(V_2 \omega _D^2 m_l \left(k_t\cdot q_t\right)-V_2 \omega _D \omega _l^2 \left(k_t\cdot q_t\right)\\
&-M V_2 \omega _D^2 \left(k_t\cdot q_t\right)-V_2 \omega _D^3 \left(k_t\cdot q_t\right)+M^2 V_2 m_l \left(k_t\cdot q_t\right)-V_2 m_l \omega _l^2 \left(k_t\cdot q_t\right)\\
&-M^3 V_2 \left(k_t\cdot q_t\right)-M^2 V_1 \omega _D m_l+2 M^2 V_2 m_D^2 m_l+2 M V_2 m_D^2 \omega _l^2+2 M V_2 m_D^2 m_l \omega _l\\
&-2 M V_1 \omega _D m_l \omega _l+2 M V_2 m_D^2 \omega _D \omega _l+M V_1 \omega _D^2 m_l-2 M V_1 m_D^2 m_l+M^2 V_2 \omega _D \left(k_t\cdot q_t\right)\\
&+2 V_1 m_D^2 \omega _l^2-V_1 \omega _D m_l \omega _l^2+2 V_2 m_D^2 \omega _D \omega _l^2+V_1 \omega _D^2 m_l \omega _l+2 V_1 m_D^2 \omega _D \omega _l+V_1 \omega _D^3 m_l\\
&+2 M k_t^2 V_2 \omega _D \omega _l-2 M V_1 \omega _D \omega _l^2-M V_1 \omega _D^2 \omega _l+k_t^2 V_2 \omega _D \omega _l^2-V_1 \omega _D \omega _l^3-V_1 \omega _D^2 \omega _l^2\\
&-V_1 \omega _D^3 \omega _l+2 M k_t^2 V_2 m_D^2+M^2 k_t^2 V_2 \omega _D-k_t^2 V_2 \omega _D^3+M^3 V_1 m_l-M^2 V_1 \omega _D \omega _l\\
&+2 M V_2 m_D^2 \left(k_t\cdot q_t\right)+M V_2 \omega _l^2 \left(k_t\cdot q_t\right)+2 V_2 m_D^2 \omega _l^3+M^2 V_1 m_l \omega _l-M V_1 m_l \omega _l^2\\
&-V_1 m_l \omega _l^3+M^3 V_1 \omega _l+M^2 V_1 \omega _l^2-M V_1 \omega _l^3-V_1 \omega _l^4-2 V_2 \omega _D^2 \omega _l \left(k_t\cdot q_t\right)\Big)\widetilde E_1(q_t)\\
&+\Big(3 V_2 \omega _l^5+V_1 \omega _l^4+6 M V_2 \omega _l^4+3 m_l V_2 \omega _l^4+3 V_2 \omega _D \omega _l^4+3 V_2 \omega _D^2 \omega _l^3+M V_1 \omega _l^3\\
&+m_l V_1 \omega _l^3+k_t^2 V_2 \omega _l^3-12 m_D^2 V_2 \omega _l^3+2 \left(k_t\cdot q_t\right) V_2 \omega _l^3+6 M m_l V_2 \omega _l^3+V_1 \omega _D \omega _l^3\\
&+9 M V_2 \omega _D \omega _l^3+3 m_l V_2 \omega _D \omega _l^3+3 V_2 \omega _D^3 \omega _l^2+V_1 \omega _D^2 \omega _l^2+6 M V_2 \omega _D^2 \omega _l^2-3 m_l V_2 \omega _D^2 \omega _l^2\\
&-M^2 V_1 \omega _l^2+k_t^2 V_1 \omega _l^2-4 m_D^2 V_1 \omega _l^2+M m_l V_1 \omega _l^2-6 M^3 V_2 \omega _l^2-3 \left(k_t\cdot q_t\right) m_l V_2 \omega _D^2\\
&-M \left(k_t\cdot q_t\right) V_2 \omega _l^2+3 \left(k_t\cdot q_t\right) m_l V_2 \omega _l^2+2 M V_1 \omega _D \omega _l^2+m_l V_1 \omega _D \omega _l^2+9 M^2 V_2 \omega _D \omega _l^2\\
&-12 m_D^2 V_2 \omega _D \omega _l^2+5 \left(k_t\cdot q_t\right) V_2 \omega _D \omega _l^2+9 M m_l V_2 \omega _D \omega _l^2+V_1 \omega _D^3 \omega _l+3 M V_2 \omega _D^3 \omega _l\\
&+M V_1 \omega _D^2 \omega _l-m_l V_1 \omega _D^2 \omega _l+3 M^2 V_2 \omega _D^2 \omega _l-k_t^2 V_2 \omega _D^2 \omega _l+4 \left(k_t\cdot q_t\right) V_2 \omega _D^2 \omega _l\\
&-M^3 V_1 \omega _l-M^2 m_l V_1 \omega _l-3 M^4 V_2 \omega _l-M^2 k_t^2 V_2 \omega _l-2 M^2 \left(k_t\cdot q_t\right) V_2 \omega _l-6 M^3 m_l V_2 \omega _l\\
&+M^2 V_1 \omega _D \omega _l-4 m_D^2 V_1 \omega _D \omega _l+2 M m_l V_1 \omega _D \omega _l+3 M^3 V_2 \omega _D \omega _l-4 M k_t^2 V_2 \omega _D \omega _l\\
&+4 M \left(k_t\cdot q_t\right) V_2 \omega _D \omega _l+9 M^2 m_l V_2 \omega _D \omega _l-m_l V_1 \omega _D^3+2 k_t^2 V_2 \omega _D^3+\left(k_t\cdot q_t\right) V_2 \omega _D^3\\
&-k_t^2 V_1 \omega _D^2-M m_l V_1 \omega _D^2-M k_t^2 V_2 \omega _D^2+M \left(k_t\cdot q_t\right) V_2 \omega _D^2-3 M^2 m_l V_2 \omega _D^2+M k_t^2 V_2 \omega _l^2\\
&-M^2 k_t^2 V_1-M^3 m_l V_1-M^3 k_t^2 V_2+M^3 \left(k_t\cdot q_t\right) V_2-3 M^4 m_l V_2-3 M^2 \left(k_t\cdot q_t\right) m_l V_2\\
&+M^2 m_l V_1 \omega _D-2 M^2 k_t^2 V_2 \omega _D-M^2 \left(k_t\cdot q_t\right) V_2 \omega _D+3 M^3 m_l V_2 \omega _D-12 M m_D^2 V_2 \omega _D \omega _l\\
&-12 M m_D^2 V_2 \omega _l^2-2 k_t^2 V_2 \omega _D \omega _l^2-3 m_l V_2 \omega _D^3 \omega _l-6 M m_l V_2 \omega _D^2 \omega _l-3 M m_l V_2 \omega _D^3\Big)\widetilde E_2(q_t)\\
\end{align*}
\begin{align*}
&+\Big(M^2 V_2 \omega _D m_l \left(k_t\cdot q_t\right)+M V_2 \omega _D^2 m_l \left(k_t\cdot q_t\right)+2 M V_2 m_D^2 \omega _l \left(k_t\cdot q_t\right)+2 M V_2 \omega _D m_l \omega _l \left(k_t\cdot q_t\right)\\
&-V_2 \omega _D^3 m_l \left(k_t\cdot q_t\right)-V_1 \omega _D^2 m_l \left(k_t\cdot q_t\right)-4 V_2 m_D^2 \omega _l^2 \left(k_t\cdot q_t\right)+V_2 \omega _D m_l \omega _l^2 \left(k_t\cdot q_t\right)\\
&+V_2 \omega _D^2 m_l \omega _l \left(k_t\cdot q_t\right)-4 V_2 m_D^2 \omega _D \omega _l \left(k_t\cdot q_t\right)-3 M V_2 \omega _D \omega _l^2 \left(k_t\cdot q_t\right)-3 M V_2 \omega _D^2 \omega _l \left(k_t\cdot q_t\right)\\
&+2 M V_1 \omega _D \omega _l \left(k_t\cdot q_t\right)-2 V_2 \omega _D \omega _l^3 \left(k_t\cdot q_t\right)-2 V_2 \omega _D^2 \omega _l^2 \left(k_t\cdot q_t\right)+2 V_1 \omega _D \omega _l^2 \left(k_t\cdot q_t\right)\\
&+V_1 \omega _D^2 \omega _l \left(k_t\cdot q_t\right)+2 M^2 V_2 m_D^2 \left(k_t\cdot q_t\right)+M^3 V_2 \omega _D \left(k_t\cdot q_t\right)-M^2 V_2 \omega _D^2 \left(k_t\cdot q_t\right)\\
&-M V_2 \omega _D^3 \left(k_t\cdot q_t\right)+M^3 V_2 m_l \left(k_t\cdot q_t\right)+M^2 V_2 m_l \omega _l \left(k_t\cdot q_t\right)-M^2 V_1 m_l \left(k_t\cdot q_t\right)\\
&-M V_2 m_l \omega _l^2 \left(k_t\cdot q_t\right)-V_2 m_l \omega _l^3 \left(k_t\cdot q_t\right)+V_1 m_l \omega _l^2 \left(k_t\cdot q_t\right)-M^3 V_2 \omega _l \left(k_t\cdot q_t\right)\\
&+M^2 V_2 \omega _l^2 \left(k_t\cdot q_t\right)-M^2 V_1 \omega _l \left(k_t\cdot q_t\right)+M V_2 \omega _l^3 \left(k_t\cdot q_t\right)+V_1 \omega _l^3 \left(k_t\cdot q_t\right)-M^4 V_2\left(k_t\cdot q_t\right)\Big)\widetilde E_3(q_t)\\
&+\Big(q_t^2 V_2 \omega _l^4-\left(k_t\cdot q_t\right) V_1 \omega _l^3+M q_t^2 V_2 \omega _l^3+q_t^2 m_l V_2 \omega _l^3+q_t^2 V_2 \omega _D \omega _l^3-\left(k_t\cdot q_t\right) V_2 \omega _D \omega _l^3\\
&+q_t^2 V_2 \omega _D^2 \omega _l^2-M \left(k_t\cdot q_t\right) V_1 \omega _l^2+\left(k_t\cdot q_t\right)^2 V_2 \omega _l^2-M^2 q_t^2 V_2 \omega _l^2-2 q_t^2 m_D^2 V_2 \omega _l^2+2 a_2 m_D^2 V_2 \omega _l^2\\
&-2 \left(k_t\cdot q_t\right) m_D^2 V_2 \omega _l^2+M q_t^2 m_l V_2 \omega _l^2-\left(k_t\cdot q_t\right) V_1 \omega _D \omega _l^2+2 M q_t^2 V_2 \omega _D \omega _l^2-2 M \left(k_t\cdot q_t\right) V_2 \omega _D \omega _l^2\\
&+q_t^2 m_l V_2 \omega _D \omega _l^2+\left(k_t\cdot q_t\right) m_l V_2 \omega _D \omega _l^2+q_t^2 V_2 \omega _D^3 \omega _l+\left(k_t\cdot q_t\right) V_2 \omega _D^3 \omega _l+\left(k_t\cdot q_t\right) V_1 \omega _D^2 \omega _l\\
&-q_t^2 m_l V_2 \omega _D^2 \omega _l+M^2 \left(k_t\cdot q_t\right) V_1 \omega _l-M^3 q_t^2 V_2 \omega _l-2 M \left(k_t\cdot q_t\right) m_D^2 V_2 \omega _l-M^2 q_t^2 m_l V_2 \omega _l\\
&+M^2 q_t^2 V_2 \omega _D \omega _l-2 q_t^2 m_D^2 V_2 \omega _D \omega _l+2 a_2 m_D^2 V_2 \omega _D \omega _l-2 \left(k_t\cdot q_t\right) m_D^2 V_2 \omega _D \omega _l-M^2 \left(k_t\cdot q_t\right) V_2 \omega _D \omega _l\\
&+2 M q_t^2 m_l V_2 \omega _D \omega _l+2 M \left(k_t\cdot q_t\right) m_l V_2 \omega _D \omega _l+\left(k_t\cdot q_t\right) V_1 \omega _D^3-q_t^2 m_l V_2 \omega _D^3-\left(k_t\cdot q_t\right) m_l V_2 \omega _D^3\\
&-\left(k_t\cdot q_t\right)^2 V_2 \omega _D^2-M q_t^2 m_l V_2 \omega _D^2-2 M \left(k_t\cdot q_t\right) m_D^2 V_1+M^3 \left(k_t\cdot q_t\right) V_1-M^2 \left(k_t\cdot q_t\right)^2 V_2\\
&-2 M^2 \left(k_t\cdot q_t\right) m_D^2 V_2-M^3 q_t^2 m_l V_2+2 M q_t^2 m_D^2 m_l V_2+2 a_2 M m_D^2 m_l V_2+2 M \left(k_t\cdot q_t\right) m_D^2 m_l V_2\\
&-M^2 \left(k_t\cdot q_t\right) V_1 \omega _D+M^2 q_t^2 m_l V_2 \omega _D+M^2 \left(k_t\cdot q_t\right) m_l V_2 \omega _D+M q_t^2 V_2 \omega _D^2 \omega _l\\
&-2 M \left(k_t\cdot q_t\right) V_1 \omega _D \omega _l+M \left(k_t\cdot q_t\right) V_1 \omega _D^2\Big)\widetilde E_4(q_t)\\
&+\Big(-3 \left(k_t\cdot q_t\right) V_2 M^4+\left(k_t\cdot q_t\right) V_1 M^3-q_t^2 m_l V_2 M^3+\left(k_t\cdot q_t\right) m_l V_2 M^3+3 \left(k_t\cdot q_t\right) V_2 \omega _D M^3\\
&-q_t^2 V_2 \omega _l M^3-5 \left(k_t\cdot q_t\right) V_2 \omega _l M^3-3 \left(k_t\cdot q_t\right) V_2 \omega _D^2 M^2-q_t^2 V_2 \omega _l^2 M^2+\left(k_t\cdot q_t\right) V_2 \omega _l^2 M^2\\
&-\left(k_t\cdot q_t\right) m_l V_1 M^2-2 \left(k_t\cdot q_t\right)^2 V_2 M^2-k_t^2 q_t^2 V_2 M^2-\left(k_t\cdot q_t\right) V_1 \omega _D M^2+q_t^2 m_l V_2 \omega _D M^2\\
&+2 \left(k_t\cdot q_t\right) m_l V_2 \omega _D M^2-q_t^2 m_l V_2 \omega _l M^2+\left(k_t\cdot q_t\right) m_l V_2 \omega _l M^2+q_t^2 V_2 \omega _D \omega _l M^2\\
&-3 \left(k_t\cdot q_t\right) V_2 \omega _D^3 M+q_t^2 V_2 \omega _l^3 M+5 \left(k_t\cdot q_t\right) V_2 \omega _l^3 M+\left(k_t\cdot q_t\right) V_1 \omega _D^2 M-q_t^2 m_l V_2 \omega _D^2 M\\
&+\left(k_t\cdot q_t\right) m_l V_2 \omega _D^2 M-\left(k_t\cdot q_t\right) V_1 \omega _l^2 M+q_t^2 m_l V_2 \omega _l^2 M-\left(k_t\cdot q_t\right) m_l V_2 \omega _l^2 M+2 q_t^2 V_2 \omega _D \omega _l^2 M\\
\end{align*}
\begin{align}\label{e1}
&+\left(k_t\cdot q_t\right) V_2 \omega _D \omega _l^2 M+4 a_2 m_D^2 m_l V_2 M+q_t^2 V_2 \omega _D^2 \omega _l M-7 \left(k_t\cdot q_t\right) V_2 \omega _D^2 \omega _l M+2 q_t^2 m_l V_2 \omega _D \omega _l M\nonumber\\
&+4 \left(k_t\cdot q_t\right) m_l V_2 \omega _D \omega _l M+q_t^2 V_2 \omega _l^4+2 \left(k_t\cdot q_t\right) V_2 \omega _l^4+\left(k_t\cdot q_t\right) V_1 \omega _D^3-q_t^2 m_l V_2 \omega _D^3\nonumber\\
&-2 \left(k_t\cdot q_t\right) m_l V_2 \omega _D^3+q_t^2 m_l V_2 \omega _l^3-\left(k_t\cdot q_t\right) m_l V_2 \omega _l^3+q_t^2 V_2 \omega _D \omega _l^3-\left(k_t\cdot q_t\right) V_2 \omega _D \omega _l^3\nonumber\\
&-\left(k_t\cdot q_t\right) m_l V_1 \omega _D^2-2 \left(k_t\cdot q_t\right)^2 V_2 \omega _D^2-k_t^2 q_t^2 V_2 \omega _D^2+q_t^2 V_2 \omega _D^2 \omega _l^2-4 \left(k_t\cdot q_t\right) V_2 \omega _D^2 \omega _l^2\nonumber\\
&+\left(k_t\cdot q_t\right) m_l V_1 \omega _l^2+2 \left(k_t\cdot q_t\right)^2 V_2 \omega _l^2+k_t^2 q_t^2 V_2 \omega _l^2-4 q_t^2 m_D^2 V_2 \omega _l^2+4 a_2 m_D^2 V_2 \omega _l^2\nonumber\\
&-8 \left(k_t\cdot q_t\right) m_D^2 V_2 \omega _l^2+\left(k_t\cdot q_t\right) V_1 \omega _D \omega _l^2+q_t^2 m_l V_2 \omega _D \omega _l^2+2 \left(k_t\cdot q_t\right) m_l V_2 \omega _D \omega _l^2\nonumber\\
&+q_t^2 V_2 \omega _D^3 \omega _l-\left(k_t\cdot q_t\right) V_2 \omega _D^3 \omega _l+2 \left(k_t\cdot q_t\right) V_1 \omega _D^2 \omega _l-q_t^2 m_l V_2 \omega _D^2 \omega _l+\left(k_t\cdot q_t\right) m_l V_2 \omega _D^2 \omega _l\nonumber\\
&-4 q_t^2 m_D^2 V_2 \omega _D \omega _l+4 a_2 m_D^2 V_2 \omega _D \omega _l-8 \left(k_t\cdot q_t\right) m_D^2 V_2 \omega _D \omega _l+5 \left(k_t\cdot q_t\right) V_2 \omega _D \omega _l M^2\Big)\widetilde E_5(q_t)\nonumber\\
&+\Big(q_t^2 V_2 \omega _l^5+2 M q_t^2 V_2 \omega _l^4+q_t^2 m_l V_2 \omega _l^4+q_t^2 V_2 \omega _D \omega _l^4+q_t^2 V_2 \omega _D^2 \omega _l^3+3 M q_t^2 m_l V_2 \omega _D \omega _l^2\nonumber\\
&+\left(k_t\cdot q_t\right)^2 V_2 \omega _l^3-2 q_t^2 m_D^2 V_2 \omega _l^3+4 a_2 m_D^2 V_2 \omega _l^3+2 M q_t^2 m_l V_2 \omega _l^3+3 M q_t^2 V_2 \omega _D \omega _l^3\nonumber\\
&+q_t^2 m_l V_2 \omega _D \omega _l^3+q_t^2 V_2 \omega _D^3 \omega _l^2+2 M q_t^2 V_2 \omega _D^2 \omega _l^2-q_t^2 m_l V_2 \omega _D^2 \omega _l^2+\left(k_t\cdot q_t\right)^2 V_1 \omega _l^2\nonumber\\
&+2 a_2 m_D^2 V_1 \omega _l^2-2 M^3 q_t^2 V_2 \omega _l^2-2 M q_t^2 m_D^2 V_2 \omega _l^2+4 a_2 M m_D^2 V_2 \omega _l^2+\left(k_t\cdot q_t\right)^2 M V_2 \omega _l^2\nonumber\\
&-\left(k_t\cdot q_t\right)^2 V_2 \omega _D \omega _l^2+3 M^2 q_t^2 V_2 \omega _D \omega _l^2-2 q_t^2 m_D^2 V_2 \omega _D \omega _l^2+4 a_2 m_D^2 V_2 \omega _D \omega _l^2+M^3 q_t^2 V_2 \omega _D \omega _l\nonumber\\
&+M q_t^2 V_2 \omega _D^3 \omega _l-q_t^2 m_l V_2 \omega _D^3 \omega _l-\left(k_t\cdot q_t\right)^2 V_2 \omega _D^2 \omega _l+M^2 q_t^2 V_2 \omega _D^2 \omega _l-2 M q_t^2 m_l V_2 \omega _D^2 \omega _l\nonumber\\
&-M^2 \left(k_t\cdot q_t\right)^2 V_2 \omega _l-M^4 q_t^2 V_2 \omega _l-2 M^3 q_t^2 m_l V_2 \omega _l+2 M q_t^2 m_D^2 m_l V_2 \omega _l+4 a_2 M m_D^2 m_l V_2 \omega _l\nonumber\\
&+2 a_2 m_D^2 V_1 \omega _D \omega _l-2 M \left(k_t\cdot q_t\right)^2 V_2 \omega _D \omega _l-2 M q_t^2 m_D^2 V_2 \omega _D \omega _l+4 a_2 M m_D^2 V_2 \omega _D \omega _l\nonumber\\
&+3 M^2 q_t^2 m_l V_2 \omega _D \omega _l+\left(k_t\cdot q_t\right)^2 V_2 \omega _D^3-M q_t^2 m_l V_2 \omega _D^3-\left(k_t\cdot q_t\right)^2 V_1 \omega _D^2-M \left(k_t\cdot q_t\right)^2 V_2 \omega _D^2\nonumber\\
&-M^2 q_t^2 m_l V_2 \omega _D^2-M^2 \left(k_t\cdot q_t\right)^2 V_1+2 a_2 M m_D^2 m_l V_1-M^3 \left(k_t\cdot q_t\right)^2 V_2-2 a_2 M k_t^2 m_D^2 V_2\nonumber\\
&-M^4 q_t^2 m_l V_2+4 a_2 M^2 m_D^2 m_l V_2+2 M^2 q_t^2 m_D^2 m_l V_2-M^2 \left(k_t\cdot q_t\right)^2 V_2 \omega _D+M^3 q_t^2 m_l V_2 \omega _D\Big)\widetilde E_6(q_t)\bigg\},
\end{align}
\begin{align*}
\widetilde E_2(k_t)=&\frac{-1}{4\omega_l\omega_D(M^2-(\omega_l+\omega_D)^2)m_D^2}\int\frac{d^3q_t}{(2\pi)^3}\bigg\{\Big(2 M^2 V_2 m_D^2 m_l+2 M V_2 m_D^2 \omega _l^2\\
&+2 M V_2 m_D^2 m_l \omega _l+2 M V_2 m_D^2 \omega _D \omega _l+2 V_2 m_D^2 \omega _l^3+2 V_2 m_D^2 \omega _D \omega _l^2\Big)\widetilde E_1(q_t)\\
&+\Big(2 M V_2 m_D^2 \left(k_t\cdot q_t\right)-6 M^2 V_2 m_D^2 m_l-6 M V_2 m_D^2 \omega _l^2-6 M V_2 m_D^2 m_l \omega _l\\
&-2 M V_1 m_D^2 m_l-6 V_2 m_D^2 \omega _l^3-2 V_1 m_D^2 \omega _l^2-6 V_2 m_D^2 \omega _D \omega _l^2-2 V_1 m_D^2 \omega _D \omega _l\\
&-6 M V_2 m_D^2 \omega _D \omega _l+4 M k_t^2 V_2 m_D^2\Big)\widetilde E_2(q_t)
\end{align*}
\begin{align}\label{e2}
&+\Big(-2 M V_2 m_D^2 m_l \left(k_t\cdot q_t\right)-2 V_2 m_D^2 \omega _l^2 \left(k_t\cdot q_t\right)-2 V_2 m_D^2 \omega _D \omega _l \left(k_t\cdot q_t\right)\Big)\widetilde E_3(q_t)\nonumber\\
&+\Big(-2 M V_2 m_D^2 \omega _l \left(k_t\cdot q_t\right)-2 M^2 V_2 m_D^2 \left(k_t\cdot q_t\right)-2 a_1 M V_2 m_D^2 m_l\nonumber\\
&-2 a_1 V_2 m_D^2 \omega _l^2-2 a_1 V_2 m_D^2 \omega _D \omega _l\Big)\widetilde E_4(q_t)\nonumber\\
&+\Big(-6 M V_2 m_D^2 \omega _l \left(k_t\cdot q_t\right)-4 M V_2 m_D^2 m_l \left(k_t\cdot q_t\right)-4 V_2 m_D^2 \omega _l^2 \left(k_t\cdot q_t\right)-4 V_2 m_D^2 \omega _D \omega _l \left(k_t\cdot q_t\right)\nonumber\\
&-6 M^2 V_2 m_D^2 \left(k_t\cdot q_t\right)+2 M V_1 m_D^2 \left(k_t\cdot q_t\right)-4 a_1 M V_2 m_D^2 m_l-4 a_1 V_2 m_D^2 \omega _l^2-4 a_1 V_2 m_D^2 \omega _D \omega _l\nonumber\\
&-2 M q_t^2 V_2 m_D^2 m_l-2 q_t^2 V_2 m_D^2 \omega _l^2-2 q_t^2 V_2 m_D^2 \omega _D \omega _l\Big)\widetilde E_5(q_t)\nonumber\\
&+\Big(2 M V_2 m_D^2 \left(k_t\cdot q_t\right)^2-4 a_1 M^2 V_2 m_D^2 m_l-4 a_1 M V_2 m_D^2 \omega _l^2-4 a_1 M V_2 m_D^2 m_l \omega _l\nonumber\\
&-4 a_1 M V_2 m_D^2 \omega _D \omega _l-2 a_1 M V_1 m_D^2 m_l-4 a_1 V_2 m_D^2 \omega _l^3-2 a_1 V_1 m_D^2 \omega _l^2-4 a_1 V_2 m_D^2 \omega _D \omega _l^2\nonumber\\
&-2 a_1 V_1 m_D^2 \omega _D \omega _l+2 a_1 M k_t^2 V_2 m_D^2\Big)\widetilde E_6(q_t)\bigg\},
\end{align}
\begin{align*}
\widetilde E_3(k_t)=&\frac{-1}{4\omega_l\omega_D(M^2-(\omega_l+\omega_D)^2)m_D^2}\int\frac{d^3q_t}{(2\pi)^3}\bigg\{\Big(-m_l \big[2 M V_2 \left(k_t^2-m_D^2\right) \left(k_t\cdot q_t\right)\nonumber\\
&+k_t^2 V_1 \left(\omega _D^2-\omega _l^2+M^2\right)\big]+V_2 \Big[\left(k_t\cdot q_t\right) \big[\left(\omega _D+\omega _l\right) \left(\omega _l \left(k_t^2-2 m_D^2\right)+k_t^2 \omega _D\right)+M^2 k_t^2\big]\nonumber\\
&+k_t^2 \big(4 M m_D^2 \left(\omega _l+M\right)-k_t^2 \left(-\omega _D+\omega _l+M\right) \left(\omega _D+\omega _l+M\right)\big)\Big]\nonumber\\
&+k_t^2 V_1 \omega _l \left(2 \omega _D \left(\omega _l+M\right)+\omega _D^2+\omega _l^2-M^2\right)\Big)\frac{1}{k_t^2}\widetilde E_1(q_t)\nonumber\\
&+\Big(V_2 \Big[k_t^2 \big(-4 m_D^2 \big[\omega _l \left(\omega _D+3 M\right)-M m_l+\omega _l^2+3 M^2\big]\nonumber\\
&+3 m_l \left(\omega _l+M\right) \left(\omega _D^2-\omega _l^2+M^2\right)-6 M^2 \omega _D \omega _l-12 M \omega _D \omega _l^2-3 M \omega _D^2 \omega _l\nonumber\\
&-6 \omega _D \omega _l^3-3 \omega _D^2 \omega _l^2-2 k_t^2 \omega _D^2+3 M^3 \omega _l+3 M^2 \omega _l^2+6 M k_t^2 \omega _l-3 M \omega _l^3\nonumber\\
&+2 k_t^2 \omega _l^2-3 \omega _l^4+4 M^2 k_t^2\big)-\left(k_t\cdot q_t\right) \big(m_l \left(4 M m_D^2-6 M k_t^2\right)+2 \omega _D \omega _l \left(3 k_t^2-2 m_D^2\right)\nonumber\\
&-4 m_D^2 \omega _l^2+k_t^2 \omega _D^2+5 k_t^2 \omega _l^2+M^2 k_t^2\big)\Big]+k_t^2 V_1 \Big[m_l \left(\omega _D^2-\omega _l^2+M^2\right)-2 M \omega _D \omega _l\nonumber\\
&-2 \omega _D \omega _l^2-\omega _D^2 \omega _l-4 M m_D^2+M^2 \omega _l-\omega _l^3+2 M k_t^2\Big]\Big)\frac{1}{k_t^2}\widetilde E_2(q_t)\nonumber\\
&+\Big((k_t\cdot q_t) \Big[m_l \big[V_2 \left(4 M m_D^2 \left(\omega _l+M\right)-k_t^2 \left(-\omega _D^2+4 M \omega _l+\omega _l^2+3 M^2\right)\right)\nonumber\\
&+2 M V_1 \left(k_t^2-m_D^2\right)\big]+V_2 \big[k_t^2 \big(2 \omega _D \omega _l \left(\omega _l+M\right)+M \omega _D^2+2 M^2 \omega _l+3 M \omega _l^2\nonumber\\
&+2 \omega _l^3+M^3\big)-2 m_D^2 \left(2 \omega _D \omega _l \left(\omega _l+M\right)+2 M \omega _l^2+2 \omega _l^3+M k_t^2\right)\big]\nonumber\\
&-2 V_1 \omega _l \left(\omega _D+\omega _l\right) \left(k_t^2-m_D^2\right)\Big]\Big)\frac{1}{k_t^2}\widetilde E_3(q_t)
\end{align*}
\begin{align}\label{e3}
&+\Big(2 M V_2 (k_t\cdot q_t)^2-(k_t\cdot q_t) \left(V_2 \left(m_l-\omega _l\right) \left(\left(\omega _l+M\right){}^2-\omega _D^2\right)+V_1 \left(\omega _D^2-\omega _l^2+M^2\right)\right)\nonumber\\
&-V_2 \Big[2 M m_D^2 \left(2 a_1+a_5 k_t^2\right)+q_t^2 \big[\omega _l \left(2 \omega _D \left(\omega _l+M\right)+\omega _D^2+\omega _l^2-M^2\right)\nonumber\\
&-m_l \big(\omega _D^2-\omega _l^2+M^2\big)\big]\Big]\Big)\widetilde E_4(q_t)\nonumber\\
&+\Big(4 M V_2 (k_t\cdot q_t)^2+(k_t\cdot q_t) \Big[V_2 \big[-2 m_l \left(-\omega _D^2+3 M \omega _l+\omega _l^2+2 M^2\right)+2 M \omega _D \omega _l+2 \omega _D \omega _l^2\nonumber\\
&+\omega _D^2 \omega _l-4 M m_D^2+3 M \omega _D^2+5 M^2 \omega _l+3 M \omega _l^2+\omega _l^3+3 M^3\big]-V_1 \big(2 \omega _D \omega _l+\omega _D^2-2 M m_l\nonumber\\
&+\omega _l^2+M^2\big)\Big]+V_2 \Big[q_t^2 \big(m_l \left(\omega _D^2-\omega _l^2+M^2\right)+\omega _l \left(-2 M \omega _D-\omega _D^2+M^2\right)-2 \omega _D \omega _l^2-\omega _l^3\nonumber\\
&+2 M k_t^2\big)-4 M m_D^2 \left(2 a_1+a_5 k_t^2+q_t^2\right)\Big]\Big)\widetilde E_5(q_t)\nonumber\\
&+\Big(\big[V_2 \left(-\omega _D^2+4 M \omega _l+\omega _l^2+3 M^2\right)+2 M V_1\big] (k_t\cdot q_t)^2-2 m_D^2 \left(2 a_1+a_5 k_t^2\right) \big[V_2 \big(\omega _l \left(\omega _D+2 M\right)\nonumber\\
&-M m_l+\omega _l^2+2 M^2\big)+M V_1\big]+q_t^2 V_2 \left(\omega _l+M\right) \big[m_l \left(\omega _D^2-\omega _l^2+M^2\right)\nonumber\\
&-\omega _l \left(2 \omega _D \left(\omega _l+M\right)+\omega _D^2+\omega _l^2-M^2\right)\big]\Big)\widetilde E_6(q_t)\bigg\},
\end{align}
\begin{align*}
\widetilde E_4(k_t)=&\frac{-1}{4\omega_l\omega_D(M^2-(\omega_l+\omega_D)^2)m_D^2}\int\frac{d^3q_t}{(2\pi)^3}\bigg\{\Big(V_2 \Big[k_t^2 \big[2 m_D^2 \left(\omega _l \left(\omega _D+M\right)+M m_l+\omega _l^2+M^2\right)\\
&+\omega _D \left(m_l+\omega _l\right) \left(\left(\omega _l+M\right){}^2-\omega _D^2\right)\big]-(k_t\cdot q_t) \big[m_l \big(\omega _D^2 \left(\omega _l+M\right)-\omega _D \left(\omega _l+M\right)^2\\
&-2 M m_D^2+\omega _D^3+\left(\omega _l+M\right)^2 \left(M-\omega _l\right)\big)-2 m_D^2 \omega _l^2-2 m_D^2 \omega _D \omega _l+M^2 \omega _D \omega _l\\
&+2 M \omega _D \omega _l^2+M \omega _D^2 \omega _l+\omega _D \omega _l^3+\omega _D^2 \omega _l^2+\omega _D^3 \omega _l-k_t^2 \omega _D^2-M^3 \omega _l-M^2 \omega _l^2+M \omega _l^3\\
&+k_t^2 \omega _l^2+\omega _l^4-M^2 k_t^2\big]\Big]+k_t^2 V_1 \Big[\omega _D^2 \left(\omega _l+M\right)-\omega _D \left(\omega _l+M\right)^2-2 M m_D^2+\omega _D^3\\
&+\left(\omega _l+M\right)^2 \left(M-\omega _l\right)\Big]\Big)\frac{1}{k_t^2}\widetilde E_1(q_t)\\
&+\Big(V_2 \Big[(k_t\cdot q_t) \big[\omega _D \omega _l \left(\left(\omega _l+M\right){}^2-4 m_D^2\right)+m_l \big(\omega _D^2 \left(\omega _l+M\right)-\omega _D \left(\omega _l+M\right)^2\\
&+\omega _D^3+\left(\omega _l+M\right){}^2 \left(M-\omega _l\right)\big)-4 m_D^2 \omega _l^2+\omega _D^2 \left(M \omega _l+\omega _l^2-3 k_t^2\right)+\omega _D^3 \omega _l\\
&-M^3 \omega _l-M^2 \omega _l^2+M \omega _l^3+3 k_t^2 \omega _l^2+\omega _l^4-3 M^2 k_t^2\big]-k_t^2 \big[\omega _D \big(\omega _l \left(8 m_D^2-5 M^2\right)\\
&-M \omega _l^2+\omega _l^3-3 M^3\big)+m_l \big(\omega _D^2 \left(\omega _l+M\right)+2 \omega _D \left(\omega _l+M\right){}^2-2 \omega _D^3\\
&+\left(\omega _l+M\right){}^2 \left(M-\omega _l\right)\big)+8 m_D^2 \omega _l^2+\omega _D^2 \left(7 M \omega _l+4 \omega _l^2+3 M^2\right)+\omega _D^3 \left(\omega _l+3 M\right)\\
&+5 M^3 \omega _l-M^2 \omega _l^2-5 M \omega _l^3-2 \omega _l^4+3 M^4\big]\Big]-k_t^2 V_1 \Big[m_l \left(\omega _D^2-\omega _l^2+M^2\right)\\
&+\omega _D \left(\omega _l^2-M^2\right)+\omega _D^2 \left(2 \omega _l+M\right)+\omega _D^3-M \omega _l^2+M^3\Big]\Big)\frac{1}{k_t^2}\widetilde E_2(q_t)
\end{align*}
\begin{align}\label{e4}
&+\Big(V_2 \big[\omega _D \left(\omega _l+M\right) \left(\omega _l \left(-2 m_D^2+M^2-k_t^2\right)+2 M \omega _l^2+\omega _l^3-M k_t^2\right)-\left(\omega _l+M\right)\nonumber\\
&\times\left(\omega _l^2 \left(2 m_D^2+M^2-k_t^2\right)+M^3 \omega _l-M \omega _l^3-\omega _l^4+M^2 k_t^2\right)+m_l \left(\omega _l+M\right) \big(\omega _D^2 \left(\omega _l+M\right)\nonumber\\
&-\omega _D \left(\omega _l+M\right){}^2-2 M m_D^2+\omega _D^3+\left(\omega _l+M\right){}^2 \left(M-\omega _l\right)\big)+\omega _D^2 \big(\omega _l \left(M^2-k_t^2\right)\nonumber\\
&+2 M \omega _l^2+\omega _l^3-M k_t^2\big)+\omega _D^3 \left(M \omega _l+\omega _l^2+k_t^2\right)\big]+k_t^2 V_1 \left(\omega _D^2-\omega _l^2+M^2\right)\Big)\frac{k_t\cdot q_t}{-p_t^2}\widetilde E_3(q_t)\nonumber\\
&+\Big(-V_2 \Big[m_l \left(\omega _D^2-\omega _l^2+M^2\right)+\omega _l \left(2 \omega _D \left(\omega _l+M\right)+\omega _D^2+\omega _l^2-M^2\right)\Big] (k_t\cdot q_t)^2\nonumber\\
&+(k_t\cdot q_t) \Big[2 m_D^2 \big[V_2 \left(\omega _D \omega _l \left(\omega _l+M\right)+M \omega _l^2+\omega _l^3+M k_t^2\right)-V_1 \omega _l \left(\omega _D+\omega _l\right)\big]\nonumber\\
&+m_l \big[V_1 \left(\omega _D^2 \left(\omega _l+M\right)-\omega _D \left(\omega _l+M\right){}^2-2 M m_D^2+\omega _D^3+\left(\omega _l+M\right)^2 \left(M-\omega _l\right)\right)\nonumber\\
&-2 M V_2 m_D^2 \left(\omega _l+M\right)\big]+M^2 V_1 \omega _D \omega _l+2 M k_t^2 V_2 \omega _D \omega _l+2 M V_1 \omega _D \omega _l^2+M V_1 \omega _D^2 \omega _l\nonumber\\
&+k_t^2 V_2 \omega _D \omega _l^2+V_1 \omega _D \omega _l^3+V_1 \omega _D^2 \omega _l^2+V_1 \omega _D^3 \omega _l+M^2 k_t^2 V_2 \omega _D-k_t^2 V_2 \omega _D^3-M^3 V_1 \omega _l\nonumber\\
&-M^2 V_1 \omega _l^2+M V_1 \omega _l^3+V_1 \omega _l^4\Big]+k_t^2 V_2 \Big[2 M \left(a_2+q_t^2\right) m_D^2-q_t^2 \big(\omega _D^2 \left(\omega _l+M\right)\nonumber\\
&-\omega _D \left(\omega _l+M\right){}^2+\omega _D^3+\left(\omega _l+M\right){}^2 \left(M-\omega _l\right)\big)\Big]\Big)\frac{1}{k_t^2}\widetilde E_4(q_t)\nonumber\\
&+\Big(-2 V_2 \Big[m_l \left(\omega _D^2-\omega _l^2+M^2\right)+\omega _l \left(2 \omega _D \left(\omega _l+M\right)+\omega _D^2+\omega _l^2-M^2\right)\Big] (k_t\cdot q_t)^2\nonumber\\
&+(k_t\cdot q_t) \Big[V_1 \big(\omega _D \omega _l \left(\left(\omega _l+M\right)^2-4 m_D^2\right)+m_l \big(\omega _D^2 \left(\omega _l+M\right)-\omega _D \left(\omega _l+M\right)^2\nonumber\\
&+\omega _D^3+\left(\omega _l+M\right){}^2 \left(M-\omega _l\right)\big)-4 m_D^2 \omega _l^2+\omega _D^2 \left(M \omega _l+\omega _l^2-k_t^2\right)+\omega _D^3 \omega _l-M^3 \omega _l\nonumber\\
&-M^2 \omega _l^2+M \omega _l^3+k_t^2 \omega _l^2+\omega _l^4-M^2 k_t^2\big)+V_2 \big(-\omega _D \left(\omega _l+M\right) \big(\omega _l \left(-12 m_D^2+3 M^2-2 k_t^2\right)\nonumber\\
&+6 M \omega _l^2+3 \omega _l^3-2 M k_t^2\big)+\left(\omega _l+M\right) \big(\omega _l^2 \left(12 m_D^2+3 M^2-k_t^2\right)+3 M^3 \omega _l-3 M \omega _l^3-3 \omega _l^4\nonumber\\
&+M^2 k_t^2\big)-3 m_l \left(\omega _l+M\right) \left(\omega _D^2 \left(\omega _l+M\right)-\omega _D \left(\omega _l+M\right){}^2+\omega _D^3+\left(\omega _l+M\right){}^2 \left(M-\omega _l\right)\right)\nonumber\\
&+\omega _D^2 \left(\omega _l \left(k_t^2-3 M^2\right)-6 M \omega _l^2-3 \omega _l^3+M k_t^2\right)+\omega _D^3 \left(-\left(3 M \omega _l+3 \omega _l^2+2 k_t^2\right)\right)\big)\Big]-k_t^2V_2\nonumber\\
&\times\Big[q_t^2 \left(m_l \left(\omega _D^2-\omega _l^2+M^2\right)+\omega _D \left(\omega _l^2-M^2\right)+\omega _D^2 \left(2 \omega _l+M\right)+\omega _D^3-M \omega _l^2+M^3\right)\nonumber\\
&-4 a_2 M m_D^2\Big]\Big)\frac{1}{k_t^2}\widetilde E_5(q_t)\nonumber\\
&+\Big(2 m_D^2 \Big[V_2 \big(\omega _l \left(\omega _D \left(a_2 k_t^2-2(k_t\cdot q_t)^2\right)+M k_t^2 \left(2 a_2+q_t^2\right)\right)+\omega _l^2 \left(a_2 k_t^2-2(k_t\cdot q_t)^2\right)\nonumber\\
&-a_2 M k_t^2 m_l+M^2 k_t^2 \left(2 a_2+q_t^2\right)\big)+a_2 M k_t^2 V_1\Big]+(k_t\cdot q_t)^2 \Big[m_l \big(V_1 \left(\omega _D^2-\omega _l^2+M^2\right)\nonumber\\
&+V_2 \left(\omega _D^2 \left(\omega _l+M\right)+\omega _D \left(\omega _l+M\right){}^2-\omega _D^3+\left(\omega _l+M\right){}^2 \left(M-\omega _l\right)\right)\big)+\omega _l \big(V_1 \big(2 \omega _D \left(\omega _l+M\right)\nonumber\\
&+\omega _D^2+\omega _l^2-M^2\big)+V_2 \left(\omega _D^2 \left(\omega _l+M\right)+3 \omega _D \left(\omega _l+M\right){}^2-\omega _D^3-\left(M-\omega _l\right) \left(\omega _l+M\right){}^2\right)\big)\Big]\nonumber\\
&+k_t^2 q_t^2 V_2 \left(\omega _l+M\right) \Big[\omega _D^2 \left(\omega _l+M\right)-\omega _D \left(\omega _l+M\right){}^2+\omega _D^3+\left(\omega _l+M\right){}^2 \left(M-\omega _l\right)\Big]\Big)\frac{1}{k_t^2}\widetilde E_6(q_t)\bigg\},
\end{align}
\begin{align}\label{e5}
\widetilde E_5(k_t)=&\frac{-1}{4\omega_l\omega_D(M^2-(\omega_l+\omega_D)^2)m_D^2}\int\frac{d^3q_t}{(2\pi)^3}\bigg\{-2 M V_2 m_D^2 \left(\omega _l+M\right)\widetilde E_1(q_t)\nonumber\\
&+\Big(2 m_D^2 \big[V_2 \big(\omega _l \left(\omega _D \left(k_t\cdot q_t+2 k_t^2\right)+3 M k_t^2\right)-M m_l \left(k_t\cdot q_t+2 k_t^2\right)\nonumber\\
&+\omega _l^2 \left(k_t\cdot q_t+2 k_t^2\right)+3 M^2 k_t^2\big)+M k_t^2 V_1\big]\Big)\frac{1}{k_t^2}\widetilde E_2(q_t)+2 M V_2 m_D^2 (k_t\cdot q_t)\widetilde E_3(q_t)\nonumber\\
&+2 V_2 m_D^2 \left(\left(\omega _l+M\right) (k_t\cdot q_t) \left(M m_l-\omega _l \left(\omega _D+\omega _l\right)\right)+a_1 M k_t^2\right)\frac{1}{k_t^2}\widetilde E_4(q_t)\nonumber\\
&+2 m_D^2 \Big((k_t\cdot q_t) \Big[V_2 \left(-3 \omega _D \omega _l \left(\omega _l+M\right)-3 M \omega _l^2-3 \omega _l^3+2 M k_t^2\right)+V_1 \omega _l \left(\omega _D+\omega _l\right)\nonumber\\
&+M m_l \left(3 V_2 \left(\omega _l+M\right)-V_1\right)\Big]+M k_t^2 V_2 \left(2 a_1+q_t^2\right)\Big)\frac{1}{k_t^2}\widetilde E_5\nonumber\\
&+2 m_D^2 \Big(V_2 \big[\omega _l \left(\omega _D \left((k_t\cdot q_t)^2+a_1 k_t^2\right)+2 a_1 M k_t^2\right)-M m_l \left((k_t\cdot q_t)^2+a_1 k_t^2\right)\nonumber\\
&+\omega _l^2 \left((k_t\cdot q_t)^2+a_1 k_t^2\right)+2 a_1 M^2 k_t^2\big]+a_1 M k_t^2 V_1\Big)\frac{1}{k_t^2}\widetilde E_6(q_t)\bigg\},
\end{align}
\begin{align*}
\widetilde E_6(k_t)=&\frac{-1}{4\omega_l\omega_D(M^2-(\omega_l+\omega_D)^2)m_D^2}\int\frac{d^3q_t}{(2\pi)^3}\bigg\{\Big(V_2 \big[(k_t\cdot q_t) \big(m_l \left(\omega _D^2-\omega _l^2+M^2\right)\\
&+2 M \omega _D \omega _l+2 \omega _D \omega _l^2+\omega _D^2 \omega _l+2 M m_D^2-M^2 \omega _l+\omega _l^3-2 M k_t^2\big)-k_t^2 \left(m_l+\omega _l\right)\\
&\times\left(\left(\omega _l+M\right){}^2-\omega _D^2\right)\big]-k_t^2 V_1 \left(\omega _D^2-\omega _l^2+M^2\right)\Big)\frac{1}{k_t^2}\widetilde E_1(q_t)\\
&+\Big(V_2 \big[k_t^2 \big(2 m_l \left(-\omega _D^2+3 M \omega _l+\omega _l^2+2 M^2\right)+2 M \omega _D \omega _l+2 \omega _D \omega _l^2+\omega _D^2 \omega _l\\
&-4 M m_D^2+3 M \omega _D^2+5 M^2 \omega _l+3 M \omega _l^2+\omega _l^3+3 M^3\big)-(k_t\cdot q_t) \big(m_l \left(\omega _D^2-\omega _l^2+M^2\right)\\
&+2 M \omega _D \omega _l+2 \omega _D \omega _l^2+\omega _D^2 \omega _l+8 M m_D^2-M^2 \omega _l+\omega _l^3-6 M k_t^2\big)\big]\\
&+k_t^2 V_1 \left(2 \omega _D \omega _l+\omega _D^2+2 M m_l+\omega _l^2+M^2\right)\Big)\frac{1}{k_t^2}\widetilde E_2(q_t)\\
&+\Big(V_2 \big[2 m_D^2 \left(\omega _l \left(\omega _D+2 M\right)+M m_l+\omega _l^2+2 M^2\right)+m_l \left(\omega _l+M\right) \left(\omega _D^2-\omega _l^2+M^2\right)\\
&+2 M^2 \omega _D \omega _l+4 M \omega _D \omega _l^2+M \omega _D^2 \omega _l+2 \omega _D \omega _l^3+\omega _D^2 \omega _l^2+k_t^2 \omega _D^2-M^3 \omega _l-M^2 \omega _l^2\\
&-4 M k_t^2 \omega _l+M \omega _l^3-k_t^2 \omega _l^2+\omega _l^4-3 M^2 k_t^2\big]+2 M V_1 \left(k_t^2-m_D^2\right)\Big)\frac{k_t\cdot q_t}{p_t^2}\widetilde E_3(q_t)\\
&+\Big(2 V_2 \left(\omega _l \left(\omega _D+\omega _l\right)+M m_l\right) (k_t\cdot q_t)^2-(k_t\cdot q_t) \big[V_1 m_l \left(\omega _D^2-\omega _l^2+M^2\right)\\
&+V_2 \left(k_t^2 \left(\left(\omega _l+M\right){}^2-\omega _D^2\right)-4 M m_D^2 \left(\omega _l+M\right)\right)+V_1 \omega _l \big(2 \omega _D \left(\omega _l+M\right)+\omega _D^2\\
&+\omega _l^2-M^2\big)\big]+k_t^2 V_2 \big[q_t^2 \left(\omega _D^2-\omega _l^2+M^2\right)-2 a_5 m_D^2 \left(\omega _l \left(\omega _D+\omega _l\right)+M m_l\right)\big]\Big)\frac{1}{k_t^2}\widetilde E_4(q_t)\\
&+\Big(4 V_2 \big[\omega _l \left(\omega _D+\omega _l\right)+M m_l\big] (k_t\cdot q_t)^2+(k_t\cdot q_t) \big[V_2 \big(4 m_D^2 \big(\omega _l \left(\omega _D+3 M\right)+M m_l
\end{align*}
\begin{align}\label{e6}
&+\omega _l^2+3 M^2\big)+3 m_l \left(\omega _l+M\right) \left(\omega _D^2-\omega _l^2+M^2\right)+6 M^2 \omega _D \omega _l+12 M \omega _D \omega _l^2+3 M \omega _D^2 \omega _l\nonumber\\
&+6 \omega _D \omega _l^3+3 \omega _D^2 \omega _l^2+2 k_t^2 \omega _D^2-3 M^3 \omega _l-3 M^2 \omega _l^2-6 M k_t^2 \omega _l+3 M \omega _l^3-2 k_t^2 \omega _l^2+3 \omega _l^4\nonumber\\
&-4 M^2 k_t^2\big)-V_1 \big(m_l \left(\omega _D^2-\omega _l^2+M^2\right)+2 M \omega _D \omega _l+2 \omega _D \omega _l^2+\omega _D^2 \omega _l+4 M m_D^2-M^2 \omega _l\nonumber\\
&+\omega _l^3-2 M k_t^2\big)\big]+k_t^2 V_2 \big[2 M m_l \left(q_t^2-2 a_5 m_D^2\right)+2 \omega _D \omega _l \left(q_t^2-2 a_5 m_D^2\right)-4 a_5 m_D^2 \omega _l^2\nonumber\\
&+q_t^2 \omega _D^2+q_t^2 \omega _l^2+M^2 q_t^2\big]\Big)\frac{1}{k_t^2}\widetilde E_5(q_t)\nonumber\\
&+\Big((k_t\cdot q_t)^2 \big[m_l \left(V_2 \left(-\omega _D^2+4 M \omega _l+\omega _l^2+3 M^2\right)+2 M V_1\right)+\omega _l \big(V_2 \big(2 \omega _D \left(\omega _l+M\right)-\omega _D^2\nonumber\\
&+4 M \omega _l+3 \omega _l^2+M^2\big)+2 V_1 \left(\omega _D+\omega _l\right)\big)-4 M V_2 m_D^2\big]+k_t^2 \big[q_t^2 V_2 \left(\omega _l+M\right) \left(\omega _D^2-\omega _l^2+M^2\right)\nonumber\\
&-2 a_5 m_D^2 \big(V_2 \left(2 \omega _D \omega _l \left(\omega _l+M\right)+2 M \omega _l^2+2 \omega _l^3-M k_t^2\right)+V_1 \omega _l \left(\omega _D+\omega _l\right)\nonumber\\
&+M m_l \left(2 V_2 \left(\omega _l+M\right)+V_1\right)\big)\big]\Big)\frac{1}{k_t^2}\widetilde E_6(q_t)\bigg\}.
\end{align}

\section{The explicit expressions for coupled integral equations for $\widetilde C_i$ $(i=1,\cdots,8)$ for spin-$\frac{3}{2}$ doubly heavy baryons}\label{appb}
\begin{align}\label{C11}
\widetilde C_1(k_t)=&\frac{1}{4\omega_l\omega_D(M^2-(\omega_l+\omega_D)^2)m_D^2}\int\frac{d^3q_t}{(2\pi)^3}\bigg\{\Big(2 m_D^2 \big[-M V_2 k_t^2 \left(k_t\cdot q_t+k_t^2\right)\nonumber\\
&+k_t^2 m_l \left(V_2 \left(\omega _D^2-\omega _l^2+M^2\right)+M V_1\right)+k_t^2 \omega _l \big(V_2 \left(2 \omega _D \left(\omega _l+M\right)+\omega _D^2+\omega _l^2-M^2\right)\nonumber\\
&+V_1 \left(\omega _D+\omega _l\right)\big)\big]\Big)\frac{1}{-k_t^2}\widetilde C_1(q_t)\nonumber\\
&+\Big(2 m_D^2 \big(k_t^2 (k_t\cdot q_t) \big[V_2 \left(-\omega _D \omega _l+\omega _D^2+M m_l-2 \omega _l^2+M^2\right)-M V_1\big]\nonumber\\
&+k_t^2 V_2 q_t^2 \left(\omega _l \left(\omega _D+\omega _l\right)+M m_l\right)\big)\Big)\frac{1}{k_t^2}\widetilde C_2(q_t)-2 a_1 V_2 m_D^2 \left(\omega _l \left(\omega _D+\omega _l\right)+M m_l\right)\widetilde C_3(q_t)\nonumber\\
&+\Big(-2 a_1 m_D^2 \big[-M V_2 k_t^2+m_l \left(V_2 \left(\omega _D^2-\omega _l^2+M^2\right)+M V_1\right)+\omega _l \big(V_2 \big(2 \omega _D \left(\omega _l+M\right)\nonumber\\
&+\omega _D^2+\omega _l^2-M^2\big)+V_1 \left(\omega _D+\omega _l\right)\big)\big]\Big)\widetilde C_4(q_t)+2 k_t^2 x_2 m_D^2 \big[V_2 \big(-\omega _D \omega _l+\omega _D^2+M m_l\nonumber\\
&-2 \omega _l^2+M^2\big)-M V_1\big]\widetilde C_5(q_t)+2 M k_t^2 V_2 x_2 m_D^2\widetilde C_6(q_t)-8 a_1 V_2 m_D^2 \big[\omega _l \left(\omega _D+\omega _l\right)\nonumber\\
&+M m_l\big]\widetilde C_7(q_t)-4 M k_t^2 V_2 x_2 m_D^2\widetilde C_8(q_t)\bigg\},
\end{align}
\begin{align*}
\widetilde C_2(k_t)=&\frac{1}{4\omega_l\omega_D(M^2-(\omega_l+\omega_D)^2)m_D^2}\int\frac{d^3q_t}{(2\pi)^3}\bigg\{2 m_D^2 \Big[V_2 \big(\omega _D \omega _l \left(k_t\cdot q_t+k_t^2\right)-M m_l\\
&\times\left(k_t\cdot q_t+k_t^2\right)+\omega _l^2 k_t\cdot q_t+k_t^2 \omega _D^2+M^2 k_t^2\big)+M k_t^2 V_1\Big]\frac{1}{-k_t^2}\widetilde C_1(q_t)
\end{align*}
\begin{align}\label{C12}
&+2 m_D^2 \Big[(k_t\cdot q_t) \big[M V_2 k_t^2+m_l \left(V_2 \left(\omega _D^2-\omega _l^2+M^2\right)-M V_1\right)+\omega _l \big(V_2 \big(-2 \omega _D \left(\omega _l+M\right)\nonumber\\
&-\omega _D^2-\omega _l^2+M^2\big)+V_1 \left(\omega _D+\omega _l\right)\big)\big]+M k_t^2 V_2 q_t^2\Big]\frac{1}{k_t^2}\widetilde C_2(q_t)-2a_1 M V_2 m_D^2\widetilde C_3(q_t)\nonumber\\
&-2 a_1 m_D^2 \Big[V_2 \left(\omega _D \omega _l+\omega _D^2-M m_l+M^2\right)+M V_1\Big]\widetilde C_4(q_t)+2x_2 m_D^2 \Big(M V_2 k_t^2+m_l\nonumber\\
&\times\left(V_2 \left(\omega _D^2-\omega _l^2+M^2\right)-M V_1\right)+\omega _l \big[V_2 \left(-2 \omega _D \left(\omega _l+M\right)-\omega _D^2-\omega _l^2+M^2\right)\nonumber\\
&+V_1 \left(\omega _D+\omega _l\right)\big]\Big)\widetilde C_5(q_t)+2 V_2 x_2 m_D^2 \Big[M m_l-\omega _l \left(\omega _D+\omega _l\right)\Big]\widetilde C_6(q_t)-8 a_1 M V_2 m_D^2\widetilde C_7(q_t)\nonumber\\
&+4 V_2 x_2 m_D^2 \big[\omega _l \left(\omega _D+\omega _l\right)-M m_l\big]\widetilde C_8(q_t)\bigg\},
\end{align}
\begin{align*}
\widetilde C_3(k_t)=&\frac{1}{4\omega_l\omega_D(M^2-(\omega_l+\omega_D)^2)m_D^2}\int\frac{d^3q_t}{(2\pi)^3}\bigg\{\Big(V_2 k_t^2 (k_t\cdot q_t) \left(\omega _D^2-\omega _l^2+M^2\right)-m_lk_t^2\\
&\times\big[V_1 \left(\omega _D^2-\omega _l^2+M^2\right)+V_2 \big(\omega _D^2 \left(\omega _l+M\right)-\omega _D \left(\omega _l+M\right)^2+\omega _D^3+\left(\omega _l+M\right)^2\\
&\times\left(M-\omega _l\right)\big)\big]-\omega_lk_t^2\big[V_1 \left(2 \omega _D \left(\omega _l+M\right)+\omega _D^2+\omega _l^2-M^2\right)\\
&+V_2 \left(\omega _D^2 \left(\omega _l+M\right)+\omega _D \left(\omega _l+M\right){}^2+\omega _D^3-\left(M-\omega _l\right) \left(\omega _l+M\right){}^2\right)\big]\Big)\frac{1}{k_t^2}\widetilde C_1(q_t)\\
&+\Big[V_2 q_t^2 \left(m_l \left(\omega _D^2-\omega _l^2+M^2\right)+\omega _l \left(2 \omega _D \left(\omega _l+M\right)+\omega _D^2+\omega _l^2-M^2\right)\right)\\
&+(k_t\cdot q_t) \big(V_2 \left(\omega _D^2 \left(\omega _l+M\right)-\omega _D \left(\omega _l+M\right)^2+\omega _D^3+\left(\omega _l+M\right)^2 \left(M-\omega _l\right)\right)\\
&-V_1 \left(\omega _D^2-\omega _l^2+M^2\right)\big)\Big]\widetilde C_2(q_t)\\
&+\Big(-V_2 \big[m_l \left(\omega _D^2-\omega _l^2+M^2\right)+\omega _l \left(2 \omega _D \left(\omega _l+M\right)+\omega _D^2+\omega _l^2-M^2\right)\big] (k_t\cdot q_t)^2\\
&+(k_t\cdot q_t) \big[2 M V_2 k_t^2 m_D^2+m_l \big(V_1 \big(\omega _D^2 \left(\omega _l+M\right)-\omega _D \left(\omega _l+M\right){}^2-2 M m_D^2+\omega _D^3\\
&+\left(\omega _l+M\right){}^2 \left(M-\omega _l\right)\big)-V_2 m_D^2 \left(\omega _D^2-\omega _l^2+M^2\right)\big)+\omega _l \big(V_2 m_D^2 \big(-2 \omega _D \left(\omega _l+M\right)\\
&-\omega _D^2-\omega _l^2+M^2\big)+V_1 \big(-\omega _l \left(2 m_D^2+M^2\right)+\omega _D \left(\left(\omega _l+M\right){}^2-2 m_D^2\right)+\omega _D^2 \left(\omega _l+M\right)\\
&+\omega _D^3+M \omega _l^2+\omega _l^3-M^3\big)\big)\big]+a_5 k_t^4 V_2 \big[-\omega _D^2 \left(\omega _l+M\right)+\omega _D \left(\omega _l+M\right){}^2+2 M m_D^2\\
&-\omega _D^3-\left(M-\omega _l\right) \left(\omega _l+M\right){}^2\big]\Big)\frac{1}{k_t^2}\widetilde C_3(q_t)\\
&+\frac{1}{k_t^2}\widetilde C_4(q_t)\Big(a_5 k_t^4 V_2 \big[m_D^2 \left(\omega _D^2-\omega _l^2+M^2\right)-\omega _D^3 \left(\omega _l+M\right)+\omega _D \left(\omega _l+M\right)^3-\omega _D^4\\
&-\left(M-\omega _l\right) \left(\omega _l+M\right){}^3\big]-(k_t\cdot q_t)^2 \big[m_l \big(V_1 \left(\omega _D^2-\omega _l^2+M^2\right)+V_2 \big(\omega _D^2 \left(\omega _l+M\right)\\
&-\omega _D \left(\omega _l+M\right){}^2+\omega _D^3+\left(\omega _l+M\right){}^2 \left(M-\omega _l\right)\big)\big)+\omega _l \big(V_1 \big(2 \omega _D \left(\omega _l+M\right)+\omega _D^2\\
&+\omega _l^2-M^2\big)+V_2 \left(\omega _D^2 \left(\omega _l+M\right)+\omega _D \left(\omega _l+M\right){}^2+\omega _D^3-\left(M-\omega _l\right) \left(\omega _l+M\right){}^2\right)\big)\big]\Big)\\
\end{align*}
\begin{align}\label{C13}
&+\Big(V_2a_5 k_t^4 \big[\omega _D^2 \left(\omega _l+M\right)-\omega _D \left(\omega _l+M\right){}^2+\omega _D^3+\left(\omega _l+M\right){}^2 \left(M-\omega _l\right)\big]\nonumber\\
&-q_t^2 \big[m_l \big(-m_D^2 \left(\omega _D^2-\omega _l^2+M^2\right)+\omega _D^3 \left(\omega _l+M\right)-\omega _D \left(\omega _l+M\right){}^3+\omega _D^4+\left(\omega _l+M\right){}^3\nonumber\\
&\times\left(M-\omega _l\right)\big)+\omega _l \big(m_D^2 \left(-2 \omega _D \left(\omega _l+M\right)-\omega _D^2-\omega _l^2+M^2\right)+\omega _D^3 \left(\omega _l+M\right)+\omega _D \left(\omega _l+M\right){}^3\nonumber\\
&+\omega _D^4-\left(M-\omega _l\right) \left(\omega _l+M\right){}^3\big)\big]-a_5 k_t^4 V_1 \left(\omega _D^2-\omega _l^2+M^2\right)\Big)\frac{k_t\cdot q_t}{k_t^2}\widetilde C_5(q_t)\nonumber\\
&+\Big(V_2 \Big[(k_t\cdot q_t) \big[a_5 k_t^4 \left(\omega _D^2-\omega _l^2+M^2\right)-q_t^2 \big(\omega _l \big(-\omega _l \left(2 m_D^2+M^2\right)+\omega _D \left(\left(\omega _l+M\right){}^2-2 m_D^2\right)\nonumber\\
&+\omega _D^2 \left(\omega _l+M\right)+\omega _D^3+M \omega _l^2+\omega _l^3-M^3\big)+m_l \big(\omega _D^2 \left(\omega _l+M\right)-\omega _D \left(\omega _l+M\right){}^2-2 M m_D^2\nonumber\\
&+\omega _D^3+\left(\omega _l+M\right){}^2 \left(M-\omega _l\right)\big)\big)\big]+a_5 k_t^4 m_D^2 \left(-2 \omega _D \omega _l+\omega _D^2+2 M m_l-3 \omega _l^2+M^2\right)\Big]\nonumber\\
&+a_5 k_t^4 V_1 \Big[\omega _D^2 \left(\omega _l+M\right)-\omega _D \left(\omega _l+M\right){}^2-2 M m_D^2+\omega _D^3+\left(\omega _l+M\right){}^2 \left(M-\omega _l\right)\Big]\Big)\frac{1}{k_t^2}\widetilde C_6(q_t)\nonumber\\
&+\Big(k_t^2 \big(-4 V_2 \big[m_l \left(\omega _D^2-\omega _l^2+M^2\right)+\omega _l \left(2 \omega _D \left(\omega _l+M\right)+\omega _D^2+\omega _l^2-M^2\right)\big] (k_t\cdot q_t)^2\nonumber\\
&+(k_t\cdot q_t) \big[m_l \big(V_2 \big(4 m_D^2 \left(\omega _D^2-\omega _l^2+M^2\right)-3 \big(\omega _D^3 \left(\omega _l+M\right)-\omega _D \left(\omega _l+M\right){}^3+\omega _D^4\nonumber\\
&+\left(\omega _l+M\right){}^3 \left(M-\omega _l\right)\big)\big)+V_1 \left(\omega _D^2 \left(\omega _l+M\right)-\omega _D \left(\omega _l+M\right){}^2+\omega _D^3+\left(\omega _l+M\right){}^2 \left(M-\omega _l\right)\right)\big)\nonumber\\
&+\omega _l \big(3 V_2 \big(2 m_D^2 \left(2 \omega _D \left(\omega _l+M\right)+\omega _D^2+\omega _l^2-M^2\right)-\omega _D^3 \left(\omega _l+M\right)-\omega _D \left(\omega _l+M\right){}^3-\omega _D^4\nonumber\\
&+\left(\omega _l+M\right){}^3 \left(M-\omega _l\right)\big)+V_1 \left(\omega _D^2 \left(\omega _l+M\right)+\omega _D \left(\omega _l+M\right){}^2+\omega _D^3-\left(M-\omega _l\right) \left(\omega _l+M\right){}^2\right)\big)\big]\nonumber\\
&-a_5 k_t^4 V_2 \big[\omega _D^2 \left(\omega _l+M\right)-\omega _D \left(\omega _l+M\right){}^2+\omega _D^3+\left(\omega _l+M\right){}^2 \left(M-\omega _l\right)\big]\big)+k_t^2 (k_t\cdot q_t)\nonumber\\
&\times\big(V_2 \big((k_t\cdot q_t) \left(m_l \left(\omega _D^2-\omega _l^2+M^2\right)-\omega _l \left(2 \omega _D \left(\omega _l+M\right)+\omega _D^2+\omega _l^2-M^2\right)\right)+k_t^2 \big(\omega _D^2 \left(\omega _l+M\right)\nonumber\\
&-\omega _D \left(\omega _l+M\right){}^2+\omega _D^3+\left(\omega _l+M\right){}^2 \left(M-\omega _l\right)\big)\big)-k_t^2 V_1 \left(\omega _D^2-\omega _l^2+M^2\right)\big)\Big)\frac{1}{k_t^4}\widetilde C_7(q_t)\nonumber\\
&+\frac{1}{k_t^2}\widetilde C_8(q_t)\Big(V_2 \big((k_t\cdot q_t) \big[q_t^2 \big(m_l \left(\omega _D^2 \left(\omega _l+M\right)-\omega _D \left(\omega _l+M\right){}^2+\omega _D^3+\left(\omega _l+M\right){}^2 \left(M-\omega _l\right)\right)\nonumber\\
&+\omega _l \left(\omega _D^2 \left(\omega _l+M\right)+\omega _D \left(\omega _l+M\right){}^2+\omega _D^3-\left(M-\omega _l\right) \left(\omega _l+M\right){}^2\right)\big)-k_t^2 q_t^2 \left(\omega _D^2-\omega _l^2+M^2\right)\nonumber\\
&-2 a_5 k_t^4 \left(\omega _D^2-\omega _l^2+M^2\right)\big]+a_5 k_t^4 \big[4 m_D^2 \left(\omega _D^2-\omega _l^2+M^2\right)+m_l \big(\omega _D^2 \left(\omega _l+M\right)-\omega _D \left(\omega _l+M\right){}^2\nonumber\\
&+\omega _D^3+\left(\omega _l+M\right){}^2 \left(M-\omega _l\right)\big)+8 M^2 \omega _D \omega _l+7 M \omega _D \omega _l^2-M \omega _D^2 \omega _l+2 \omega _D \omega _l^3-\omega _D^2 \omega _l^2-4 \omega _D^3 \omega _l\nonumber\\
&+3 M^3 \omega _D-3 M \omega _D^3-3 \omega _D^4-5 M^3 \omega _l+M^2 \omega _l^2+5 M \omega _l^3+2 \omega _l^4-3 M^4\big]\big)-a_5 k_t^4 V_1 \big(m_l \big(\omega _D^2-\omega _l^2\nonumber\\
&+M^2\big)-\omega _D \left(4 M \omega _l+3 \omega _l^2+M^2\right)+M \omega _D^2+\omega _D^3+2 M^2 \omega _l-M \omega _l^2-2 \omega _l^3+M^3\big)\Big)\bigg\},
\end{align}
\begin{align*}
\widetilde C_4(k_t)=&\frac{1}{4\omega_l\omega_D(M^2-(\omega_l+\omega_D)^2)m_D^2}\int\frac{d^3q_t}{(2\pi)^3}\bigg\{\Big(-2 M V_2 (k_t\cdot q_t)+\big[\omega _l \big(V_2 \big(2 \omega _D \left(\omega _l+M\right)+\omega _D^2\\
&+\omega _l^2-M^2\big)+2 V_1 \left(\omega _D+\omega _l\right)\big)-4 M V_2 m_D^2\big]+m_l \left(V_2 \left(\omega _D^2-\omega _l^2+M^2\right)+2 M V_1\right)\Big)\widetilde C_1(q_t)
\end{align*}
\begin{align}\label{C14}
&+\Big((k_t\cdot q_t) \left(2 M V_1-V_2 \left(\omega _D^2-\omega _l^2+M^2\right)\right)-2V_2 q_t^2 \left(\omega _l \left(\omega _D+\omega _l\right)+M m_l\right)\Big)\widetilde C_2(q_t)\nonumber\\
&+\Big(2 V_2 \big[\omega _l \left(\omega _D+\omega _l\right)+M m_l\big] (k_t\cdot q_t)^2-(k_t\cdot q_t) \big[-2 V_2 m_D^2 \left(\omega _D^2-\omega _l^2+M^2\right)\nonumber\\
&+V_1 m_l \left(\omega _D^2-\omega _l^2+M^2\right)+V_1 \omega _l \left(2 \omega _D \left(\omega _l+M\right)+\omega _D^2+\omega _l^2-M^2\right)\big]+a_5 k_t^2 V_2 \nonumber\\
&\times\big[k_t^2 \left(\omega _D^2-\omega _l^2+M^2\right)-2 m_D^2 \left(\omega _l \left(\omega _D+\omega _l\right)+M m_l\right)\big]\Big)\frac{1}{k_t^2}\widetilde C_3(q_t)\nonumber\\
&+\Big((k_t\cdot q_t)^2 \big(m_l \big[V_2 \left(\omega _D^2-\omega _l^2+M^2\right)+2 M V_1\big]+\omega _l \big[V_2 \left(2 \omega _D \left(\omega _l+M\right)+\omega _D^2+\omega _l^2-M^2\right)\nonumber\\
&+2 V_1 \left(\omega _D+\omega _l\right)\big]-4 M V_2 m_D^2\big)+a_5 k_t^2 \big(k_t^2 V_2 \big[\omega _D^2 \left(\omega _l+M\right)-\omega _D \left(\omega _l+M\right){}^2\nonumber\\
&+\omega _D^3+\left(\omega _l+M\right){}^2 \left(M-\omega _l\right)\big]-2 m_D^2 \big(-M V_2 k_t^2+m_l \big[V_2 \left(\omega _D^2-\omega _l^2+M^2\right)+M V_1\big]\nonumber\\
&+\omega _l \left(V_2 \left(2 \omega _D \left(\omega _l+M\right)+\omega _D^2+\omega _l^2-M^2\right)+V_1 \left(\omega _D+\omega _l\right)\right)\big)\big)\Big)\frac{1}{k_t^2}\widetilde C_4(q_t)\nonumber\\
&+\Big((k_t\cdot q_t) \Big[V_2 \big(q_t^2 \big(m_l \left(\omega _D^2 \left(\omega _l+M\right)-\omega _D \left(\omega _l+M\right){}^2+\omega _D^3+\left(\omega _l+M\right){}^2 \left(M-\omega _l\right)\right)\nonumber\\
&+\omega _l \left(\omega _D^2 \left(\omega _l+M\right)+\omega _D \left(\omega _l+M\right){}^2+\omega _D^3-\left(M-\omega _l\right) \left(\omega _l+M\right){}^2\right)\big)-a_5 k_t^4 \left(\omega _D^2-\omega _l^2+M^2\right)\big)\nonumber\\
&+2 a_5 M k_t^4 V_1\Big]+2 k_t^4 x_{10} m_D^2 \left(V_2 \left(-\omega _D \omega _l+\omega _D^2+M m_l-2 \omega _l^2+M^2\right)-M V_1\right)\Big)\frac{1}{k_t^2}\widetilde C_5(q_t)\nonumber\\
&+\Big(V_2 \big[(k_t\cdot q_t) \big(q_t^2 \left(m_l \left(\omega _D^2-\omega _l^2+M^2\right)+\omega _l \left(2 \omega _D \left(\omega _l+M\right)+\omega _D^2+\omega _l^2-M^2\right)\right)-2 a_5 M k_t^4\big)\nonumber\\
&+2 M k_t^4 x_{10} m_D^2\big]-a_5 k_t^4 V_1 \left(\omega _D^2-\omega _l^2+M^2\right)\Big)\frac{1}{p_t^2}\widetilde C_6(q_t)\nonumber\\
&+\Big(4 V_2 (k_t\cdot q_t)^2 \big[M m_l \left(m_D^2+2 k_t^2\right)+\omega _l \left(\omega _D+\omega _l\right) \left(2 k_t^2-m_D^2\right)\big]+k_t^2 (k_t\cdot q_t)\nonumber\\
&\times\big[-2 m_D^2 \left(V_2 \left(2 \omega _D \omega _l+\omega _D^2-2 M m_l+\omega _l^2+M^2\right)+2 M V_1\right)+m_l \big(3 V_2 \big(\omega _D^2 \left(\omega _l+M\right)\nonumber\\
&-\omega _D \left(\omega _l+M\right){}^2+\omega _D^3+\left(\omega _l+M\right){}^2 \left(M-\omega _l\right)\big)-V_1 \left(\omega _D^2-\omega _l^2+M^2\right)\big)\nonumber\\
&+\omega _l \big(V_1 \left(-2 \omega _D \left(\omega _l+M\right)-\omega _D^2-\omega _l^2+M^2\right)+3 V_2 \big(\omega _D^2 \left(\omega _l+M\right)+\omega _D \left(\omega _l+M\right){}^2+\omega _D^3\nonumber\\
&-\left(M-\omega _l\right) \left(\omega _l+M\right){}^2\big)\big)\big]+k_t^2 (k_t\cdot q_t) \big[V_2 \big(2 (k_t\cdot q_t) \left(\omega _l \left(\omega _D+\omega _l\right)-M m_l\right)\nonumber\\
&-k_t^2 \left(\omega _D^2-\omega _l^2+M^2\right)\big)+2 M k_t^2 V_1\big]+a_5 k_t^4 V_2 \big[k_t^2 \left(\omega _D^2-\omega _l^2+M^2\right)\nonumber\\
&-8 m_D^2 \left(\omega _l \left(\omega _D+\omega _l\right)+M m_l\right)\big]\Big)\frac{1}{k_t^4}\widetilde C_7(q_t)\nonumber\\
&+\Big(V_2 \big((k_t\cdot q_t) \big[-q_t^2 \big(m_l \left(\omega _D^2-\omega _l^2+M^2\right)+\omega _l \left(2 \omega _D \left(\omega _l+M\right)+\omega _D^2+\omega _l^2-M^2\right)+4 M m_D^2\big)\nonumber\\
&+2 M k_t^2 q_t^2+4 a_5 M k_t^4\big]+k_t^4 \big[a_5 \big(-m_l \left(\omega _D^2-\omega _l^2+M^2\right)-\omega _D \left(4 M \omega _l+\omega _l^2+3 M^2\right)\nonumber\\
&+\omega _D^2 \left(4 \omega _l+3 M\right)+3 \omega _D^3+2 M^2 \omega _l-3 M \omega _l^2-2 \omega _l^3+3 M^3\big)-4 M (a_5+x_{10}) m_D^2\big]\big)\nonumber\\
&+a_5 k_t^4 V_1 \left(-2 \omega _D \omega _l+\omega _D^2+2 M m_l-3 \omega _l^2+M^2\right)\Big)\frac{1}{k_t^2}\widetilde C_8(q_t)\bigg\},
\end{align}
\begin{align*}
\widetilde C_5(k_t)=&\frac{1}{4\omega_l\omega_D(M^2-(\omega_l+\omega_D)^2)m_D^2}\int\frac{d^3q_t}{(2\pi)^3}\bigg\{\Big(V_2 \big[2 (k_t\cdot q_t) \left(\omega _l \left(\omega _D+\omega _l\right)-M m_l\right)\nonumber\\
&+k_t^2 \left(\omega _D^2-\omega _l^2+M^2\right)\big]+2 M k_t^2 V_1\Big)\frac{1}{k_t^2}\widetilde C_1(q_t)\nonumber\\
&+\Big((k_t\cdot q_t) \big(m_l \big[2 M V_1-V_2 \left(\omega _D^2-\omega _l^2+M^2\right)\big]+\omega _l \big[V_2 \big(2 \omega _D \left(\omega _l+M\right)+\omega _D^2+\omega _l^2\nonumber\\
&-M^2\big)-2 V_1 \left(\omega _D+\omega _l\right)\big]-4 M V_2 m_D^2\big)-2 M k_t^2 V_2 q_t^2\Big)\frac{1}{k_t^2}\widetilde C_2(q_t)\nonumber\\
&+\Big(2 M V_2 (k_t\cdot q_t)^2-V_1 (k_t\cdot q_t) \left(\omega _D^2-\omega _l^2+M^2\right)+a_5 k_t^2 V_2 \big[m_l \left(\omega _D^2-\omega _l^2+M^2\right)\nonumber\\
&-\omega _l \left(2 \omega _D \left(\omega _l+M\right)+\omega _D^2+\omega _l^2-M^2\right)-2 M m_D^2\big]\Big)\frac{1}{k_t^2}\widetilde C_3(q_t)\nonumber\\
&+\Big(\big[V_2 \left(\omega _D^2-\omega _l^2+M^2\right)+2 M V_1\big] (k_t\cdot q_t)^2+a_5k_t^2 \big[V_2 \big(m_l \big(\omega _D^2 \left(\omega _l+M\right)+\omega _D^3\nonumber\\
&-\omega _D \left(\omega _l+M\right)^2+\left(\omega _l+M\right){}^2 \left(M-\omega _l\right)\big)-\omega _l \big(\omega _D^2 \left(\omega _l+M\right)+\omega _D \left(\omega _l+M\right)^2+\omega _D^3\nonumber\\
&-\left(M-\omega _l\right) \left(\omega _l+M\right){}^2\big)\big)-2 m_D^2 \left(V_2 \left(\omega _D \omega _l+\omega _D^2-M m_l+M^2\right)+M V_1\right)\big]\Big)\frac{1}{k_t^2}\widetilde C_4(q_t)\nonumber\\
&+\Big(-4 M V_2 m_D^2 (k_t\cdot q_t)^3+k_t^2 (k_t\cdot q_t) \big[V_2 q_t^2 \big(\omega _D^2 \left(\omega _l+M\right)-\omega _D \left(\omega _l+M\right){}^2+\omega _D^3\nonumber\\
&+\left(\omega _l+M\right){}^2 \left(M-\omega _l\right)\big)+a_5 k_t^2 m_l \left(2 M V_1-V_2 \left(\omega _D^2-\omega _l^2+M^2\right)\right)+a_5 k_t^2 \omega _l \nonumber\\
&\times\left(V_2 \left(2 \omega _D \left(\omega _l+M\right)+\omega _D^2+\omega _l^2-M^2\right)-2 V_1 \left(\omega _D+\omega _l\right)\right)\big]+2 k_t^4 x_{10} m_D^2 \nonumber\\
&\times\big[M V_2 k_t^2+m_l \left(V_2 \left(\omega _D^2-\omega _l^2+M^2\right)-M V_1\right)+\omega _l \big(V_2 \big(-2 \omega _D \left(\omega _l+M\right)\nonumber\\
&-\omega _D^2-\omega _l^2+M^2\big)+V_1 \left(\omega _D+\omega _l\right)\big)\big]\Big)\frac{1}{k_t^4}\widetilde C_5(q_t)\nonumber\\
&+\Big(2 V_2 m_D^2 \left(\omega _D^2-\omega _l^2+M^2\right) (k_t\cdot q_t)^2-m_lk_t^4\big[2 M V_2 \left(a_5 (k_t\cdot q_t)-x_{10} m_D^2\right)\nonumber\\
&+a_5 V_1 \left(\omega _D^2-\omega _l^2+M^2\right)\big]+k_t^2 V_2 (k_t\cdot q_t) \big[q_t^2 \left(\omega _D^2-\omega _l^2+M^2\right)+2 a_5 k_t^2 \omega _l \left(\omega _D+\omega _l\right)\big]\nonumber\\
&+k_t^4 \omega _l \left(a_5 V_1 \left(2 \omega _D \left(\omega _l+M\right)+\omega _D^2+\omega _l^2-M^2\right)-2 V_2 x_{10} m_D^2 \left(\omega _D+\omega _l\right)\right)\Big)\frac{1}{k_t^4}\widetilde C_6(q_t)\nonumber\\
&+\Big(4 M V_2 (k_t\cdot q_t)^2+a_5 k_t^2 V_2 \big[2 M k_t^2+\omega _D^2 m_l-2 M \omega _D \omega _l-2 \omega _D \omega _l^2-\omega _D^2 \omega _l-8 M m_D^2\nonumber\\
&+M^2 m_l-m_l \omega _l^2+M^2 \omega _l-\omega _l^3\big]+(k_t\cdot q_t) \big[V_2 \big(-m_l \left(\omega _D^2-\omega _l^2+M^2\right)-4 M \omega _D \omega _l\nonumber\\
&-\omega _D \omega _l^2+4 \omega _D^2 \omega _l-4 M m_D^2-3 M^2 \omega _D+3 M \omega _D^2+3 \omega _D^3+2 M^2 \omega _l-3 M \omega _l^2-2 \omega _l^3\nonumber\\
&+3 M^3\big)-V_1 \left(2 \omega _D \omega _l+\omega _D^2-2 M m_l+\omega _l^2+M^2\right)\big]\Big)\widetilde C_7(q_t)\nonumber\\
&+\frac{1}{k_t^4}\widetilde C_8(q_t)\Big(-2 m_D^2 (k_t\cdot q_t)^2 \big[V_2 \left(2 \omega _D \omega _l+3 \omega _D^2-2 M m_l-\omega _l^2+3 M^2\right)+2 M V_1\big]\nonumber\\
&-k_t^2 V_2 (k_t\cdot q_t) \big[q_t^2 \left(2 \omega _D \omega _l+\omega _D^2-2 M m_l+\omega _l^2+M^2\right)+4 a_5 k_t^2 \left(\omega _l \left(\omega _D+\omega _l\right)-M m_l\right)\big]\nonumber\\
&+a_5 k_t^4 k_t^2 \big[V_2 \left(\omega _D^2-\omega _l^2+M^2\right)+2 M V_1\big]+k_t^4 \big[m_l \big(V_2 \big(3 a_5 \big(\omega _D^2 \left(\omega _l+M\right)+\omega _D^3
\end{align*}
\begin{align}\label{C15}
&-\omega _D \left(\omega _l+M\right){}^2+\left(\omega _l+M\right){}^2 \left(M-\omega _l\right)\big)-4 M x_{10} m_D^2\big)+a_5 V_1 \left(\omega _D^2-\omega _l^2+M^2\right)\big)\nonumber\\
&-\omega _l \big(V_2 \big(-\omega _l \left(3 a_5 M^2+4 x_{10} m_D^2\right)+\omega _D \left(3 a_5 \left(\omega _l+M\right){}^2-4 x_{10} m_D^2\right)+3 a_5 \omega _D^2 \left(\omega _l+M\right)\nonumber\\
&+3 a_5 \omega _D^3+3 a_5 M \omega _l^2+3 a_5 \omega _l^3-3 a_5 M^3\big)+a_5 V_1 \left(2 \omega _D \left(\omega _l+M\right)+\omega _D^2+\omega _l^2-M^2\right)\big)\big]\Big)\bigg\},
\end{align}
\begin{align*}
\widetilde C_6(k_t)=&\frac{1}{4\omega_l\omega_D(M^2-(\omega_l+\omega_D)^2)m_D^2}\int\frac{d^3q_t}{(2\pi)^3}\bigg\{\Big(V_2 \big((k_t\cdot q_t) \big[\omega _l \big(2 \omega _D \left(\omega _l+M\right)+\omega _D^2\nonumber\\
&+\omega _l^2-M^2\big)-m_l \left(\omega _D^2-\omega _l^2+M^2\right)\big]+k_t^2 \big[\omega _D^2 \left(\omega _l+M\right)-\omega _D \left(\omega _l+M\right){}^2+\omega _D^3\nonumber\\
&+\left(\omega _l+M\right){}^2 \left(M-\omega _l\right)\big]\big)+k_t^2 V_1 \left(\omega _D^2-\omega _l^2+M^2\right)\Big)\frac{1}{-k_t^2}\widetilde C_1(q_t)\nonumber\\
&+\Big((k_t\cdot q_t) \big(m_l \big[V_2 \left(\omega _D^2 \left(\omega _l+M\right)-\omega _D \left(\omega _l+M\right){}^2+\omega _D^3+\left(\omega _l+M\right){}^2 \left(M-\omega _l\right)\right)\nonumber\\
&-V_1 \left(\omega _D^2-\omega _l^2+M^2\right)\big]-\omega _l \big[V_1 \left(-2 \omega _D \left(\omega _l+M\right)-\omega _D^2-\omega _l^2+M^2\right)\nonumber\\
&+V_2 \left(\omega _D^2 \left(\omega _l+M\right)+\omega _D \left(\omega _l+M\right){}^2+\omega _D^3-\left(M-\omega _l\right) \left(\omega _l+M\right){}^2\right)\big]\big)\nonumber\\
&+k_t^2 V_2 q_t^2 \left(\omega _D^2-\omega _l^2+M^2\right)\Big)\frac{1}{k_t^2}\widetilde C_2(q_t)\nonumber\\
&+\Big(-V_2 \left(\omega _D^2-\omega _l^2+M^2\right)(k_t\cdot q_t)^2+(k_t\cdot q_t) \big[V_1 \big(\omega _D^2 \left(\omega _l+M\right)-\omega _D \left(\omega _l+M\right){}^2\nonumber\\
&-2 M m_D^2+\omega _D^3+\left(\omega _l+M\right){}^2 \left(M-\omega _l\right)\big)-V_2 m_D^2 \big(2 \omega _D \omega _l+\omega _D^2-2 M m_l+\omega _l^2\nonumber\\
&+M^2\big)\big]+a_5 k_t^2 V_2 \big[\omega _l \big(-\omega _l \left(2 m_D^2+M^2\right)+\omega _D \left(\left(\omega _l+M\right){}^2-2 m_D^2\right)+\omega _D^2 \left(\omega _l+M\right)\nonumber\\
&+\omega _D^3+M \omega _l^2+\omega _l^3-M^3\big)+m_l \big(-\omega _D^2 \left(\omega _l+M\right)+\omega _D \left(\omega _l+M\right){}^2+2 M m_D^2-\omega _D^3\nonumber\\
&-\left(M-\omega _l\right) \left(\omega _l+M\right){}^2\big)\big]\Big)\frac{1}{p_t^2}\widetilde C_3(q_t)\nonumber\\
&+\Big(a_5 k_t^2 V_2 \big(m_l \big[m_D^2 \left(\omega _D^2-\omega _l^2+M^2\right)-\omega _D^3 \left(\omega _l+M\right)+\omega _D \left(\omega _l+M\right){}^3-\omega _D^4\nonumber\\
&-\left(M-\omega _l\right) \left(\omega _l+M\right){}^3\big]+\omega _l \big[m_D^2 \left(-2 \omega _D \left(\omega _l+M\right)-\omega _D^2-\omega _l^2+M^2\right)\nonumber\\
&+\omega _D^3 \left(\omega _l+M\right)+\omega _D \left(\omega _l+M\right){}^3+\omega _D^4-\left(M-\omega _l\right) \left(\omega _l+M\right){}^3\big]\big)\nonumber\\
&-\big(V_1 \left(\omega _D^2-\omega _l^2+M^2\right)+V_2 \big[\omega _D^2 \left(\omega _l+M\right)-\omega _D \left(\omega _l+M\right){}^2+\omega _D^3\nonumber\\
&+\left(\omega _l+M\right){}^2 \left(M-\omega _l\right)\big]\big) (k_t\cdot q_t)^2\Big)\frac{1}{k_t^2}\widetilde C_4(q_t)\nonumber\\
&+\Big(V_2 q_t^2 \big[m_D^2 \left(\omega _D^2-\omega _l^2+M^2\right)-\omega _D^3 \left(\omega _l+M\right)+\omega _D \left(\omega _l+M\right){}^3-\omega _D^4\nonumber\\
&-\left(M-\omega _l\right) \left(\omega _l+M\right){}^3\big]+a_5 k_t^2 m_l \big[V_2 \big(\omega _D^2 \left(\omega _l+M\right)-\omega _D \left(\omega _l+M\right){}^2\nonumber\\
\end{align*}
\begin{align}\label{C16}
&+\omega _D^3+\left(\omega _l+M\right){}^2 \left(M-\omega _l\right)\big)-V_1 \left(\omega _D^2-\omega _l^2+M^2\right)\big]-a_5 k_t^2 \omega _l \big[V_1 \big(-2 \omega _D \left(\omega _l+M\right)-\omega _D^2\nonumber\\
&-\omega _l^2+M^2\big)+V_2 \left(\omega _D^2 \left(\omega _l+M\right)+\omega _D \left(\omega _l+M\right){}^2+\omega _D^3-\left(M-\omega _l\right) \left(\omega _l+M\right){}^2\right)\big]\Big)\frac{k_t\cdot q_t}{k_t^2}\widetilde C_5(q_t)\nonumber\\
&+\Big(a_5 k_t^2 m_l \big[V_2 \left(\omega _D^2-\omega _l^2+M^2\right) \left((k_t\cdot q_t)+m_D^2\right)+V_1 \big(\omega _D^2 \left(\omega _l+M\right)-\omega _D \left(\omega _l+M\right){}^2\nonumber\\
&-2 M m_D^2+\omega _D^3+\left(\omega _l+M\right){}^2 \left(M-\omega _l\right)\big)\big]-V_2 (k_t\cdot q_t) \big[q_t^2 \big(\omega _D^2 \left(\omega _l+M\right)-\omega _D \left(\omega _l+M\right){}^2\nonumber\\
&-2 M m_D^2+\omega _D^3+\left(\omega _l+M\right){}^2 \left(M-\omega _l\right)\big)+a_5 k_t^2 \omega _l \left(2 \omega _D \left(\omega _l+M\right)+\omega _D^2+\omega _l^2-M^2\right)\big]\nonumber\\
&+a_5 k_t^2 \big[2 M V_2 k_t^2 m_D^2-\omega _l \big(V_2 m_D^2 \left(2 \omega _D \left(\omega _l+M\right)+\omega _D^2+\omega _l^2-M^2\right)+V_1 \big(-\omega _l \left(2 m_D^2+M^2\right)\nonumber\\
&+\omega _D \left(\left(\omega _l+M\right){}^2-2 m_D^2\right)+\omega _D^2 \left(\omega _l+M\right)+\omega _D^3+M \omega _l^2+\omega _l^3-M^3\big)\big)\big]\Big)\frac{1}{k_t^2}\widetilde C_6(q_t)\nonumber\\
&+\Big(-2 V_2 \left(\omega _D^2-\omega _l^2+M^2\right) (k_t\cdot q_t)^2+a_5 k_t^2 V_2 \big[-k_t^2\left(\omega _D^2-\omega _l^2+M^2\right)-m_l \big(\omega _D^2 \left(\omega _l+M\right)\nonumber\\
&-\omega _D \left(\omega _l+M\right){}^2+\omega _D^3+\left(\omega _l+M\right){}^2 \left(M-\omega _l\right)\big)+\omega _l \big(\omega _D^2 \left(\omega _l+M\right)+\omega _D \left(\omega _l+M\right){}^2+\omega _D^3\nonumber\\
&-\left(M-\omega _l\right) \left(\omega _l+M\right)^2\big)\big]+(k_t\cdot q_t) \big[V_1 \big(-m_l \left(\omega _D^2-\omega _l^2+M^2\right)+\omega _D \left(\omega _l^2-M^2\right)\nonumber\\
&+\omega _D^2 \left(2 \omega _l+M\right)+\omega _D^3-M \omega _l^2+M^3\big)+V_2 \big(4 m_D^2 \left(\omega _D^2-\omega _l^2+M^2\right)+m_l \big(\omega _D^2 \left(\omega _l+M\right)\nonumber\\
&-\omega _D \left(\omega _l+M\right){}^2+\omega _D^3+\left(\omega _l+M\right){}^2 \left(M-\omega _l\right)\big)+8 M^2 \omega _D \omega _l+7 M \omega _D \omega _l^2-M \omega _D^2 \omega _l\nonumber\\
&+2 \omega _D \omega _l^3-\omega _D^2 \omega _l^2-4 \omega _D^3 \omega _l+3 M^3 \omega _D-3 M \omega _D^3-3 \omega _D^4-5 M^3 \omega _l+M^2 \omega _l^2+5 M \omega _l^3\nonumber\\
&+2 \omega _l^4-3 M^4\big)\big]\Big)\frac{1}{k_t^2}\widetilde C_7(q_t)+\Big(3 a_5 k_t^2 V_2 \omega _l^5+a_5 k_t^2 V_1 \omega _l^4+6 a_5 M k_t^2 V_2 \omega _l^4+3 a_5 k_t^2 V_2 \omega _D \omega _l^4\nonumber\\
&+a_5 M k_t^2 V_1 \omega _l^3-6 a_5 k_t^2 m_D^2 V_2 \omega _l^3+4 a_5 M k_t^2 (k_t\cdot q_t) V_2 \omega _D \omega _l+(k_t\cdot q_t) q_t^2 V_2 \omega _D^3\nonumber\\
&+2 a_5 k_t^2 (k_t\cdot q_t) V_2 \omega _l^3+a_5 k_t^2 V_1 \omega _D \omega _l^3+9 a_5 M k_t^2 V_2 \omega _D \omega _l^3+3 a_5 k_t^2 V_2 \omega _D^3 \omega _l^2+a_5 k_t^2 V_1 \omega _D^2 \omega _l^2\nonumber\\
&-a_5 M^2 k_t^2 V_1 \omega _l^2-6 a_5 M^3 k_t^2 V_2 \omega _l^2-M (k_t\cdot q_t) q_t^2 V_2 \omega _l^2+2 a_5 M k_t^2 V_1 \omega _D \omega _l^2+9 a_5 M^2 k_t^2 V_2 \omega _D \omega _l^2\nonumber\\
&+(k_t\cdot q_t) q_t^2 V_2 \omega _D \omega _l^2-12 a_5 k_t^2 m_D^2 V_2 \omega _D \omega _l^2+4 a_5 k_t^2 (k_t\cdot q_t) V_2 \omega _D \omega _l^2+3 a_5 k_t^2 V_2 \omega _D^4 \omega _l\nonumber\\
&+a_5 k_t^2 V_1 \omega _D^3 \omega _l+3 a_5 M k_t^2 V_2 \omega _D^3 \omega _l+a_5 M k_t^2 V_1 \omega _D^2 \omega _l+2 (k_t\cdot q_t) q_t^2 V_2 \omega _D^2 \omega _l-6 a_5 k_t^2 m_D^2 V_2 \omega _D^2 \omega _l\nonumber\\
&+2 a_5 k_t^2 (k_t\cdot q_t) V_2 \omega _D^2 \omega _l-a_5 M^3 k_t^2 V_1 \omega _l-3 a_5 M^4 k_t^2 V_2 \omega _l+6 a_5 M^2 k_t^2 m_D^2 V_2 \omega _l\nonumber\\
&-2 a_5 M^2 k_t^2 (k_t\cdot q_t) V_2 \omega _l+a_5 M^2 k_t^2 V_1 \omega _D \omega _l+3 a_5 M^3 k_t^2 V_2 \omega _D \omega _l-12 a_5 M k_t^2 m_D^2 V_2 \omega _D \omega _l\nonumber\\
&+M^3 (k_t\cdot q_t) q_t^2 V_2+M (k_t\cdot q_t) q_t^2 V_2 \omega _D^2-M^2 (k_t\cdot q_t) q_t^2 V_2 \omega _D-a_5 k_t^2 k_t^2 \big[V_1 \left(M^2+\omega _D^2-\omega _l^2\right)\nonumber\\
&+V_2 \big(\omega _D^3+\left(M+\omega _l\right) \omega _D^2-\left(M+\omega _l\right){}^2 \omega _D+\left(M-\omega _l\right) \left(M+\omega _l\right){}^2\big)\big]\nonumber\\
&-m_l \big[a_5 V_1 \big(\omega _D^3+\left(M+\omega _l\right) \omega _D^2-\left(M+\omega _l\right){}^2 \omega _D+\left(M-\omega _l\right) \left(M+\omega _l\right){}^2\big) k_t^2\nonumber\\
&+V_2 \big(a_5 \big(3 \big(\omega _D^4+\left(M+\omega _l\right) \omega _D^3-\left(M+\omega _l\right){}^3 \omega _D+\left(M-\omega _l\right) \left(M+\omega _l\right){}^3\big)\nonumber\\
&-4 m_D^2 \left(M^2+\omega _D^2-\omega _l^2\right)\big) k_t^2+(k_t\cdot q_t) \left(2 a_5 k_t^2+q_t^2\right)\left(M^2+\omega _D^2-\omega _l^2\right)\big)\big]\Big)\frac{1}{k_t^2}\widetilde C_8(q_t)\bigg\},
\end{align}
\begin{align}\label{C17}
\widetilde C_7(k_t)=&\frac{1}{4\omega_l\omega_D(M^2-(\omega_l+\omega_D)^2)m_D^2}\int\frac{d^3q_t}{(2\pi)^3}\bigg\{-2 V_2 m_D^2 \big[\omega _l \left(\omega _D+\omega _l\right)+M m_l\big]\widetilde C_1(q_t)\nonumber\\
&+2 M V_2 m_D^2 (k_t\cdot q_t)\widetilde C_2(q_t)+\Big(V_2 m_D^2 (k_t\cdot q_t) \big[m_l \left(\omega _D^2-\omega _l^2+M^2\right)\nonumber\\
&+\omega _l \left(2 \omega _D \left(\omega _l+M\right)+\omega _D^2+\omega _l^2-M^2\right)\big]\Big)\frac{1}{k_t^2}\widetilde C_3(q_t)\nonumber\\
&-2 V_2 m_D^2 \big[\omega _l \left(\omega _D+\omega _l\right)+M m_l\big]\frac{(k_t\cdot q_t)^2}{k_t^2}\widetilde C_4(q_t)+2 M V_2 m_D^2 (k_t\cdot q_t)^3\frac{1}{k_t^2}\widetilde C_5(q_t)\nonumber\\
&-V_2 m_D^2 \left(\omega _D^2-\omega _l^2+M^2\right)\frac{(k_t\cdot q_t)^2}{k_t^2}\widetilde C_6(q_t)-m_D^2 (k_t\cdot q_t) \big[-2 M V_2 k_t^2 \left(k_t\cdot q_t+2 k_t^2\right)\nonumber\\
&+k_t^2 m_l \left(V_2 \left(\omega _D^2-\omega _l^2+M^2\right)+2 M V_1\right)+k_t^2 \omega _l \big(3 V_2 \big(2 \omega _D \left(\omega _l+M\right)+\omega _D^2+\omega _l^2\nonumber\\
&-M^2\big)+2 V_1 \left(\omega _D+\omega _l\right)\big)\big]\frac{1}{k_t^4}\widetilde C_7(q_t)+m_D^2 (k_t\cdot q_t) \big[k_t^2 (k_t\cdot q_t) \big(V_2 \big(3 \left(\omega _D^2-\omega _l^2+M^2\right)\nonumber\\
&-4 M m_l\big)+2 M V_1\big)-2 k_t^2 V_2 q_t^2 \left(\omega _l \left(\omega _D+\omega _l\right)+M m_l\right)\big]\frac{1}{k_t^4}\widetilde C_8(q_t)\bigg\},
\end{align}
\begin{align}\label{C18}
\widetilde C_8(k_t)=&\frac{1}{4\omega_l\omega_D(M^2-(\omega_l+\omega_D)^2)m_D^2}\int\frac{d^3q_t}{(2\pi)^3}\bigg\{2 M V_2 m_D^2\widetilde C_1(q_t)-2 V_2 m_D^2\big[\omega _l \left(\omega _D+\omega _l\right)\nonumber\\
&+M m_l\big]\frac{k_t\cdot q_t}{k_t^2}\widetilde C_2(q_t)-V_2 m_D^2 \left(\omega _D^2-\omega _l^2+M^2\right)\frac{k_t\cdot q_t}{k_t^2}\widetilde C_3(q_t)+2 M V_2 m_D^2\frac{(k_t\cdot q_t)^2}{k_t^2}\widetilde C_4(q_t)\nonumber\\
&-2 V_2 m_D^2 \big[\omega _l \left(\omega _D+\omega _l\right)+M m_l\big] (k_t\cdot q_t)^3\frac{1}{k_t^4}\widetilde C_5(q_t)+\Big(V_2 m_D^2 \big[m_l \left(\omega _D^2-\omega _l^2+M^2\right)\nonumber\\
&+\omega _l \left(2 \omega _D \left(\omega _l+M\right)+\omega _D^2+\omega _l^2-M^2\right)\big] (k_t\cdot q_t)^2\Big)\frac{1}{k_t^4}\widetilde C_6(q_t)\nonumber\\
&+m_D^2 (k_t\cdot q_t) \Big[V_2 \big(2 \omega _D \omega _l (k_t\cdot q_t)-2 M m_l \left(k_t\cdot q_t+2 k_t^2\right)+2 \omega _l^2 (k_t\cdot q_t)+k_t^2 \omega _D^2\nonumber\\
&-k_t^2 \omega _l^2+M^2 k_t^2\big)+2 M k_t^2 V_1\Big]\frac{1}{k_t^4}\widetilde C_7(q_t)-m_D^2 (k_t\cdot q_t) \big((k_t\cdot q_t) \big(-4 M V_2 k_t^2\nonumber\\
&+m_l \big[3 V_2 \left(\omega _D^2-\omega _l^2+M^2\right)+2 M V_1\big]+\omega _l \big[3 V_2 \big(2 \omega _D \left(\omega _l+M\right)+\omega _D^2+\omega _l^2-M^2\big)\nonumber\\
&+2 V_1 \left(\omega _D+\omega _l\right)\big]\big)-2 M k_t^2 V_2 q_t^2\big)\frac{1}{k_t^4}\widetilde C_8(q_t)\bigg\}.
\end{align}

\end{appendix}

\end{document}